\documentclass[11pt]{article}
\pdfoutput=1

\usepackage{euscript}
\usepackage{amsfonts}
\usepackage{amsbsy}
\usepackage{epsfig}
\usepackage{amsthm}
\usepackage{amscd}
\usepackage{amstext}
\usepackage{verbatim}
\usepackage{cancel}
\usepackage{capt-of}
\usepackage{empheq}

\usepackage[nosort]{cite}
\usepackage{bm}
\usepackage{authblk}
\usepackage{esint}
\usepackage[T1]{fontenc}
\usepackage{mathdots}
\usepackage[centertableaux]{ytableau}

\usepackage[utf8]{inputenc}
\usepackage{graphicx}
\usepackage{physics}
\usepackage{mathtools}
\usepackage{bbold}

\usepackage{setspace}
\usepackage{colortbl}
\usepackage{xcolor}

\usepackage{amsmath}
\makeatletter
\renewcommand*\env@matrix[1][\arraystretch]{%
  \edef\arraystretch{#1}%
  \hskip -\arraycolsep
  \let\@ifnextchar\new@ifnextchar
  \array{*\c@MaxMatrixCols c}}
\makeatother
\usepackage{amssymb}
\usepackage{mathrsfs}
\usepackage{mathtools}
\usepackage{physics}
\usepackage{booktabs}
\usepackage{bbm}
\usepackage{float}

\usepackage{pgfplots}
\pgfplotsset{compat=1.15}
\usepackage{tikz}
\usetikzlibrary{external}
\usetikzlibrary{arrows}

\usepackage{subcaption}
\usepackage{graphicx}
\usepackage{mathtools}
\usepackage[many]{tcolorbox}
\tcbset{shield externalize}
\usepackage{xcolor, colortbl}
\usepackage[a4paper]{geometry}
\usepackage[hidelinks,linktocpage]{hyperref}
\hypersetup{
    colorlinks=true,
    linkcolor=blue,
    citecolor=blue,
    }

\usepackage{mathrsfs}

\def\ben{\begin{equation}}
\def\een{\end{equation}}

\let\a=\alpha \let\b=\beta  \let\d=\delta

 \let\G=\Gamma

\let\pa=\partial
\def\be{\begin{equation}}
\def\ee{\end{equation}}
\def\beq{\begin{equation}}
\def\eeq{\end{equation}}
\def\ba{\begin{array}}
\def\ea{\end{array}}

\def\dalemb#1#2{{\vbox{\hrule height .#2pt
       \hbox{\vrule width.#2pt height#1pt \kern#1pt
               \vrule width.#2pt}
       \hrule height.#2pt}}}

\newcommand{\bea}{\begin{eqnarray}}
\newcommand{\eea}{\end{eqnarray}}

\makeatletter
\newcommand*\bigcdot{\mathpalette\bigcdot@{.5}}
\newcommand*\bigcdot@[2]{\mathbin{\vcenter{\hbox{\scalebox{#2}{$\m@th#1\bullet$}}}}}
\makeatother

\renewcommand{\Bar}[1]{\overline{#1}}

\def\R{{{\mathbb R}}}

\title{The Polarised IKKT Matrix Model}
\author{Sean~A.~Hartnoll and Jun~Liu}
\affil{\it Department of Applied Mathematics and Theoretical Physics, \\
\it University of Cambridge, Cambridge CB3 0WA, UK
}
\date{}

\begin{document}

\maketitle

\begin{abstract}

We establish a correspondence between a supersymmetric mass deformation of the IKKT matrix integral at large $N$ and a background of Euclidean type IIB string theory. Both sides have sixteen supersymmetries and an $SO(3)\times SO(7)$ symmetry. In the limit of large mass the integral is dominated by a fuzzy sphere saddle point. This saddle corresponds to a Euclidean $D1$-brane in a finite, Euclidean, ellipsoidal cavity. The cavity is supported by three-form NSNS flux that polarises $N$ $D$-instantons into the $D1$-brane. We furthermore use supersymmetric localisation to show that the deformed matrix integral can be reduced to a moduli space integral, allowing exact results away from the large mass limit. At small mass the $D1$-branes can backreact on the geometry, and we discuss
the possible formulation of a `timeless' holography in such regimes.

\end{abstract}

\newpage

\tableofcontents

\section{Introduction}

The quantum mechanics of supersymmetric large $N$ matrices underpins the holographic emergence of semiclassical gravity \cite{tHooft:1973alw, deWit:1988wri, Banks:1996vh, Maldacena:1997re}. These models come with a conventional quantum mechanical time, which might appear to be in tension with the absence of a preferred time in gravity. In the best-understood constructions this tension is resolved by appending a non-gravitating boundary to the system. The quantum mechanical clock is placed at this boundary. The gravitational Hilbert space is then reconstructed from the boundary one, see e.g.~\cite{Witten:2022xxp}.

An asymptotic boundary is not always available. For example, for cosmologies or observers infalling into black holes. Furthermore, away from the special features of asymptotically anti-de Sitter spacetimes, incorporating boundaries consistently into theories of gravity is a subtle issue that remains under active investigation, e.g.~\cite{Anderson:2006lqb, Witten:2018lgb, Anninos:2023epi, Liu:2024ymn}. One may wish, in parallel, to consider frameworks in which the microscopic theory of supersymmetric large $N$ matrices does not come with its own time. This is the IKKT model \cite{Ishibashi:1996xs}, a simple integral over ten hermitian $N\times N$ bosonic matrices and their supersymmetric partners.

The absence of a time or a Hilbert space in the IKKT model has meant that it has not yet, in our opinion, found its home within the contemporary understanding of holographic dualities. There do exist, however, holographic models of timeless theories, such as the matrix model duals for two dimensional quantum gravity coupled to $c<1$ matter \cite{Kazakov:1986hu}, see \cite{Anninos:2020ccj} for a recent review.

The Euclidean IKKT model can be thought of as the `world-point' theory of $N$ $D$-instantons.
$D$-instanton calculus in IIB string theory \cite{Green:1997tn} motivated the exact evaluation of the Euclidean IKKT integral by supersymmetric localisation \cite{Moore:1998et}, as well as by numerical Monte Carlo simulations \cite{Krauth:1998xh}. $D$-instantons may also be described as 
a Euclidean background of type IIB string theory \cite{Gibbons:1995vg, Bergshoeff:1998ry}. This background admits a scaling similarity transformation with some resemblance to more established models of holography \cite{Biggs:2023sqw}. Despite these many promising results, it has remained unclear how a holographic dictionary could be formulated.

In this paper we will gather some data points that, we argue, can guide us towards a timeless holography for type IIB string theory. We do this by studying a mass deformation of the IKKT model that preserves its supersymmetries. The deformed model was introduced in an appendix of \cite{Bonelli:2002mb} twenty years ago and, to our knowledge, only studied in one other work since \cite{Kumar:2022giw}. This deformation is a close cousin of the BMN mass deformation \cite{Berenstein:2002jq} of the supersymmetric BFSS matrix quantum mechanics \cite{Banks:1996vh}. It is also closely related to the ${\mathcal N} = 1^\ast$ mass deformation of ${\mathcal N} = 4$ SYM theory \cite{Vafa:1994tf, Polchinski:2000uf}. The spacetime physics of this class of worldvolume mass terms is well established, they capture the polarisation of $D$-branes in the presence of a background flux \cite{Myers:1999ps, Polchinski:2000uf, Berenstein:2002jq, Lin:2004kw}.
We will see that this is also the case in our setting; for this reason we refer to the deformed theory as the polarised IKKT model. We define the polarised model in \S\ref{sec:model} below.

The IKKT model has no tunable coupling constant. The mass deformation introduces a dimensionless coupling $\Omega$. The large $\Omega$ matrix integral can be performed by saddle point, which we do in \S\ref{semicl} to obtain (\ref{pertresult}). The dominant saddle is a fuzzy sphere. This fuzzy sphere can equivalently be viewed as a spherical probe Euclidean $D1$-brane, carrying $N$ units of $D$-instanton charge, in a certain Euclidean background of type IIB string theory. We construct the background and the probe brane explicitly in \S\ref{sec:grav}, the setup is illustrated in Fig.~\ref{fig:cavity}, these preserve the same supersymmetries as the polarised IKKT model. In this way, the large $\Omega$ limit is a handle on the system that allows elements of a dictionary to be established between the matrix and spacetime descriptions. In particular, $\Omega$ corresponds to the strength of a background NSNS three-form flux according to (\ref{eq:match}).

The power of holographic dualities \cite{Maldacena:1997re} lies in taking a different limit, in which the matrices become strongly fluctuating and, simultaneously, the $D$-branes backreact on the spacetime. We take a first step in this direction by mapping out the physics of the IIB background as a function of the strength of the three-form flux and of $g_\text{s} N$, which controls the strength of gravitational backreaction ($g_\text{s}$ is the string coupling). This is the content of Fig.~\ref{fig:phases} below. The probe $D1$-brane backreacts when $g_\text{s} N$ becomes large or the three-form flux becomes small. The backreacting regimes necessarily correspond to small $\Omega$ in the matrix integral, see Fig.~\ref{fig:phasediag}, and are hence indeed in the strongly fluctuating regime. We have not constructed the backreacted geometries in this paper. Given that they will have a large amount of symmetry and supersymmetry, this should be achievable in the near future.

A promising aspect of IKKT holography is that the matrix integral is far more tractable than the quantum mechanical path integrals of other setups. In particular, it may be feasible to perform the mass deformed integral at all values of $\Omega$. One can hope to match such results with the backreacted geometries. To this end, in \S\ref{sec:loc} we set up a framework for supersymmetric localisation of the polarised IKKT integral. We show that the full matrix integral can be reduced to a moduli space integral over fuzzy sphere configurations together with a `collapsed' matrix configuration. We demonstrate the correctness of the method by re-deriving our previous large $\Omega$ results, including the precise one-loop correction about the saddle. We furthermore obtain the $N=2$ fuzzy sphere contribution at all $\Omega$ in (\ref{eq:N2}). There is much future work to be done here.

In the final \S\ref{sec:disc} we comment on several conceptual and technical questions raised by our work. A key point is the interpretation of the backreaction scale $g_\text{s} N$ in the matrix integral. In conventional holographic models this quantity defines an energy scale in the dual field theory, allowing different bulk radial locations to be identified with different points in the field theory RG \cite{Itzhaki:1998dd}.
In our timeless model we argue that instead of zooming in on different energy scales on the matrix side, which no longer exist, the value of $g_\text{s} N$ instruct us to zoom in on a certain range of the matrix integral coordinates. We go on to discuss the use of timeless holography as a framework for quantum cosmology, the possibility of spontaneous symmetry breaking, and possible connections between the fact that our IIB background is a finite cavity, see Fig.~\ref{fig:cavity}, and fluctuations in the $D$-instanton number.

\section{The polarised IKKT model and its supersymmetries}
\label{sec:model}

The action of the Euclidean IKKT matrix model is given by a trace over $N\times N$ matrices,
\begin{equation}\label{eq:ikkt}
    S_\text{IKKT} = \text{Tr} \left( -\frac{1}{4} [X_\mu, X_\nu] [X^\mu, X^\nu] + \frac{1}{2} \Psi_\alpha  \left(\mathcal{C} \Gamma^\mu\right)_{\alpha \beta} [X_\mu, \Psi_\beta] \right) \,.
\end{equation}
Here $X_\mu$ are ten hermitian matrices transforming as a vector under $SO(10)$. There is no difference between raised and lowered indices. The $\Psi_\alpha$ are matrices of 32-component Weyl spinors, satisfying $\Gamma^{11}_{\a\b} \Psi_\b = \Psi_\a$. Our conventions for the gamma matrices $\Gamma^\mu$ and charge conjugation matrix $\mathcal{C}$ are described in appendix \ref{gam}. To obtain a supersymmetric theory it is important that the hermitian conjugate of $\Psi_\alpha$ does not appear in the action. This is a Euclidean manifestation of the fact that the Lorentzian theory, which has an $SO(1,9)$ symmetry, has Majorana-Weyl spinors.
In Euclidean signature one cannot impose a reality condition because $SO(10)$ does not have a Majorana-Weyl representation. Following previous literature (cf. \cite{Pestun:2007rz, Krauth:1998xh, Moore:1998et}) the fact that only $\Psi_\a$ appears in the action \eqref{eq:ikkt}, without its hermitian conjugate, means that it is possible to restrict the integral over $\Psi_\a$ to hermitian matrices. In the chiral basis, each component of these hermitian matrices consists of 16 non-zero Grassmann-odd numbers.

The model (\ref{eq:ikkt}) can be viewed as the complete dimensional reduction of ten-dimensional super Yang-Mills (SYM) theory to zero dimensions. It inherits sixteen supersymmetries, under which the matrices transform as
\begin{equation}\label{ikktsusy}
    \delta X^\mu = - \Psi_\alpha (\mathcal{C} \Gamma^\mu)_{\alpha \beta} \epsilon_\beta \,, \qquad \delta \Psi_\alpha = \frac{1}{2} \Gamma^{\mu \nu}_{\alpha \beta} [X_\mu, X_\nu] \epsilon_\beta \,. 
\end{equation}
The infinitesimal spinor parameters $\epsilon_\alpha$ obey the same chirality condition as the $\Psi_\alpha$ matrices. They are taken to be Grassmann-even, and hence $\delta$ is a Grassmann-odd transformation. We take the matrices to be valued in $SU(N)$ and hence traceless.\footnote{For the case of $U(N)$ there are additional symmetries acting on the trace. The kinematic supersymmetries act as a shift of the fermions proportional to the identity matrix, $\delta X^\mu = 0\,, \delta \Psi_\alpha = \eta_\alpha \mathbb{1}$. The corresponding bosonic symmetries shift the $D$-instanton collective coordinates, $\delta X_\mu = x_\mu \mathbb{1} \,, \delta \Psi_\alpha = 0$.} As noted above, one cannot impose a reality condition on Euclidean chiral spinors. We have restricted the integral to hermitian matrices, but the supersymmetry transformations above do not preserve this hermiticity. We will see that this is not a problem for the two uses we make of supersymmetry: finding the dual geometry in \S\ref{sec:grav} and performing a localisation computation \S\ref{sec:loc}.

The polarised IKKT model is an $SO(3) \times SO(7)$ invariant mass deformation of (\ref{eq:ikkt}), written down by \cite{Bonelli:2002mb}:
\begin{equation}\label{eq:fullS}
    S_\Omega = S_\text{IKKT} + \text{Tr}\left( \frac{\Omega^2}{4^3} X_A^2 + \frac{3\Omega^2}{4^3} X_a^2 + i \Omega X_8 [X_9, X_{10}] + \frac{i \Omega}{8} \Psi_\alpha (\mathcal{C} \hat{N})_{\alpha \beta} \Psi_\beta \right),
\end{equation}
where $A = 1,\ldots,7$, $a = 8,9,10$ and $\hat{N} = - \Gamma^8 \Gamma^9 \Gamma^{10}$.
The deformation \eqref{eq:fullS} preserves the sixteen supersymmetries in \eqref{ikktsusy}, which also get deformed so that 
\begin{equation}\label{eq:masssusy}
    \delta X^\mu = - \Psi_\alpha (\mathcal{C} \Gamma^\mu)_{\alpha \beta} \epsilon_\beta \,, \qquad
    \delta \Psi_\alpha = \frac{1}{2} \Gamma^{\mu \nu}_{\alpha \beta} [X_\mu, X_\nu] \epsilon_\beta + \frac{i \Omega }{8} T^\mu_{\alpha \beta} X_\mu \epsilon_\beta \,,
\end{equation}
where $T^\mu = \G^\mu \hat{N} + 2 \hat{N} \G^\mu$. As with \eqref{ikktsusy} previously, these transformations leave the action invariant but only close on-shell. Off-shell supersymmetry will be discussed in \S\ref{sec:loc} below.

This paper will be concerned with the matrix integral
\be\label{eq:Z}
Z_N[\Omega] = \int dX_\mu d\Psi_\alpha e^{-S_\Omega} \,.
\ee
An overall coupling of $1/g^2$ in the exponent can be introduced by rescaling $X \to X/g^{1/2}$, $\Psi \to \Psi/g^{3/4}$ and $\Omega \to \Omega/g^{1/2}$. The measure is then rescaled by $dX_\mu d\Psi_\alpha \to g^{7(N^2-1)} dX_\mu d\Psi_\alpha$, where the ten bosonic matrices and sixteen fermionic matrices contribute $-10 \times \frac{1}{2} + 16 \times \frac{3}{4} = 7$, each with $N^2-1$ real entries. For our purposes there is no reason to scale in a redundant variable, so we will work without this factor. The absence of a meaningful Yang-Mills coupling constant is a key difference between the present zero-dimensional theory and quantum mechanical theories that have a worldvolume time and, in general, also space dimensions. In those cases the Yang-Mills coupling is either dimensionless (in four spacetime dimensions) or defines an energy scale in the theory. This energy scale is used to map different regions of the bulk spacetime, which experience varying redshift, to different energies in the matrix theory \cite{Itzhaki:1998dd}. In
\S\ref{sec:disc} we will suggest a way to encode the bulk spacetime in our timeless matrix theory.

\section{Solution of the model at large mass}\label{semicl}

In the large mass limit the matrix integral (\ref{eq:Z}) is dominated by small fluctuations about a saddle point. In this section we
will evaluate both the saddle point action and the `one-loop' contribution of Gaussian fluctuations.
The saddle point action will be matched with a dual gravitational computation in \S\ref{sec:grav} while the one-loop part will be recovered using supersymmetric localisation in \S\ref{sec:loc}, offering a nontrivial check of all three computations.

The remainder of this section is somewhat technical, even with many details moved to the appendices. The final one-loop result for the matrix integral at large mass is
\begin{equation}\label{pertresult}
\begin{aligned}
    Z_{\Omega\to \infty}
    = 2^{3(N^2-1)}\frac{(2\pi)^{\frac{(10N + 11)(N-1)}{2}}}{N^{\frac{3}{2}} G(N+1)}\prod_{l=1}^{N-1} \left(\frac{3l+2}{3l+1} \right)^3 e^{\frac{9 \Omega^4}{2^{15}} N (N^2 - 1)} \,.
\end{aligned}
\end{equation}
This expression follows from combining equations \eqref{S0}, \eqref{Z7}, \eqref{Z3}, and \eqref{Zf} below. The Barnes $G$ function $G(n) = \prod_{m=1}^{n-2} m!$. The remaining product over $l$ in \eqref{pertresult} can be written in terms of gamma functions but this does not increase the clarity of the expression. We may note that there are no factors of $\Omega$ in the coefficient of the exponential.

The large $\Omega$ perturbation theory leading to (\ref{pertresult}) is valid for
\be\label{eq:masslimit}
\Omega^4 \gg N \,.
\ee
This condition, requiring the large $\Omega$ limit to be taken before the large $N$ limit, is similar to that allowing large mass perturbation theory in BMN matrix quantum mechanics \cite{Dasgupta:2002hx}.
We demonstrate (\ref{eq:masslimit}) in appendix \ref{app:fieldtheory} by mapping the fluctuations about the fuzzy sphere saddle to a noncommutative Maxwell theory on a sphere. The noncommutativity introduces higher derivative interactions, whose contribution to the partition function is controlled by the limit (\ref{eq:masslimit}).

\subsection{Saddle point action}

The saddle point equations of motion are
\begin{align}
    [X_\nu, [X^\nu, X_A]] + \frac{2 \Omega^2}{4^3} X_A &= 0\,, \qquad A \in \{1,\ldots,7\} \,, \\
    [X_\nu, [X^\nu, X_a]] + \frac{6 \Omega^2}{4^3} X_a + i \frac{\Omega}{2} \epsilon_{abc}[X_b,X_c] &= 0\,, \qquad a, b, c \in \{8, 9, 10\}\,,   
\end{align}
where $\epsilon_{abc}$ is the totally antisymmetric Levi-Civita symbol with $\epsilon_{8910} = 1$. The first equation has only trivial solutions $X_A = 0$. This can be seen by multiplying through by $X_A$ (no summation over $A$) and taking the trace, which gives
\begin{align}\label{eq:Aeq}
  - \text{Tr}\left([X_\nu, X_A][X^\nu, X_A]\right) + \frac{2 \Omega^2}{4^3} \text{Tr}\left(X_A^2 \right) &= 0 \,.
\end{align}
Since the $X$'s are hermitian, both terms in \eqref{eq:Aeq} are positive semi-definite. The only solution is then if both terms vanish, which requires $X_A = 0$. The equations of motion become
\begin{gather}
   X_A = 0 \,,\\
   [X_b, [X_b, X_a]] + \frac{6 \Omega^2}{4^3} X_a + i \frac{\Omega}{2} \epsilon_{abc}[X_b,X_c] = 0 \,.  \label{eq:eom1}
\end{gather}
Equation \eqref{eq:eom1} can be multiplied by $X_a$, traced and factorised to obtain
\be\label{eq:factor}
\Tr \left[\left([X_a, X_b] - i \frac{3 \Omega}{8} \epsilon_{abc} X_c \right)\left([X_a, X_b] - i \frac{\Omega}{8} \epsilon_{abd} X_d \right)\right] = 0 \,.
\ee
This expression makes two solutions manifest, these are respectively minima and maxima of the action. We believe that the only nontrivial solutions to (\ref{eq:eom1}) that are minima of the action are matrices obeying the commutation relations
\begin{equation}\label{eom}
    [X_a, X_b] = i \frac{3 \Omega}{8} \epsilon_{abc} X_c \,.
\end{equation}
This assumption is consistent with the results in \S\ref{sec:loc} below, as the localisation computation only gets contributions from matrices obeying (\ref{eom}), together with a `collapsed' contribution.

The commutation relation in \eqref{eom} is solved by
\begin{equation}\label{eq:background}
    X_a = \frac{3 \Omega}{8} J_a \,,
\end{equation}
where $J_a$ form an $N$-dimensional representation of the $SU(2)$ Lie algebra. In (\ref{eq:background}) the index `$a$' is conflated between $8,9,10$ on the left and $1,2,3$ on the right. We may write $J_a$ as a direct sum of irreducible representations with dimensions $d_k$, with the constraint $\sum_k d_k = N$. The action \eqref{eq:fullS} evaluated on such a saddle point is 
\begin{equation}
    S_\Omega^{(0)} = - \frac{\Omega^2}{128} \text{Tr}(X_a^2) = - \frac{9 \Omega^4}{32768} \sum_{k} d_k (d_k^2 - 1) \,.
  \label{eq:act0}
\end{equation}
In the last line we have used the fact that on the $k$th irreducible representation $J_a^2 = \frac{1}{4}(d_k^2-1) \mathbb{1}_{d_k}$. The action \eqref{eq:act0} is minimised when there is a single $N$-dimensional irreducible representation.
In the large $\Omega$ limit, the contribution from all other saddles, which are reducible representations, will be exponentially suppressed. Therefore we consider only the irreducible one, wherein
\begin{equation}\label{S0}
    S_\Omega^{(0)} = - \frac{9 \Omega^4}{2^{15}} N (N^2 - 1) \,.
\end{equation}

The saddle we have just described is known as the maximal fuzzy sphere. We will now obtain the contribution to the matrix integral \eqref{eq:Z} from quadratic fluctuations about this saddle.

\subsection{Quadratic fluctuations}\label{sec:quad}

The quadratic action for fluctuations around the maximal fuzzy sphere takes the form
\begin{equation}
    S^{(2)} = S^{(2)}_7(\delta X_A) + S^{(2)}_3 (\delta X_a) + S^{(2)}_{f}(\theta_
    \alpha) \,,
\end{equation}
where $S^{(2)}_7, S^{(2)}_3, S^{(2)}_f$ are, respectively, quadratic actions for the seven transverse bosonic matrices, the three bosonic directions on the fuzzy sphere, and the sixteen fermionic matrices.  Explicitly,
\begin{gather}
    S^{(2)}_7 =  \text{Tr} \left( - \frac{1}{2}[\delta X_A, X_a]^2 + \frac{\Omega^2}{4^3} \delta X_A^2 \right), \label{S7} \\ 
    S^{(2)}_3 = \text{Tr} \left( -\frac{1}{2}( [ X_a, \delta X_b]^2 + [\delta X_a, X_b][X_a, \delta X_b] ) + i \frac{5 \Omega}{16} \epsilon_{abc} X_a [\delta X_b, \delta X_c] + \frac{3 \Omega^2}{4^3} \delta X_a^2 \right), \label{S3} \\ 
    S^{(2)}_f = \text{Tr} \left( \frac{1}{2} \theta_\alpha \Bar{\sigma}^a_{\alpha \beta} [X_a, \theta_\beta] - i \frac{\Omega}{8} \theta_\alpha (\Bar{\sigma}^8 \sigma^9 \Bar{\sigma}^{10} )_{\alpha \beta} \theta_\beta \right), \label{Sf}
\end{gather}
with $X_a = 3 \Omega / 8 \, J_a$. The spinor $\theta$ and the $\sigma, \Bar{\sigma}$ matrices are defined in appendix \ref{gam}.

The next step is to diagonalise the kinetic operators in \eqref{S7}, \eqref{S3} and \eqref{Sf} to obtain independent Gaussian integrals. A convenient basis of matrices with which to do this are matrix spherical harmonics. We will summarise the results in the main text and leave details to appendix \ref{app:modes}. One fact that will be important is that the quantum numbers labelling the matrix harmonics are integers $l,m$ with
\be
1 \leq l \leq N - 1 \,, \qquad -l \leq m \leq l \,.
\ee
Thus $N$ is a cutoff on the angular momentum, this is why the sphere is `fuzzy'. We may note that $l=0$ is the identity and hence does not contribute to traceless perturbations.

Computations along similar lines to those described below have previously been performed for matrix integrals over three bosonic matrices, with and without supersymmetric partners, with a fuzzy sphere saddle in \cite{Iso:2001mg, Azuma:2004zq, Anagnostopoulos:2005cy}.

\subsubsection{Transverse bosonic modes}

From the quadratic action \eqref{S7}, the normal modes of the seven transverse directions satisfy
\begin{equation} \label{S7eigen}
    \frac{1}{2}\left(\frac{3 \Omega}{8}\right)^2 [J_a, [J_a, \hat{H}_{lm}] + \frac{\Omega^2}{4^3} \hat{H}_{lm} = \omega_{l} \hat{H}_{lm}.
\end{equation}
The hats are to remind us that the $\hat H$ are matrices. We have suppressed the spacetime $A$ index on $\hat{H}$ since, from the point of view of the fuzzy sphere worldvolume, these transverse matrices are scalars. In appendix \ref{app:modes} we show that the eigenvalues of \eqref{S7eigen} are
\begin{equation}\label{eq:tranmodes}
    \omega_l = \frac{1}{2} \left(\frac{3 \Omega}{8}\right)^2 l (l+1) + \frac{\Omega^2}{4^3}, \qquad \text{with degeneracy} \;\; 2l+1 \,.
\end{equation}

We must be careful to specify the overall normalisation of the integral (\ref{eq:Z}). The integral is defined to be over coefficients $x_A^M$, such that $X_A = x_A^M T^M$. Here $T^M$ are the $N^2-1$ orthonormal generators of $SU(N)$, in the sense that $\text{Tr}(T^M T^N) = 2 \delta^{MN}$. Each of the $N^2-1$ Gaussian integrals thus comes with a factor of $\sqrt{\pi/2}$.
The `one-loop' contribution from the seven transverse scalar bosons to the partition function is therefore
\begin{equation}\label{Z7}
    Z^{\text{(1-loop)}}_7 = \left[ \left(\frac{\pi}{2} \cdot \frac{64}{\Omega^2}\right)^{(N^2-1)/2} \prod_{l=1}^{N-1} \left(\frac{9}{2} l(l+1) + 1 \right)^{-\frac{(2l+1)}{2}} \right]^7.
\end{equation}

\subsubsection{Bosonic modes on the fuzzy sphere}

From the quadratic action \eqref{S3}, eigenmodes on the fuzzy sphere obey
\begin{equation}\label{S3eigen}
    \frac{1}{2}\left(\frac{3 \Omega}{8}\right)^2 \left([J_b, [J_b, \hat{H}_{ia}]] - [J_b, [J_a, \hat{H}_{ib}]]\right) + i \frac{15 \Omega^2}{128} \epsilon_{abc}[J_b, \hat{H}_{ic}] + \frac{3 \Omega^2}{4^3} \hat{H}_{ia} = \omega_i \hat{H}_{ia}. 
\end{equation}
The normal modes now carry a spacetime index $a$ in the fuzzy sphere directions and an index $i$ (no summation) which labels the $3 (N^2-1)$ modes. The index $i$ will run over three sets of modes, each of which carry quantum numbers $l,m$.

Equation \eqref{S3eigen} will have $N^2-1$ zero modes corresponding to $SU(N)$ conjugations of $X_{8,9,10}$. We will refer to these as gauge transformations. In the partition function, these modes will produce a group volume which we compute exactly below. In \eqref{S3eigen}, the zero modes are
\be
\hat H^{\text{(zero)}}_a = i [J_a, \Lambda] \,,
\ee
with $\Lambda$ a hermitian matrix.

The remaining non-zero modes obey the constraint
\begin{equation}\label{gauge}
    [J_a, \hat H^{\text{(nonzero)}}_a] = 0. 
\end{equation}
Physically \eqref{gauge} is reminiscent of the Lorentz gauge condition that selects out the modes orthogonal to gauge orbits. One can prove from equation \eqref{S3eigen} that all modes with non-zero $\omega_i$ satisfy the constraint (\ref{gauge}); taking the commutator with $J_a$ on both sides of \eqref{S3eigen}, using the Jacobi identity and simplifying, one obtains $0 = \omega_i [J_a, \hat{H}_{ia}]$. We may therefore impose \eqref{gauge} in \eqref{S3eigen} to simplify the calculation of the non-zero spectrum, and at the end check that the modes we obtain satisfy the constraint. Imposing the constraint in \eqref{S3eigen}, and simplifying, we have that for non-zero modes
\begin{equation}\label{simpeigen}
    \frac{3}{2} [J_b, [J_b, \hat{H}_{ia}]] + i \epsilon_{abc}[J_b, \hat{H}_{ic}]  + \hat{H}_{ia} = \lambda_i \hat{H}_{ia} ,
\end{equation}
where we set $\omega_i = 3\Omega^2  \lambda_i /4^3.$

We solve the eigenvalue problem \eqref{simpeigen} in appendix \ref{app:modes}. The eigenvalues are again labelled by the angular momentum quantum number $l$. The result for the spectrum is
\begin{equation}\label{eq:lamresult}
\lambda =
    \begin{cases}
     \frac{1}{2} l(3l+1) \,, & \text{  with degeneracy $2l-1$}\\
      \frac{1}{2} (l+1)(3l+2) \,, & \text{  with degeneracy $2l+3$}
    \end{cases} \,.
\end{equation}
At each $l$ these are to be combined with $2l+1$ zero modes, which gives the expected total of $3(N^2-1)$ modes overall.

The integral over the zero modes must be performed exactly. These parametrise $SU(N)/\mathbb{Z}_N$. The quotient is because the transformation
\begin{equation}\label{eq:resid}
    X_\mu \to U X_\mu U^\dagger , \qquad U \in SU(N) \,,
\end{equation}
leaves $X_\mu$ unchanged if $U$ is $z \mathbb{1}$, where $z^N$ = 1. This means that elements of $SU(N)$ related by a $\mathbb{Z}_N$ transformation do not correspond to different values of the fields and should be identified. Thus we must only integrate over $SU(N)/\mathbb{Z}_N$.

There are thus three contributions to $Z^{\text{(1-loop)}}_3$. Firstly, the contribution from the $2(N^2-1)$ non-zero modes is analogous to that of the transverse modes discussed previously and leads to
\be
    Z^{\text{(1-loop)}}_{3,\text{nonzero}} = \left(\frac{\pi}{2} \cdot \frac{128}{3 \Omega^2}\right)^{N^2-1} \prod_{l=1}^{N-1} \Big(l(3l+1) \Big)^{-\frac{(2l-1)}{2}}\Big((l+1)(3l+2) \Big)^{-\frac{(2l+3)}{2}} \,.
\ee
Secondly there is the contribution of the zero modes, which is a factor of the volume \cite{Boya:2003km}
\be\label{eq:vol}
Z^{\text{(1-loop)}}_{3,\text{zero}} = \text{vol}\left(SU(N)/\mathbb{Z}_N\right) = \frac{2^{(N-1)/2} \pi^{(N-1)(N+2)/2}}{\sqrt{N}} \prod_{k=1}^{N-1} \frac{1}{k!} \,.
\ee
The quotient by $\mathbb{Z}_N$ simply amounts to dividing by $N$.
Finally, there is a Jacobian term that is needed to transform the integral over zero modes to the integral over angles on $SU(N)$. The zero modes $\hat H_a^{\text{(zero)},J}$ are constructed explicitly in appendix \ref{app:modes}, with the orthogonality relation $\sum_a \Tr \left(\hat H_a^{\text{(zero)},I} \hat H_a^{\text{(zero)},J} \right) = N \delta^{IJ}$. We wish to effect the coordinate transformation
\begin{equation}
    \delta X_a = \sum_{J} \delta h^J {\textstyle \sqrt{\frac{2}{N}}} \hat H_a^{\text{(zero)},J} = i \sum_M \delta\theta^M [T^M, X_a] \,.
\end{equation}
Here $X_a = \frac{3 \Omega}{8} J_a$ and $\theta^M$ are the angles on $SU(N)$. The volume \eqref{eq:vol} is simply the integral over these angles. The factor of $\sqrt{\frac{2}{N}}$ matches the normalisation of the $\hat H_a^{\text{(zero)},J}$ with the generators $T^M$.
From the orthogonality of $\hat H_a^{\text{(zero)},J}$,
\begin{equation}
    \delta h^J = \frac{i}{\sqrt{2N}} \frac{3\Omega}{8} \sum_{a,M} \delta \theta^M \text{Tr} \left([T^M,J_a]\hat H_a^{\text{(zero)},J}\right) \,,
\end{equation}
Using the explicit form of the $\hat H_a^{\text{(zero)},J}$, the Jacobian can be evaluated as
\begin{equation}
    Z^{\text{(1-loop)}}_{3,\text{jac}} = \left|\text{det}_{JM} \left[\frac{i}{\sqrt{2N}} \frac{3\Omega}{8} \sum_{a} \text{Tr} \left([T^M,J_a]\hat H_a^{\text{(zero)},J} \right)\right] \right| = \left(\frac{3 \Omega}{8}\right)^{N^2-1} N^{N-1/2} \prod_{k=1}^{N-1} k^{2k} \,.
\end{equation}
Assembling the above results we obtain
\begin{align}
Z^{\text{(1-loop)}}_{3} & = Z^{\text{(1-loop)}}_{3,\text{nonzero}}Z^{\text{(1-loop)}}_{3,\text{zero}}Z^{\text{(1-loop)}}_{3,\text{jac}} \\
 & = \frac{2^{(7+6N)(N-1)/2} \pi^{(4+3N)(N-1)/2} N^{N-1} \Omega^{1 - N^2}}{\sqrt{\prod_{l=1}^{N-1} \Big(l(3l+1) \Big)^{2l-1}\Big((l+1)(3l+2) \Big)^{2l+3}}} \prod_{k=1}^{N-1} \frac{k^{2k}}{k!} \,. \label{Z3}
\end{align}

\subsubsection{Fermions}

The matrices appearing in the spinor action \eqref{Sf} are seen in appendix \ref{gam} to take the form
\begin{equation}
    \Bar{\sigma}^8 = 
    \begin{pmatrix}
        0 & 1 \\
        1 & 0
    \end{pmatrix}
    \otimes \mathbb{1}_8, \quad
    \Bar{\sigma}^9 = 
    \begin{pmatrix}
        1 & 0 \\
        0 & -1
    \end{pmatrix}
    \otimes \mathbb{1}_8, \quad
    \Bar{\sigma}^{10} = 
    \begin{pmatrix}
        -i & 0 \\
        0 & -i
    \end{pmatrix}
    \otimes \mathbb{1}_8, \quad
    \Bar{\sigma}^{8}\sigma^9 \Bar{\sigma}^{10} = 
    \begin{pmatrix}
        0 & i \\
        -i & 0
    \end{pmatrix}
    \otimes \mathbb{1}_8.
\end{equation}
This motivates a factorisation of the spinor as
\begin{equation}
    \theta = \psi \otimes \xi \,,
\end{equation}
where $\psi$ is a two-component spinor and $\xi$ has eight components.
In this way we have eight copies of two-component spinors, since the eight-component is only acted on by the identity matrix. The partition function will then become the eighth power of that of a single two-component spinor. From the spinor action \eqref{Sf}, the modes of the hermitian matrix $\hat \psi$ of these two-component spinors about the background \eqref{eq:background} obey
\begin{equation}\label{fermioneigen}
    \frac{1}{2} \frac{3 \Omega}{8} 
    \begin{pmatrix}
        \text{Ad}(J_-) & \text{Ad}(J_3)  \\
        \text{Ad}(J_3)  & -\text{Ad}(J_+)  
    \end{pmatrix} \hat{\psi}
        + \frac{\Omega}{8}
        \begin{pmatrix}
            0 & 1 \\
            -1 & 0
        \end{pmatrix} \hat{\psi}  = \omega 
    \begin{pmatrix}
        0 & 1 \\
        -1 & 0
    \end{pmatrix} \hat{\psi} \,.
\end{equation}
Here the adjoint action $\text{Ad}(X) = [X, \cdot \,]$ and $J_\pm = J_1 \pm i J_2$. The antisymmetric matrix on the right hand side is necessary to account for the anticommuting nature of the spinors in the quadratic action \eqref{Sf}.

The equations \eqref{fermioneigen} are solved in appendix \ref{app:modes}. The eigenvalues are found to be 
\begin{equation}
\omega = 
    \frac{\Omega}{16} \begin{cases}
    - (3l+1) & \text{with degeneracy $2l$}\\
      3l+2 & \text{with degeneracy $2l+2$}
    \end{cases} \,.
\end{equation}
Here a degeneracy of $2l$ means that there are $l$ pairs of modes with the given eigenvalue. That is, if the matrices are expanded in modes as $\hat \psi = \sum_{I} \alpha_I \hat{\psi}_{I}$ then on a given eigen-pair, with coefficients $\alpha_I$ and $\alpha'_{I}$, the quadratic action takes the form $S^{(2)}_f = 2 \omega_I (\alpha_I \a'_I - \a'_I \a_I) = 4 \omega_I \a_I \a'_I$. 
The first factor of 2 here comes from the trace normalisation of the $SU(N)$ generators $\Tr(T^M T^N) = 2\delta^{MN}$. There are $N^2-1$ such pairs and hence the one-loop contribution from fermions is given by
\begin{align}\label{Zf}
    Z_{f}^{\text{1-loop}} 
    = \left(\frac{\Omega}{4}\right)^{8(N^2-1)} \left[\prod_{l=1}^{N-1} \left(3l+1\right)^{l} \left(3l+2\right)^{l+1}\right]^8.
\end{align}
The overall power of $8$ is for the eight copies of two-component fermions.

\section{The dual supergravity background}
\label{sec:grav}

In this section we will show that the large $\Omega$ fuzzy sphere saddle has a corresponding dual supergravity background. The simplification of the large $\Omega$ limit is that the gravitational description can be split into two parts. We will firstly find a supersymmetric Euclidean background of type IIB supergravity that has the symmetries of the deformed matrix integral. We will then find a probe Euclidean $D1$-brane in this background that preserves the symmetries and supersymmetries. This probe $D1$-brane will recover the matrix fuzzy sphere in detail. The physics here is very much analogous to that of the mass deformed Polchinksi-Strassler background \cite{Polchinski:2000uf} and, even closer, the gravitational dual of the mass deformed BMN matrix quantum mechanics \cite{Berenstein:2002jq, Lin:2004kw}.
In the final \S\ref{sec:backreact} we discuss the backreaction of the $D1$-brane on the geometry as $\Omega$ is lowered.

\subsection{Summary of the solution}

To find the solution will require grappling with some subtleties concerning Euclidean supersymmetry. It is therefore useful to present the solution beforehand. All conventions will be given below. The Einstein frame metric is flat,
\be\label{eq:ds2}
ds^2 = \sum_{\mu=1}^{10} dx_\mu^2 \,,
\ee
the dilaton $\phi$ and axion $C_0$ are given by\footnote{The dilaton at the origin is, in general, set by a second constant of integration, the `string coupling' $g_\text{s}$. The factors of $g_\text{s}$ can be removed from the solution and incorporated into the action instead by rescaling the fields. We have done this, so that $g_\text{s}$ appears explicitly in the actions (\ref{eq:iibaction}) and (\ref{eq:d1}) below.\label{foot:rescale}}
\be\label{eq:dilaton}
e^{\phi} = - \frac{1}{C_0} = 1 - \frac{\mu^2}{32}\left(\sum_{A = 1}^7 x_A^2 + 3 \sum_{a=8}^{10} x_a^2 \right)  \,,
\ee
where $\mu$ is a constant, while the NSNS three-form field strength is
\be\label{eq:H}
H = \mu \, dx^8 \wedge dx^{9} \wedge dx^{10} \,.
\ee
The remaining RR field strengths are zero. This background is very similar to the type IIA solution obtained in \cite{Dasgupta:2002hx} by dimensional reduction of the eleven dimensional BMN plane wave \cite{Berenstein:2002jq}.
We may further recall that the $D$-instanton geometry is also flat in Einstein frame \cite{Gibbons:1995vg}.

The axion in (\ref{eq:dilaton}) is seen to diverge at a critical radius. Furthermore, the string frame metric,
\be
ds^2_\text{str} = e^{\frac{1}{2}\phi} ds^2 \,,
\ee
collapses at that radius, with a diverging Ricci scalar. The geometry therefore has finite extent and volume, with
\be
0 \leq \sum_{A = 1}^7 x_A^2 + 3 \sum_{a=8}^{10} x_a^2 \leq \frac{32}{\mu^2} \,.
\ee
We will discuss the physical meaning of a bounded geometry below, most immediately we will need to ensure that the probe $D1$-brane remains in a weakly curved region of the background. The string coupling remains bounded over the entire space.

The background we have just presented does not carry $D$-instanton charge. This can be added in the form of $N$ probe $D$-instantons. Because of the background flux, the $D$-instantons are able to lower their action by polarising into a spherical $D1$-brane, in the spirit of \cite{Emparan:1997rt, Myers:1999ps}. We will find that there is a supersymmetric Euclidean $D1$-brane sitting at
\be
\sum_{a=8}^{10} x_a^2 = \frac{9 \mu^2 (\pi \alpha')^2 N^2}{16} \,.
\ee
Here we have taken the limit, described below, of small $\pi \alpha' N \mu^2$. The $D1$-brane then avoids the region of large curvatures. The $D1$-brane carries a worldvolume Maxwell field $\mathcal{F}$, which induces $N$ units of $D$-instanton charge via the worldvolume coupling $\mathcal{C}_0 \mathcal{F}$. In (\ref{eq:match}) below, the
spherical $D1$-brane is identified with the fuzzy sphere saddle of the large $\Omega$ matrix integral, with $\Omega \propto \mu$.

\begin{figure}[h]
    \centering
    \includegraphics[width=0.28\linewidth]{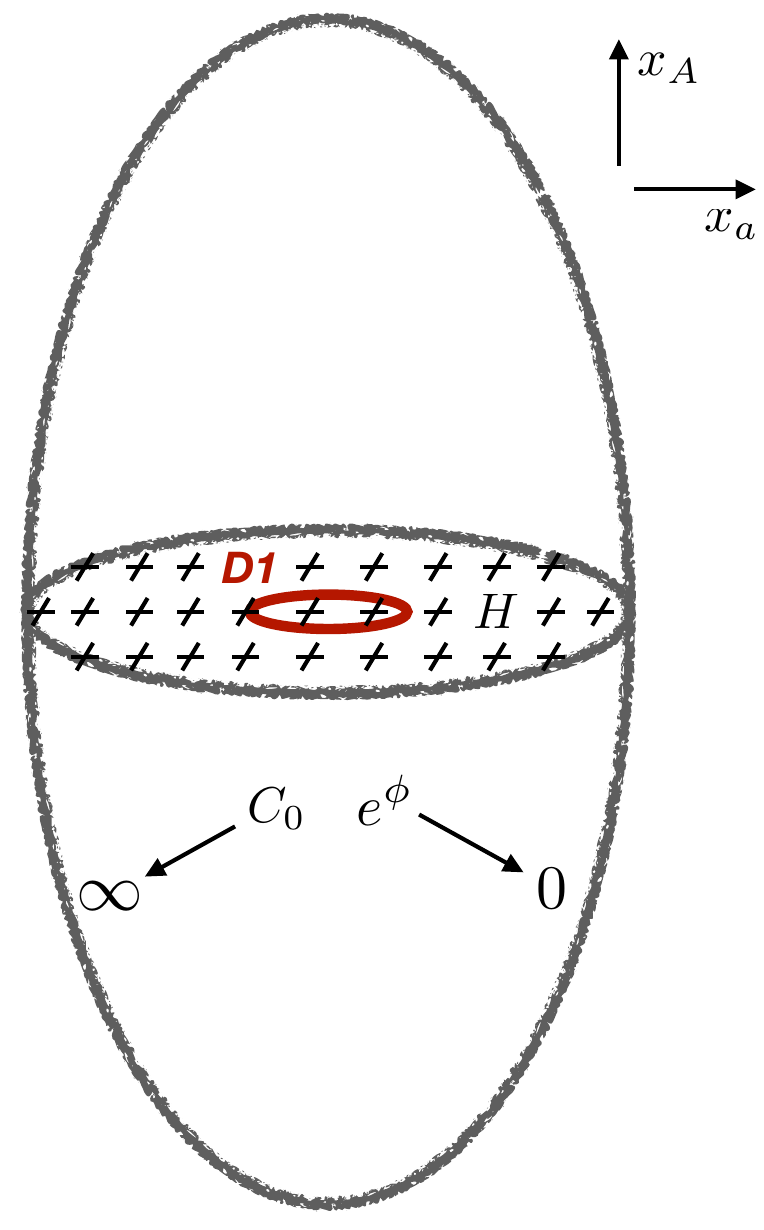}
    \caption{The Euclidean type IIB background is an ellipsoidal cavity, supported by three-form NSNS flux $H$, dilaton $e^\phi$ and axion $C_0$. The dilaton goes to zero at the boundary of the cavity while the axion diverges. A spherical $D1$-brane polarises due to the background flux and carries $N$ units of $D$-instanton charge.}
    \label{fig:cavity}
\end{figure}

A cartoon of the background we have just outlined is shown in Fig.~\ref{fig:cavity}. In the remainder of this section we will obtain the solution and describe its properties in more detail.
One way to motivate the presence of a singular exterior boundary at finite distance is to recall that the scaling dimension of a scalar field becomes negative below two spacetime dimensions. This can be expected to invert the dual holographic renormalisation group flow, so that the gap induced by the mass term cuts off the geometry at large distance instead of in the interior. While there is no notion of energy scale in the matrix integral (see \S\ref{sec:disc}), an analogous cavity is seen in the IIA geometry dual to BMN matrix quantum mechanics \cite{Dasgupta:2002hx}.

\subsection{Euclidean type IIB theory}
\label{sec:euc}

Our backgrounds have vanishing RR three-form, and hence the RR five-form can consistently be set to zero. The Euclidean type IIB supergravity equations of motion, in Einstein frame, are those derived from the action
\begin{equation}\label{eq:iibaction}
    S_\text{IIB} =  \frac{1}{(2\pi)^7 \alpha^{\prime4} g_\text{s}^2}\int\, d^{10}x \sqrt{g}\, \left[- R + \frac{1}{2}(\nabla \phi)^2 - \frac{1}{2} e^{2 \phi} (\nabla C_0)^2 + \frac{1}{12} e^{-\phi} H^2 - \frac{1}{12}e^{\phi} (F - C_0 H)^2 \right] \,.
\end{equation}
Here $\phi$ is the dilaton, $C_0$ the RR scalar (the axion), $H = dB$ the NSNS three-form, and $F = d C_2$ the RR three-form. We sometimes write $\tilde{F} = F - C_0 H$. It is important to note the `wrong' sign kinetic terms for both the RR scalar and three-form in \eqref{eq:iibaction}. The axion is a pseudo-scalar and hence acquires a factor of $i$ in passing to Euclidean signature \cite{Gibbons:1995vg, Bergshoeff:1998ry}. This factor of $i$ leads to the opposite-sign kinetic term for $C_0$, as $C_0$ is real in (\ref{eq:iibaction}). For the same reason, the three-form $F$ acquires a negative kinetic term while the three-form $H$ does not.\footnote{The final term in \eqref{eq:iibaction} requires that one, but not both, of $F$ and $H$ must compensate for the factor of $i$ in $C_0$ under Wick rotation. Indeed, it is $C_2$ that couples to worldvolume Maxwell fields in an analogous way to $C_0$.} Both $F$ and $H$ are real in \eqref{eq:iibaction}. The equations of motion following from the action \eqref{eq:iibaction} are given in appendix \ref{app:sugra}.

In Euclidean signature the $SL(2,\R)$ covariant $\tau$-parameter and its conjugate are \cite{Bergshoeff:1998ry}
\begin{equation}
   \tau = C_0 +  e^{-\phi}, \qquad \bar{\tau} = C_0 -  e^{-\phi} \,. 
\end{equation}
Under $SL(2, \R)$ they transform as
\begin{equation}\label{tautrans}
    \tau \to \frac{a \tau + b}{c \tau + d}, \qquad \bar \tau \to \frac{a \bar{\tau} + b}{c \bar{\tau} + d}, \qquad ad - bc = 1 \,.
\end{equation}
The two three-form field strengths transform in a doublet as
\begin{equation}
    \begin{pmatrix}
        H \\
        F
    \end{pmatrix}
    \to\begin{pmatrix}
        d & c \\
        b & a 
    \end{pmatrix}
    \begin{pmatrix}
        H \\
        F
    \end{pmatrix} \,.
\end{equation}
The action \eqref{eq:iibaction} may then be written in a manifestly $SL(2,\R)$ invariant form as
\begin{equation}
    S =  - \frac{1}{(2\pi)^7 \alpha^{\prime4} g_\text{s}^2}\int d^{10}x \sqrt{g} \,\left[ R +  \frac{2\partial_\mu \tau \partial^\mu \bar{\tau}}{(\tau - \bar{\tau})^2} + \frac{1}{6} \frac{1}{(\tau - \bar{\tau})} (F - \tau H)_{\mu \nu \rho} (F - \bar{\tau} H)^{\mu \nu \rho} \right] \,.
\end{equation}

To find the background we will impose that it preserve sixteen supersymmetries, in correspondence with the mass deformation of the IKKT model in \S\ref{sec:model}.
Euclidean supersymmetry transformations for the dilaton and axion have been written down in \cite{Gibbons:1995vg, Bergshoeff:1998ry}. We believe that there is a minus sign off in the dilatino transformation in \cite{Bergshoeff:1998ry}, as the transformations given there are not $SL(2,\R)$ covariant. The axion-dilaton sector of the transformations we write shortly agrees with that in 
\cite{Gibbons:1995vg} upon letting their $e \to \pm 1$, and with an overall sign redefinition of $\lambda$ below. To extend the transformations to include $H$ and $F$ we have started from the Lorentzian supersymmetry transformations in \cite{Bergshoeff:1996ui}, which are $SL(2,\R)$ covariant. Almost all relative signs in the Euclidean transformations are fixed by requiring this covariance to carry over to Euclidean signature. All remaining signs are then fixed by requiring the Euclidean flat $D1$ brane solution to preserve half of the supersymmetries of flat space. A final check of our transformations will be that
the background we find by imposing supersymmetry is indeed a solution to the full equations of motion.
All told, we will use the following Euclidean supersymmetry transformations for the dilatino and gravitino (the $\pm$ labels are explained shortly),
\begin{equation}\label{eq:susy}
\begin{aligned}
    \delta \lambda^{\pm} &= \frac{1}{4}\Gamma^{\mu}\epsilon^{\mp} \left(\partial_\mu \phi \mp e^\phi \partial_\mu C_0 \right)  \pm \frac{1}{8} e^{\phi/2} \Gamma^{(3)} \epsilon^\pm  \left(\tilde{F} \mp e^{-\phi} H\right) \,, \\
    \delta \psi_\mu^{\pm} &= \nabla_\mu \epsilon^\pm \mp \frac{1}{4} e^\phi \partial_\mu C_0 \, \epsilon^\pm \pm \frac{1}{16} e^{\phi/2}\left(\Gamma_\mu \Gamma^{(3)} + 2 \Gamma^{(3)} \Gamma_\mu \right) \epsilon^\mp \left(\tilde{F} \mp e^{-\phi} H \right) \,.\\
\end{aligned}
\end{equation}
where $\epsilon^\pm$ are $SO(10)$ chiral spinors satisfying $\Gamma_{11} \epsilon^{\pm} = \epsilon^\pm$. We may recall here that in type IIB theory, the supersymmetries all have the same chirality. In (\ref{eq:susy}) we have used the 
shorthand $\Gamma^{(n)} A = \frac{1}{n!} \Gamma^{\mu_1\ldots\mu_n} A_{\mu_1\ldots\mu_n}$. The transformations (\ref{eq:susy}) are covariant under $SL(2,\R)$ with
\be\label{eq:sl2r1}
    \delta \lambda^\pm \to \left(\frac{d + c \bar{\tau}}{d + c \tau}\right)^{\pm 3/4} \delta \lambda^\pm \,, \quad 
    \delta \psi_\mu^\pm \to \left(\frac{d + c \bar{\tau}}{d + c \tau}\right)^{\pm 1/4} \delta \psi_\mu^\pm \,, \quad
    \epsilon^\pm \to \left(\frac{d + c \bar{\tau}}{d + c \tau}\right)^{\pm 1/4} \epsilon^\pm \,.
\ee

The Wick rotation can be viewed as turning Lorentzian spinors $\lambda, \psi, \epsilon$ and their complex conjugates $\lambda^*, \psi^*, \epsilon^*$ into $\lambda^+, \psi^+, \epsilon^+$ and $\lambda^-, \psi^-, \epsilon^-$, respectively. More explicitly, this is done in three steps. Firstly, the real and imaginary parts of a complex spinor in Lorentzian signature are each an $SO(1,9)$ Majorana-Weyl spinor. That is, we may write $\lambda = \lambda_1 + i\lambda_2$, $\psi = \psi_1 + i \psi_2$ and $\epsilon = \epsilon_1 + i \epsilon_2$ with $\lambda_{1,2}$, $\psi_{1,2}$ and $\epsilon_{1,2}$ being Majorana-Weyl. Secondly, in parallel with the matrix model, as described in \S \ref{sec:model}, the Wick rotated spinors cannot satisfy any Majorana condition, but only themselves appear in the Euclidean action while their hermitian conjugates do not. As in the matrix model, the path integral is restricted to only the spinor fields themselves, and hence the number of degrees of freedom remains the same as in the Lorentzian theory.
Thirdly, we define linear combinations $\lambda^\pm = \lambda_1 \pm \lambda_2$, $\psi^\pm = \psi_1 \pm \psi_2$, $\epsilon^\pm = \epsilon_1 \pm \epsilon_2$. 

In appendix \ref{app:sugra} we obtain the background (\ref{eq:ds2}), (\ref{eq:dilaton}) and (\ref{eq:H}) by making an Ansatz with $SO(3) \times SO(7)$ symmetry and using (\ref{eq:susy}) to impose that it preserves sixteen supersymmetries. As in the matrix model, these supersymmetries are not consistent with any reality conditions on the spinor fields.

\subsection{Spherical probe $D$1-brane}\label{sec:probe}

We will now show that $N$ probe $D$-instantons in the background can lower their action by polarising into a probe Euclidean $D1$-brane. The Euclidean worldvolume action of a single $D1$-brane is given by
\begin{equation}\label{eq:d1}
    S_{D1} = \frac{T_1}{g_\text{s}}\int d^2 \sigma e^{-\phi} \sqrt{ \text{det}(\mathcal{G}_{ij} + \mathcal{B}_{ij} - 2 \pi \alpha^\prime \mathcal{F}_{ij})} + \frac{T_1}{g_\text{s}} \int \mathcal{C}_0 \left( \mathcal{B} - 2 \pi \alpha^\prime \mathcal{F} \right) \,,
\end{equation}
where $\mathcal{G}$, $\mathcal{B}$ and $\mathcal{C}_0$ are the pull-back of the string-frame metric, $B$ and $C_0$ onto the worldvolume, respectively, while $\mathcal{F}$ is the worldvolume Maxwell field strength, $T_1 = 1/(2\pi \alpha')$ is the $D1$-brane tension \cite{Polchinski:1996na} and $g_\text{s}$ is the string coupling (see also footnote \ref{foot:rescale} above). In this section and appendix \ref{app:fieldtheory}, only, we will use $i,j$ to denote the worldvolume index. The final term in (\ref{eq:d1}) does not come with a factor of $i$ as, we may recall from \S\ref{sec:euc}, the axion acquires a factor of $i$ upon Wick rotation that will cancel the factor of $i$ coming from Wick rotating the time coordinate.

We look for spherical solutions where the $D1$-brane is at the origin in the $x_A$ directions, with $A = 1, \ldots, 7$, and forms a two-sphere in the remaining directions. This configuration preserves the $SO(7)\times SO(3)$ symmetry of the background. We would like to find the radius $r$, with
\be\label{eq:r22}
r^2 = \sum_{a=8}^{10} x_a^2 \,,
\ee
of the sphere. On such spherical configurations, and in the background described by (\ref{eq:ds2}), (\ref{eq:dilaton}) and (\ref{eq:H}), the $D1$-brane action (\ref{eq:d1}) is
\begin{equation}\label{eq:D1mini}
    S_{D1} = \frac{4 \pi T_1}{g_\text{s} \left(1 - \frac{3 \mu^2 r^2}{32} \right)}
    \left[\sqrt{\left( 1 - \frac{3 \mu^2 r^2}{32}\right) r^4 + \left(\frac{\mu r^3}{3}  - \pi \alpha^\prime N \right)^2} - \left(\frac{\mu r^3}{3} - \pi \alpha^\prime N \right) \right].
\end{equation}
Here the worldvolume Maxwell field strength has been taken to be
\be\label{eq:maxwell}
\mathcal{F} = \frac{N}{2} \text{vol}_{S^2} \,,
\ee
where the volume form on $S^2$ is normalised such that $\int \text{vol}_{S^2} = 4 \pi$. Moving the probe away from the origin in the $x_A$ directions increases the action.

The $N$ in the Maxwell flux (\ref{eq:maxwell}) is the $D$-instanton number. We may verify this as follows. The $D$-instanton number is the Noether charge associated to shifts of the axion \cite{Gibbons:1995vg}. From the $D1$-brane action (\ref{eq:d1}) the charge is seen to be\footnote{The term coupling the axion to $\mathcal{B}$ in (\ref{eq:d1}) does not contribute to the Noether charge because the shift of this term is compensated for by a corresponding shift in the $C_2$ coupling, not shown in (\ref{eq:d1}).}
\begin{equation}\label{eq:matchcharge}
    2 \pi \alpha^\prime  T_1 \int_{D1} \mathcal{F} = T_{-1} N,
\end{equation}
where $T_{-1} = 4 \pi^2 \alpha^\prime T_1 = 2\pi$ is the `tension' of a $D$-instanton \cite{Polchinski:1996na, Green:1997tv}. Therefore we see that adding this probe $D1$-brane to the background is equivalent, in terms of quantum numbers, to adding $N$ $D$-instantons (we will see below that in fact, with the conventions we are using, we have added $D$-anti-instantons). This is a familiar aspect of the Myers effect \cite{Emparan:1997rt, Myers:1999ps}. We may identify $N$ in (\ref{eq:maxwell}) with the rank $N$ of the matrices in the deformed IKKT model.

The minimum action configuration is found by solving
\be
\frac{dS_{D1}}{dr} = 0 \,. \label{eq:reom}
\ee
This can be done with the full expression (\ref{eq:D1mini}), and the exact supersymmetric minimum is given in appendix \ref{app:susyembed}. However, in order for the classical probe brane action to be valid, we should keep away from the high curvature boundary of space where $e^\phi$ vanishes. This can be done transparently by introducing a rescaled radius $y$ as
\begin{equation}\label{eq:rescale}
    r = \frac{\eta y}{\mu}, \qquad \eta \equiv \pi \alpha^\prime N \mu^2 \,.
\end{equation}
The dilaton becomes
\begin{equation}\label{eq:dilagain}
    e^\phi = 1 - \frac{3}{32} \eta^2 y^2  \,.
\end{equation}
We will take the limit $\eta \to 0$ with $y$ held fixed. 
The solutions for $y$ will be finite in this limit, so that $r$ is well away from the singular boundary and 
$e^{\phi} \sim 1$ in (\ref{eq:dilagain}).

We may expand the action in small $\eta$ to obtain
\begin{equation}\label{eq:actionexpand}
S_{D1}  = \frac{4 \pi T_1}{g_\text{s} \mu^2} \left[ 2 \eta + \frac{\left(24 y^4-32 y^3+9 y^2\right)}{48} \eta^3 + \mathcal{O}\left(\eta^5\right) \right] \,.
\end{equation}
The stationary points \eqref{eq:reom} are thus found, to leading order in small $\eta$, to be
\begin{equation}\label{eq:ysol}
    y = 0, \ \frac{1}{4}, \ \frac{3}{4} \,.
\end{equation}
These are two minima separated by a maximum. The ratio of $3$ between the radii at the maximum and the minimum agrees with that found in (\ref{eq:factor}) for the fuzzy sphere. As with the fuzzy sphere, the nontrivial minimum has the lowest action. These facts strongly suggest that the stable polarised $D1$-brane can be identified with the stable fuzzy sphere matrix configuration. To further support this identification, in appendix \ref{app:susyembed} we establish that the stable $D1$-brane configuration preserves the sixteen supersymmeteries of the background. In appendix \ref{app:fieldtheory} we verify that the fluctuations about the polarised brane match the matrix fluctuations about the fuzzy sphere.

To match the absolute units of length, i.e.~the radius of the stable sphere, between the two sides of the correspondence we can equate the actions at the minima. From (\ref{eq:actionexpand}) we have
\be\label{eq:onshell}
S^\text{min}_{D1} = \frac{4 \pi N}{g_\text{s}} - \frac{9}{2^{8}} \frac{\pi^3 \alpha'^2 N^3 \mu^4}{g_\text{s}} \,.
\ee
In both terms we used $T_1 = 1/(2\pi \alpha')$. The first term in (\ref{eq:onshell}) is $N$ times the action of a $D$-(anti)-instanton, i.e.~$2 \pi \bar \tau/g_\text{s}$ \cite{Green:1997tv}, in a Euclidean background with $e^{-\phi} = - C_0 = 1$.\footnote{One may also find solutions with the opposite charge, describing $N$ $D$-instantons. These are backgrounds in which the dilaton is the same as in (\ref{eq:dilaton}) but $C_0 = e^{-\phi}$, with no relative minus sign, and $\mu \to -\mu$ in (\ref{eq:H}). This solution is also seen to be supersymmetric, upon exchanging $\epsilon^+ \leftrightarrow \epsilon^-$. A spherical probe $D$1-brane on this background has the same action as \eqref{eq:D1mini}, and hence the same stable solutions and on-shell action.} The second term in (\ref{eq:onshell}) captures the lowering of the action due to the polarisation of the instantons into a Euclidean $D$1-brane.

Equating the second term in (\ref{eq:onshell}) with the fuzzy sphere action (\ref{S0}) gives, at large $N \gg 1$,
\be\label{eq:match}
\frac{\Omega^4}{2^4} = \frac{(2\pi)^3 \alpha'^{2}}{g_\text{s}} \mu^4 \,.
\ee
In (\ref{eq:match}) we have obtained an entry in the dictionary between the matrix and spacetime descriptions. The first term in (\ref{eq:onshell}) is an overall normalisation of the matrix integral. It can be accounted for by adding a constant term to the matrix action in (\ref{eq:fullS}),
\be\label{eq:DeltaS}
\Delta S = \frac{2 \pi \bar \tau}{g_\text{s}} \Tr \left(\mathbb 1 \right) \,.
\ee
Here $\bar \tau = 2$. This term is precisely the expected weighting of $D$-anti-instanton contributions to string theoretic processes \cite{Green:1997tv}. Similar terms have also been considered in other contexts, including in the original IKKT paper \cite{Ishibashi:1996xs}.

With the mapping (\ref{eq:match}) at hand, we may check the compatibility of the limits that have been considered on either side of the correspondence. On the matrix side we have assumed $\Omega^4 \gg N$ in order to treat the fuzzy sphere by saddle point. On the string theory side we have required $\eta \ll 1$ in order for the $D$1-brane geometry to be away from the strongly curved boundary of spacetime. Imposing both conditions requires
\be\label{eq:conds}
g_\text{s} N \ll \a'^2 \mu^4 \ll \frac{1}{N^2}  \qquad \Rightarrow \qquad g_\text{s} N^3 \ll 1 \,.
\ee
While $N$ is large, we will also want $g_\text{s} \ll 1$ and therefore the final condition in (\ref{eq:conds}) can be satisfied. In the regime (\ref{eq:conds}) the polarised $D$-instantons are simultaneously described by the spherical $D1$-brane and by the fuzzy sphere saddle of the deformed matrix integral. If the first inequality in (\ref{eq:conds}) does not hold, then the $D1$-brane must be treated as a fluctuating stringy object, see the discussion in appendix \ref{app:fieldtheory}. In particular, the $D1$-brane remembers its noncommutative origins. If the second inequality in (\ref{eq:conds}) does not hold, then higher powers of $X$ need to be added to the matrix action. This is because $S_\Omega + \Delta S$, from (\ref{eq:fullS}) and (\ref{eq:DeltaS}), is the action for $N$ $D$-instantons in a flat Euclidean IIB background with a constant NSNS three-form, dilaton and axion. This is indeed the correct background when $\eta \ll 1$. However once $\eta \sim 1$ the background axion, dilaton and string frame metric undergo large changes as a function of radius. These strong changes to the background induce changes to the instanton world-point action.

\subsection{Beyond the probe limit}
\label{sec:backreact}

We have seen that the fuzzy sphere saddle point of the matrix integral is in correspondence with a probe $D1$-brane in a certain Euclidean type IIB background. This gives two perspectives on the same physical system. The essence of holographic duality \cite{Maldacena:1997re} is then to take the system to a different regime of parameter space, where the $D$-branes now backreact strongly on the spacetime. We may expect that in such regimes the corresponding matrix integral is strongly coupled. In this subsection we will see how this works out for our background. We will not construct the backreacted geometries in this paper, but we will estimate the parameter regimes in which backreaction occurs.

The $D1$-brane is located at a radius $r$, defined in (\ref{eq:r22}). We would like to determine, parametrically, the transverse lengthscale $\ell \lesssim r$ over which the brane backreacts significantly. This backreaction lengthscale can be estimated by equating the action of the probe $D1$-brane (\ref{eq:d1}) with the change in individual terms in the IIB action (\ref{eq:iibaction})
\be\label{eq:equal}
\frac{r^2 \ell^8}{\a'^4 g_\text{s}^2} \frac{1}{\ell^2} \sim \left. \Delta S_\text{IIB} \right|_\text{single term}  \sim S^\text{min}_{D1} \sim \frac{N}{g_\text{s}} \,.
\ee
In the first expression $r^2 \ell^8$ is the spacetime volume of the backreaction region and $1/\ell^2$ is the scale over which the fields vary. We are working in the regime $\eta \lesssim 1$, wherein the brane action is given by the first term in (\ref{eq:onshell}) and the background supergravity fields are constant. In this regime $e^\phi \sim 1$ and so lengthscales in the string and Einstein frame are the same. Introducing the string length $l_\text{s} \equiv \sqrt{\alpha'}$, and with $r$ given by (\ref{eq:rescale}) and (\ref{eq:ysol}), we have from (\ref{eq:equal}) that
\be\label{eq:lrat}
\frac{\ell^6}{l_\text{s}^6} \sim \frac{g_\text{s} N}{(\mu \, l_\text{s} N)^2} \,. 
\ee
Backreaction produces a large (in string units) region of curved spacetime when the ratio (\ref{eq:lrat}) is large. That is
\be\label{eq:back1}
g_\text{s} N \gg (\mu \, l_\text{s} N)^2 \,.
\ee
The inequality (\ref{eq:back1}) can be understood intuitively. The standard condition for $N$ coincident $D$-branes to backreact is $g_\text{s} N \gg 1$ \cite{Maldacena:1997re}. Here, however, the $D$-instantons are diluted over a sphere of area $r^2 \sim (\mu \, l_\text{s}^2 N)^2$. The $D$-instanton density is therefore $N/r^2$, which will backreact strongly when (\ref{eq:back1}) is obeyed. Thus, in the regime (\ref{eq:back1}) the full geometry should look like a spherical `domain wall' of thickness $\ell$ separating an interior flat region and the exterior cavity geometry.

As the radius of the sphere $r \sim \mu \, l_\text{s} N$ shrinks in (\ref{eq:lrat}), the backreaction lengthscale grows and eventually reaches $\ell \sim r$ when, using (\ref{eq:lrat}) and (\ref{eq:rescale}),
\be\label{eq:dins}
g_\text{s} N \sim (\mu \, l_\text{s} N)^8 \,, \qquad \frac{\ell^8}{l_\text{s}^8} \sim g_\text{s} N \,.
\ee
The lengthscale $\ell$ here is precisely that over which $N$ coincident $D$-instantons backreact on the geometry, putting $p=-1$ in \cite{Itzhaki:1998dd}. This is a sensible result: once $\ell \sim r$ the backreaction occurs over the same lengthscale as the polarisation. The backreacted geometry therefore starts to approximate the $SO(10)$ symmetric background of $N$ clumped $D$-instantons \cite{Gibbons:1995vg, Bergshoeff:1998ry}, perturbed by a three-form flux. In the first instance this perturbation will ultimately lead to a cavity singularity enclosing the clumped $D$-instantons. However, if the backreaction lengthscale extends up to the cavity boundary, there may be other possible scenarios.

\begin{figure}[h]
    \centering
    \includegraphics[width=0.8\linewidth]{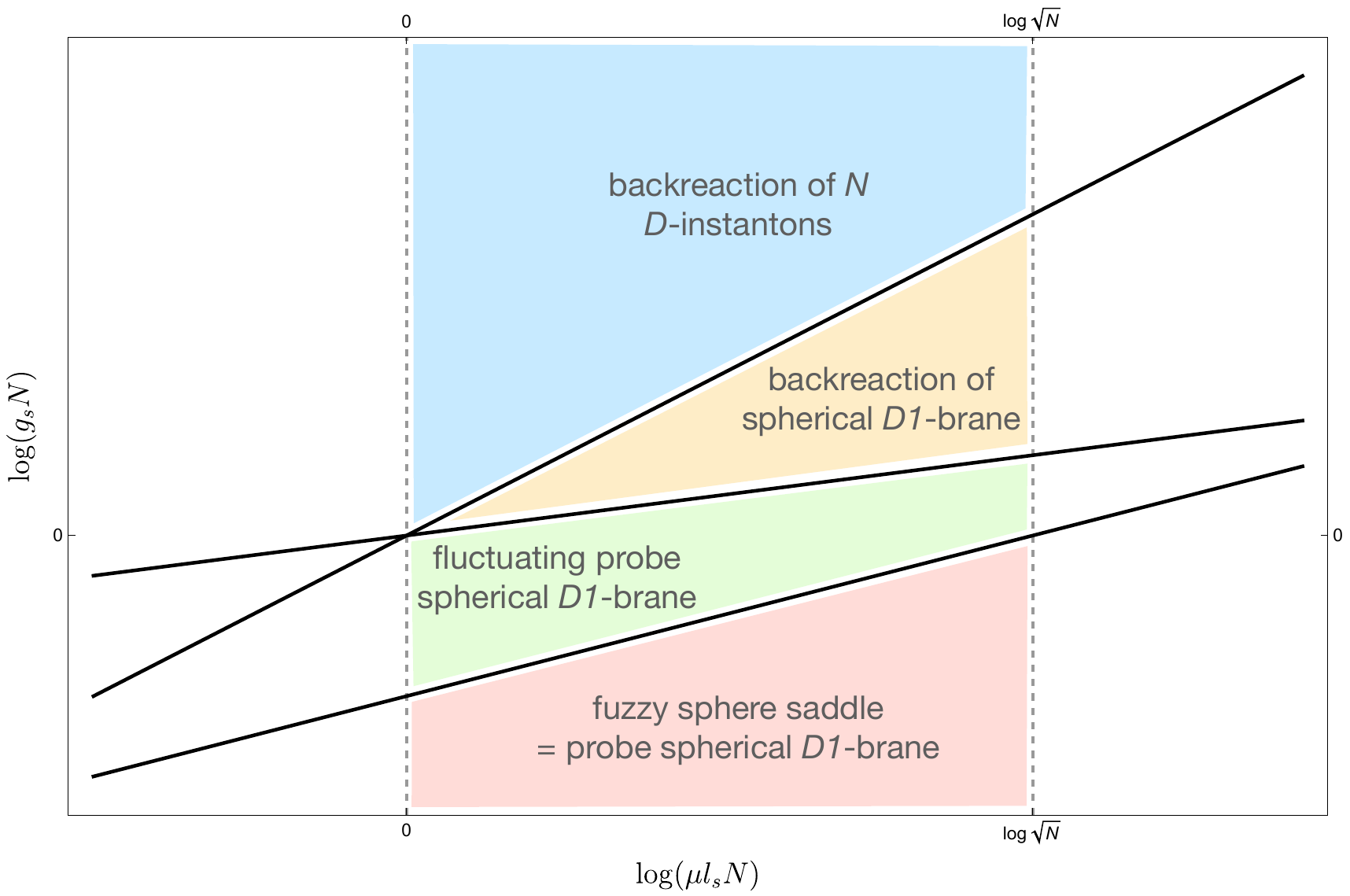}
    \caption{The expected type IIB phase diagram. The $x$ axis is (the logarithm of) the radius of the $D1$-brane in string units and the $y$ axis is the strength of gravitational backreaction. The vertical dashed line on the right corresponds to the $D1$-brane radius reaching the edge of the cavity. The dashed line on the left corresponds to the $D1$-brane radius reaching the string scale. Our analysis does not apply beyond these dashed lines. From top to bottom, the solid lines denote the conditions (\ref{eq:dins}), (\ref{eq:back1}) and (\ref{eq:conds}). In the top shaded region the background is given by the backreacted geometry of $N$ $D$-instantons, perturbed by a three-form NSNS flux. The region below this one is described by a backreacted spherical $D1$-brane inside the cavity. The bottom shaded region correponds to a probe $D1$-brane in the cavity background. In the region immediately above this one, the $D1$-brane remains a probe but has stringy fluctuations.}
    \label{fig:phases}
\end{figure}

The different regions we have just described are summarised in Fig.~\ref{fig:phases}.
The figure also indicates the regime of validity, from (\ref{eq:conds}), of the fuzzy sphere saddle of the matrix integral. A basic consistency check in Fig.~\ref{fig:phases} is that the regime where the $D1$-brane backreacts on the geometry (at the top of the figure) has no overlap with the perturbative fuzzy sphere regime of the matrix integral (at the bottom of the figure). If the matrix integral can provide a holographically dual description of the backreacted geometry, it will therefore be in the strongly fluctuating, small $\Omega$ regime. We will discuss this possibility further in \S\ref{sec:disc} below. As we have already noted in \S\ref{sec:model}, the absence of time in the matrix theory complicates matters because it removes the conventional connection between a matrix theory energy scale and the emergent holographic radial coordinate.

\section{Localisation}
\label{sec:loc}

In this section we will set up a supersymmetric localisation approach to the mass-deformed integral (\ref{eq:Z}). This allows results for the integral that are exact in $\Omega$. We show that the integral is given exactly by the one-loop fluctuations about two classes of matrix configurations. The first class are the stable fuzzy spheres considered in \S\ref{semicl}, while the second are configurations where the fuzzy sphere has collapsed to the origin.

The maximal fuzzy sphere gives the leading contribution to the integral at large $\Omega$. In \S\ref{sec:alln} we obtain the large $\Omega$ behavior of the maximal fuzzy sphere contribution for all $N$, recovering our previous formula (\ref{pertresult}) via a different integral. We furthermore obtain an exact-in-$\Omega$ expression for the maximal fuzzy sphere contribution for $N=2$, in
\S\ref{sec:n2}.  To capture the small $\Omega$ limit in its entirety it is necessary to understand, in addition, the contribution of the collapsed configuration. This involves some additional subtleties and will be presented in future work.

The localisation computations below follow familiar steps. In \S \ref{sec:offshellsusy} we write down off-shell supersymmetries for the polarised IKKT model.
We obtain these by dimensional reduction of the results for ten dimensional SYM theory \cite{Berkovits_1993}, extended to include the mass deformation. A similar procedure was outlined in \cite{Pestun:2007rz,Pestun:2009nn}. Next, in \S \ref{sec:locdef} we make a choice of a particular supercharge and localising action. Similarly to \cite{Pestun:2007rz, Pestun:2009nn}, while the individual off-shell supersymmetry transformations square to a gauge transformation and a rotation, we choose a particular one that squares to zero. This allows us to choose any localizing deformation. Our choice, \eqref{locV} below, is analogous to that in \cite{Pestun:2007rz, Pestun:2009nn}. A similar procedure has also been applied to the BMN matrix quantum mechanics in \cite{Asano:2012zt}. At the end of this subsection, we introduce a field redefinition and contour shift for the auxiliary matrices, aligned with definitions in \cite{Moore:1998et, Austing:2000rm}. In \S \ref{sec:moduli}, we explain our parameterisation of the localisation moduli space. The starting point here is a simplified set of localisation equations of motion \eqref{moduli}. The derivation of these equations is left to appendix \ref{app:moduli}; they are most easily derived along similar lines to \cite{Moore:1998et, Austing:2000rm}. Despite some similarities, our localisation differs from those in \cite{Moore:1998et, Austing:2000rm} not only by a different choice of localising action, but also by our not twisting the supersymmetries nor requiring a contour regularisation at the end. Eventually, in \S \ref{sec:locresult} we present the results advertised above. Numerical factors from normalisation of the integral are computed in appendix \ref{app:normalization}.

\subsection{Off-shell supersymmetries}\label{sec:offshellsusy}

The supersymmetry transformations in \eqref{eq:masssusy} only close on-shell. That is, the supercharges do not square to a bosonic symmetry of the action unless the equations of motion are imposed. For localisation to work we will need the supersymmetries to also close off-shell. This requires adding in seven auxiliary bosonic hermitian traceless matrices $D_A, (A=1,\ldots,7)$, with an extra term in the action \cite{Berkovits_1993, Pestun:2007rz, Pestun:2009nn}
\begin{equation}\label{eq:actionwithaux}
    S_\Omega \to S'_{\Omega} \equiv S_\Omega + 2 \text{Tr} \left( D_A D_A \right). 
\end{equation}
Integrating out $D_A$ trivially gives back the original partition function \eqref{eq:Z}, up to a constant multiplicative factor. We account for this normalisation factor in appendix \ref{app:normalization}.

In this section we will work in the chiral basis outlined in appendix \ref{gam}. In this basis only the 
top sixteen components of the spinor are nonzero. As above, see also \eqref{eq:chiral} in the appendix, we denote the nonzero components by $\theta$. Similarly, in the formulae below the supersymmetry parameter $\epsilon$ refers to the top sixteen (i.e.~nonzero) components of the thirty-two component spinor in \eqref{eq:masssusy}. We may now
follow \cite{Berkovits_1993} and write the off-shell supersymmetry transformations as
\begin{equation}
\begin{aligned}
    &\delta X^\mu = \theta_\alpha \bar{\sigma}^\mu_{\alpha \beta} \epsilon_\beta,\\
    &\delta \theta_\alpha = \frac{1}{2} \sigma^{\mu \nu}_{\alpha \beta} [X_\mu, X_\nu] \epsilon_\beta + v^A_\alpha D_A + \frac{ i \Omega}{8} X^\mu (T_\mu)_{\alpha \beta} \epsilon_\beta, \\
    &\delta D_A = - \frac{1}{4} (v_A)_\alpha \bar{\sigma}^{\mu}_{\alpha \beta} [X_\mu, \theta_\beta] + \frac{i \Omega}{4^2} (v_A)_\alpha K_{\alpha \beta} \theta_\beta,
\end{aligned}\label{eq:newsusy}
\end{equation}
where 
\begin{equation}
    \sigma^{\mu \nu} = \frac{1}{2}\left(\sigma^\mu \bar{\sigma}^\nu - \sigma^\nu \bar{\sigma}^\mu \right) \,, \qquad K = \sigma^8 \bar{\sigma}^9 \sigma^{10} = - \bar{\sigma}^8 \sigma^9 \bar{\sigma}^{10} = - \bar{K} \,, \qquad T^\mu = \sigma^\mu \bar{K} + 2 K \bar{\sigma}^\mu \,,
\end{equation}
and the $v_A$ are seven spinors satisfying 
\begin{equation}\label{eq:vconstraint}
v_A^T \bar{\sigma}^\mu \epsilon = 0 \,, \qquad \frac{1}{2} \left(\epsilon^T \bar{\sigma}^\mu \epsilon \right) (\sigma_\mu)_{\alpha \beta} = \epsilon_\alpha \epsilon_\beta + \frac{1}{4} (v_A)_\alpha (v^A)_\alpha \,, \qquad \frac{1}{4} v_A^T \Bar{\sigma}^\mu v_B = \delta_{AB} \epsilon^T \bar{\sigma}^\mu \epsilon \,.
\end{equation}
These equations determine $v_A$ in terms of $\epsilon$, up to $SO(7)$ rotations acting on the $A,B$ indices \cite{Pestun:2007rz}.

Acting with the supersymmetry transformations (\ref{eq:newsusy}) twice yields an $SU(N)$ rotation and an $SO(3) \times SO(7)$ rotation,
\begin{equation}\label{eq:susysq}
\begin{aligned}
    \delta^2 X^\mu &= - \xi^\nu[X_\nu, X^\mu] + \frac{i \Omega}{8} \left[\frac{1}{2} \omega_{AB} (M^{AB})^\mu_{\ \nu} + \frac{1}{2}\omega_{ab} (M^{ab})^\mu_{\ \nu}\right] X^\nu,\\
    \delta^2 \theta_\alpha &= - \xi^\nu[X_\nu, \theta_\alpha] + \frac{i \Omega}{8} \left[\frac{1}{2} \omega_{AB}(\sigma^{AB})_{\alpha \beta} + \frac{1}{2}\omega_{ab} (\sigma^{ab})_{\alpha \beta}\right] \theta_\beta,\\
    \delta^2 D_A &= - \xi^\nu [X_\nu, D_A] + \frac{i \Omega}{8} \left[\frac{1}{2} v_A^T K v_B \right] D_B,
\end{aligned}
\end{equation}
where, as before, we have $A,B = 1,\ldots,7$ while $a,b = 8,9,10$. We have also defined $\xi^\nu = \epsilon^T \bar{\sigma}^\nu \epsilon$, $\omega_{AB} = \epsilon^T \bar{\sigma}_A K \bar{\sigma}_B \epsilon$, and $\omega_{ab} = 2 \epsilon^T \bar{\sigma}_a K \bar{\sigma}_b \epsilon$. The matrices $M$ and $\sigma$ are generators of $SO(10)$ in the vector and Weyl spinor representation, respectively, given by
\begin{equation}
   \left( M^{\rho \sigma} \right)^\mu_{\ \nu} = \delta^{\rho \mu} \delta^{\sigma}_\nu - \delta^{\sigma \mu} \delta^\rho_\nu \, , \qquad
   \left(\sigma^{\mu \nu}\right)_{\alpha \beta} = \frac{1}{2}\left(\sigma^\mu \bar{\sigma}^\nu - \sigma^\nu \bar{\sigma}^\mu \right)_{\alpha \beta} \,.
\end{equation}
Rotations mixing the $8,9,10$ directions with $1,\ldots ,7$ are not present in (\ref{eq:susysq}). These would not be symmetries of the action.

\subsection{Deformation of the action}\label{sec:locdef}

With the off-shell supersymmetries in hand, the strategy is to use these to deform the action in a way that does not change the value of the integral. We then take a limit of the deformation that localises the integral to certain special configurations.

Let $\delta_\epsilon$, for some particular $\epsilon$ to be specified below, be one of the supersymmetries.
We deform the action by a $\delta_\epsilon$-exact term, such that the partition function (\ref{eq:Z}) becomes
\begin{equation}\label{eq:Zt}
    Z(t) = \int dX_\mu d\Psi_\alpha dD_A e^{- S'_\Omega - t \delta_\epsilon V} \,,
\end{equation}
where $t$ is a deformation parameter and $V$ is any fermionic operator satisfying 
\begin{equation}\label{Q2}
    \delta_\epsilon^2 V = 0 \,.
\end{equation}
It is crucial to use the off-shell supersymmetries here, in order to achieve (\ref{Q2}) inside the integral, which includes configurations that do not obey the equations of motion. The partition function does not depend on this deformation because
\begin{equation}\label{dZdt}
\begin{aligned}
    \frac{d}{dt}Z(t) &= - \int dX_\mu d\Psi_\alpha dD_A \delta_\epsilon V e^{- S'_\Omega - t \delta_\epsilon V} \\
    & = \int dX_\mu d\Psi_\alpha dD_A \delta_\epsilon \left(V e^{- S'_\Omega - t \delta_\epsilon V}\right)\\
    & = 0 \,.
\end{aligned}
\end{equation}
In the second line we used that $\delta_\epsilon$ is a supersymmetry, so $\delta_\epsilon S'_\Omega = 0$, and also \eqref{Q2}. The last step is a result of the integrand being $\delta_\epsilon$-exact. The fields do not appear on the right hand side of their own supersymmetry transformation in (\ref{eq:newsusy}), and therefore a $\delta_\epsilon$-exact term is a total derivative. It follows from (\ref{dZdt}) that the partition function is independent of the deformation. In particular, it is equal to the original undeformed partition function, $Z(t) = Z$.

Given that (\ref{eq:Zt}) is independent of $t$, we may calculate it in the limit $t \to \infty$, where only stationary points of $\delta_\epsilon V$ contribute. In general there will be a moduli space $\mathcal{M}$ worth of stationary points, so that the partition function becomes
\begin{equation}\label{eq:Mint}
    Z = \int_{\mathcal{M}} e^{-S^\prime_\Omega(\mathcal{M})} \, Z^{\text{1-loop}}(\mathcal{M}) \,.
\end{equation}
Here $S'_\Omega(\mathcal{M})$ is the action evaluated on the moduli space and $Z^{\text{1-loop}}(\mathcal{M})$ is the one-loop determinant calculated from the quadratic fluctuations around each point on the moduli space. We will see shortly that the stationary points of 
$\delta_\epsilon V$ coincide with its zeros, which is why $\delta_\epsilon V$ does not appear in the exponent in (\ref{eq:Mint}).

We have seen in (\ref{eq:susysq}) that the supercharges in general square to a symmetry transformation. However, we can find linear combinations that square to zero. In such cases \eqref{Q2} is satisfied for arbitrary $V$. One such choice is parameterized by the sixteen-component spinor
\begin{equation}\label{eq:loccharge}
    \epsilon  = 
    \begin{pmatrix}
        0 & \ldots & 0 & i & 1
    \end{pmatrix}^T \,,
\end{equation}
and, correspondingly, a representative solution to \eqref{eq:vconstraint} is \cite{Pestun:2009nn},
\begin{equation}
    v_A = 2 \sigma^8 \Bar{{\sigma}}_A \epsilon \,.
\end{equation}
We should emphasise that a complex $\epsilon$ in (\ref{eq:loccharge}) does not pose any problems.
As argued in \cite{Pestun:2007rz}, the action should be thought of as a holomorphic function on the complex space of $X_\mu$, $D_A$, and $\theta_\alpha$. As we have explained in \S\ref{sec:model}, especially in relation to the spinor matrices, the integral is a holomorphic integral over half of the coordinates on this complex space. The supersymmetries are then holomorphic transformations on the complex space that leave the action invariant. For the localisation argument above, it does not matter whether the coefficients that appear in these holomorphic transformations are real or complex. The important point is that the transformations do not depend on the hermitian conjugates of the matrices. This is sufficient for the above localisation argument to work.

Given that $\delta_\epsilon^2 = 0$ with the choice (\ref{eq:loccharge}), we may choose $V$ arbitrarily. A convenient choice is 
\begin{equation}\label{locV}
\begin{aligned}
    V &= \text{Tr} \left(\theta_\alpha ( \delta_\epsilon \theta_\alpha)^\dagger\right) \,, \\
    \delta_\epsilon V &= \text{Tr} \left( (\delta_\epsilon \theta_\alpha) (\delta_\epsilon \theta_\alpha)^\dagger -  \theta_\alpha \delta_\epsilon(\delta_\epsilon \theta_\alpha)^\dagger \right) \,.
\end{aligned}
\end{equation}
This choice has the benefit that the first term in $\delta_\epsilon V$, the bosonic part, is a sum of squares. Hence it is positive semi-definite and so the integral is well-behaved as $t \to \infty$. The stationary points are just the minima of $\delta_\epsilon V$, which are also its zeros. The moduli space is therefore 
\begin{equation}\label{moduliformal}
    \mathcal{M} = \left\{X_\mu, \ D_A \ |\  \delta_\epsilon \theta_\alpha = 0, \, \alpha = 1,\ldots,16 \right\}. 
\end{equation}
The fermionic part of $\delta_\epsilon V$ in \eqref{locV}, the second term, goes fully into the one-loop part of (\ref{eq:Mint}). 

We should clarify the meaning of the daggers in (\ref{locV}). Consistently with our discussion above, the deformation $\delta_\epsilon V$ of the action is defined to be a holomorphic function of $X_\mu$, $H_A$, and $\theta_\alpha$. The dagger in (\ref{locV}) is defined to take the conjugate of each element in $\delta_\epsilon \theta_\a$ while leaving $X_\mu$ and $H_A$ unchanged \cite{Pestun:2007rz}. This is equivalent to treating the matrices as hermitian.

In what follows, we use a convenient field re-definition
\begin{equation}\label{eq:DH}
    D_A = H_A + \frac{1}{2}\left([X_A, X_8] - \frac{1}{2} C_{ABC} [X_B, X_C] \right) \,, 
\end{equation}
where $C_{ABC}$ are the octonion structure constants, which have also appeared in appendix \ref{gam}. In the path integral, (\ref{eq:DH}) means that we shift the contour of the Gaussian integration over hermitian $D_A$ to an integration over hermitian $H_A$ instead. This shift does not change the value of the integral, as usual for Gaussian integrals. It has the benefit that the localisation moduli space has a simpler form. A similar contour deformation is performed in \cite{Moore:1998et, Austing:2000rm}. 

\subsection{Moduli space}\label{sec:moduli}

The equations $\delta_\epsilon \theta_\alpha = 0$ in \eqref{moduliformal} impose conditions on the bosonic matrices. In appendix \ref{app:moduli} we show that introducing the quantities,
\begin{equation}\label{redef}
A_1 = X_1 + i X_6 \,, \qquad  A_2 = X_2 - i X_5 \,, \qquad A_3 = X_3 - i X_4 \,, \qquad \phi = X_9 + i X_{10} \,,
\end{equation}
the moduli space is the space of solutions to the following equations:
\begin{equation}\label{moduli}
\begin{aligned}
\left.
\begin{aligned}
    [A_i, A_j] = 0 & \\  [\phi, A_i]  = 0 &
\end{aligned} \right\}
    & \qquad i,j \in \{1,2,3\} \,, \\
\left.    
\begin{aligned}  
    [X_7, X_a] = 0 & \\ [X_a, X_b] = i \frac{3 \Omega}{8} \epsilon_{abc} X_c &
\end{aligned} \right\} 
    &  \qquad a,b,c \in \{8,9,10\} \,, \\
    H_A = -\frac{\Omega}{16} X_A & \qquad A = 1,\ldots,7 \,.
\end{aligned}
\end{equation}
We see in (\ref{moduli}) that the stationary points organise themselves into precisely the stable fuzzy spheres in $X_{8,9,10}$ that we have encountered previously. The important difference is that now there is also a moduli space for $X_{1,\ldots,7}$ that incorporates all non-perturbative corrections to the more direct saddle point evaluation that we performed in \S\ref{semicl}. In addition to dealing with this moduli space for each fuzzy sphere, the full answer must include the contribution from the `collapsed' sphere with $X_{8,9,10} = 0$.

We will limit our attention in the remainder of this work to the contribution of the maximal fuzzy sphere saddle, i.e.~the $N$-dimensional irreducible representation of $SU(2)$. We have seen in \S\ref{semicl} that this saddle dominates the partition function at large $\Omega$. In this way we will recover our expression (\ref{pertresult}) from localisation, including the precise prefactor, giving a nontrivial check on our computations. We will furthermore evaluate this contribution at all $\Omega$ for $N=2$. Further results will be presented elsewhere.

A convenient gauge choice for solving \eqref{moduli} is to diagonalize $X_8$. That is, we may use the $SU(N)/\mathbb{Z}_N$ zero modes to set
\begin{equation}
    X_8 = \frac{3 \Omega}{8} J_3 \,,
\end{equation}
up to $S_N$ permutations of the diagonal entries of $J_3$. Recall that $J_a$ was defined below (\ref{eq:background}). Then, the fuzzy sphere equation in \eqref{moduli} fixes also 
\begin{equation}
    X_9 = \frac{3 \Omega}{8} J_1, \qquad X_{10} = \frac{3 \Omega}{8} J_2 \,,
\end{equation}
up to $U(1)^{N-1}/\mathbb{Z}^N \subset SU(N)/\mathbb{Z}^N$ residual gauge transformations that leave $X_8$ invariant. 

At this saddle, $X_7$ commuting with $X_{8,9,10}$ in \eqref{moduli} implies that
\begin{equation}
    X_7 = 0.
\end{equation}
Furthermore, from \eqref{redef} we see that $\phi$ is proportional to $J_+$. Therefore the statement in \eqref{moduli} that $\phi$ commutes with $A_i$ implies that the $A_i$ must have highest weight. Highest weight matrices are naturally written in terms of matrix spherical harmonics as
\begin{equation}\label{moduliparm}
    A_i = \sum_{l=1}^{N-1} a_i^l \, \hat{Y}_{ll} \,,  \qquad a_i^l \in \mathbb{C}.
\end{equation}
Details of matrix spherical harmonics are given in appendix \ref{app:modes}. The intuitive point here is that the highest weight condition implies that only the harmonics with quantum number $m=l$ appear.

We may set $a_i^l = u_i^l + i v_i^l$, with $u_i^l, v_i^l \in \mathbb{R}$.
With these $6(N-1)$ real coordinates on the moduli space, the full bosonic action in \eqref{eq:actionwithaux} evaluates to
\begin{equation}\label{oriaction}
    \left.S^\prime_\Omega\right|_{\mathcal{M}} = S^\prime_\Omega(u_i^l, v_i^l) = \frac{3\Omega^2}{2^7} N  \sum_{l=1}^{N-1} \sum_{i=1}^{3} (3l+1) \left[(u_i^l)^2 + (v_i^l)^2\right] - \frac{9 \Omega^4}{2^{15}} N(N^2-1) \,.
\end{equation}
It might appear surprising that the action is exactly quadratic in $(u_i^l, v_i^l)$. This occurs because the shift of the auxiliary field in \eqref{eq:DH} precisely cancels the quartic terms among $X_{1,\ldots,6}$ in the original action (\ref{eq:fullS}). We recognise the final term in (\ref{oriaction}) as the fuzzy sphere saddle point action (\ref{S0}). The first term is the additional action due to nonzero $A_i$ matrices. Reassuringly, this extra contribution is positive.
While (\ref{oriaction}) has a rotational symmetry between the different $l$ terms, this is not shared by the one-loop contributions.

To summarise, we are focusing on the maximal irreducible fuzzy sphere saddle. We fix the gauge by diagonalizing $X_8$, so the saddle point is, up to permutations and residual gauge freedoms, $X_{8,9,10} = 3\Omega/8 \, J_{3,1,2}$. This allows us to parameterize the moduli space for the other matrices by coordinates $\{u_i^l,v_i^l\},\, l = 1,\ldots,N-1$. This series of operations introduces a Vandermonde determinant and gauge volume, multiplicity of permutations, residual gauge volume, and a volume form on the moduli space, which is the Jacobian for transforming into the $\{u_i^l, v_i^l\}$ coordinates. These will all be discussed in appendix \ref{app:normalization}.

\subsection{One-loop determinants and moduli space integral}\label{sec:locresult}

The one-loop determinant in \eqref{eq:Mint} consists of a bosonic part and a fermionic part.
In the large $t$ limit the contribution from fluctuations in the $t \delta_\epsilon V$ term in the exponent in (\ref{eq:Zt}) is in principle dominant over fluctuations from the $S'_\Omega$ term. This is the case for the bosonic modes. However, the fermionic part of $t \delta_\epsilon V$ has zero modes that are not zero modes of $S'_\Omega$. We therefore need to keep track of variations of $S'_\Omega$, and we have
\begin{equation}\label{1looploc}
    Z^{\text{1-loop}}(\mathcal{M}) = 2^{3(N-1)} (2\pi)^{(8N + 5)(N-1)} \lim_{t\to\infty}  \frac{\text{Pf}\left[S^{(2)}_f + t (\delta_\epsilon V)^{(2)}_f\right]}{\sqrt{\text{det}'\left[t (\delta_\epsilon V)^{(2)}_b\right]}}.
\end{equation}
Here $(\delta_\epsilon V)^{(2)}_{b,f}$ are the bosonic and fermionic parts of the localising deformation at quadratic level, $S^{(2)}_{f}$ is the quadratic fermionic action of polarised IKKT, and $\text{det}'$ is the determinant over the subspace of non-zero bosonic modes (the zero modes are the moduli and gauge transformations). The powers of $2$ in (\ref{1looploc}), which add up to $8(N^2-1)$, arise from the $16(N^2-1)$ fermion integrals, as our fermions are real. The power of $\pi$ is given by half the number of non-zero modes in the bosonic part of $\delta_\epsilon V$. There are $17(N^2-1)$ bosonic integrals, over ($X_\mu, H_A$), but also a $6(N-1)$-dimensional moduli space and $N^2-1$ gauge transformations. Hence in the end there are $2(8N+5)(N-1)$ non-zero modes. 

The localisation argument in \eqref{dZdt} implies that the ratio of determinants in (\ref{1looploc}) should be finite and nonzero in the $t\to\infty$ limit. Indeed, the fermionic Pfaffian will be a polynomial in $t$ with degree $(\text{Rank}[(\delta_\epsilon V)^{(2)}_f])^{1/2}$ and it can be verified that $\text{Rank}[(\delta_\epsilon V)^{(2)}_f]$ is also the number of non-zero bosonic modes. Therefore, the powers of $t$ from the bosonic determinant in the denominator of \eqref{1looploc} exactly cancel those from the fermionic Pfaffian in the numerator.



The bosonic part of the quadratic localising action is
\begin{equation}
    (\delta_\epsilon V)^{(2)}_b = \text{Tr} \left((\delta_\epsilon \theta_\alpha)^{(1)\dagger}(\delta_\epsilon \theta_\alpha)^{(1)} \right) \,.
\end{equation}
Writing all bosonic matrices as a background value on the moduli space plus fluctuations,
\begin{equation}
    X_\mu = X_\mu^{(0)} + \delta X_\mu, \hspace{10pt} H_A = H_A^{(0)} + \delta H_A \,,
\end{equation}
we have that
\begin{equation}\label{bosloc1}
    (\delta_\epsilon \theta_\alpha)^{(1)} = (\sigma^{\mu\nu}\epsilon)_\alpha [X^{(0)}_\mu, \delta X_\nu] + (v_A)_\alpha (\delta H_A + \frac{1}{2} \delta \mathcal{E}_A) + i \frac{\Omega}{8} (T^\mu \epsilon)_\alpha \delta X_\mu,
\end{equation}
where 
\begin{equation}
    \delta \mathcal{E}_A = [\delta X_A, X_8^{(0)}] + [X^{(0)}_A, \delta X_8] - C_{ABC}[X^{(0)}_B, \delta X_C].
\end{equation}
As previously mentioned, a number of these fluctuations are zero-modes. The zero-modes must be removed from the fluctuations and instead be included in the moduli space integral or gauge volumes. In appendix \ref{app:factors} we write down the modes in a basis of matrix spherical harmonics and explicitly separate out the zero modes and gauge modes.

The fermionic part of both the localising deformation and the original action are quadratic when the bosonic matrices are evaluated on the moduli. The localising deformation is given by
\begin{equation}\label{locferm}
\begin{aligned}
    (\delta_\epsilon V)^{(2)}_f & = - \theta_\alpha \delta_\epsilon (\delta_\epsilon \theta_\alpha)^\dagger\\
    & = (\theta^T \sigma^{\mu\nu} \epsilon)^* [ \theta^T \Bar{\sigma}_\mu \epsilon, X^{(0)}_\nu] + \frac{1}{4} (\theta^T v_A^*)(v_A^T \Bar{\sigma}^\mu [X^{(0)}_\mu, \theta]) \\
    & \; - i \frac{\Omega}{16}(\theta^T v_A^*)(v_A^T K \theta) - i \frac{\Omega}{8} (\theta^T \Bar{\sigma}^\mu \epsilon)(\theta^T T_\mu \epsilon)^* \\
    & \; - \theta^T v_A^* \left([X^{(0)}_8, \theta^T \Bar{\sigma}^A \epsilon] + [\theta^T \Bar{\sigma}^8 \epsilon, X^{A(0)}]+ C_{ABC} [\theta^T \Bar{\sigma}^B \epsilon, X^{C(0)}] \right). 
\end{aligned}
\end{equation}
Recall from our discussion above that the complex conjugation in this expression does not act on $\theta$. The fermionic part of the original action is
\begin{equation}\label{fermoriginal}
    S_f^{(2)} = \frac{1}{2} \theta^T \Bar{\sigma}^\mu [X_\mu^{(0)}, \theta] - i \frac{\Omega}{8} \theta^T K \theta \,.
\end{equation}

The localisation procedure introduced a number of gauge volumes and Jacobians. We compute these normalisation constants in appendix \ref{app:factors}. The upshot is that the integral should be multiplied by 
\begin{equation}\label{norm}
    \mathcal{N} = \Omega^{N^2-1}\, 2^{7(N-1)(2N+1)/2} \, 3^{N^2-1} (2\pi)^{-(6N+5)(N-1)/2}\,   \frac{N \, \Gamma(N)^{2(N+1)}}{G(N+2)} \, \prod_{l=1}^{N-1} \frac{1}{l^{2l}} \,.
\end{equation}

\subsection{Results from localisation}

\subsubsection{The exact fuzzy sphere saddle for $N=2$}\label{sec:n2}

This simplest application of the formalism above is to compute the exact fuzzy sphere saddle at $N=2$. For $N=2$ the maximal fuzzy sphere is the only non-collapsed saddle.

To leading order as $t\to\infty$, the $N=2$ fermionic Pfaffian and bosonic determinant are 
\begin{equation}
\begin{aligned}
    \text{Pf}[S^{(2)}_f + t (\delta_\epsilon V)^{(2)}_f] &= t^{21} \frac{3^{13} \, 5\, \Omega^{16}}{2^{22}} \left(9 \Omega^2 + 128 u^2 \right)^2 \left(45 \Omega^2 + 128 u^2\right)^2 \,, \\
    \text{det}'[t (\delta_\epsilon V)^{(2)}_b] &= t^{42} \, 2^{12} \, 3^{26} \, \Omega^{26} \left(9 \Omega^2 + 128 u^2 \right)^8 \,.
\end{aligned}
\end{equation}
Here $u$ is a radial coordinate on the moduli space given by
\begin{equation}\label{eq:radial}
    u^2 = \sum_{i=1}^{3} \left[ (u_i^1)^2 + (v_i^1)^2 \right] \,.
\end{equation} 
The parameterisation of the moduli space in terms of a single modulus $u$
is possible due to an $SU(3)$ symmetry, $A_i \to U_{ij} A_j$ with $U \in SU(3)$, of both the original action and the deformation. The ratio of determinants in \eqref{1looploc} is then
\begin{equation}
    Z^{\text{1-loop}}(u) = \frac{5 \pi^{21} \, \Omega^3}{2^{4}}\frac{\left(45 \Omega^2 + 128 u^2\right)^2}{\left(9 \Omega^2 + 128 u^2 \right)^2} \,,
\end{equation}
and the moduli space integral becomes
\begin{align}\label{eq:N2}
    Z^\text{fuzzy}_{N=2}  =  \mathcal{N} \frac{5 \pi^{21} \, \Omega^3}{2^{4}} \int_0^{\infty} du \pi^3 u^5 \frac{\left(45 \Omega^2 + 128 u^2\right)^2}{\left(9 \Omega^2 + 128 u^2 \right)^2} \, \text{exp}\left(-\frac{3\Omega^2}{16} u^2 + \frac{27 \Omega^4}{2^{14}}\right) \,.
\end{align}
Here we have $\mathcal{N} = 2^9\, 3^3\, \Omega^3\, \pi^{-17/2}$, setting $N=2$ in (\ref{norm}). There is an additional factor of $\pi^3$ in (\ref{eq:N2}) from the conversion to radial coordinates in (\ref{eq:radial}).

At large $\Omega$, the fuzzy sphere is the dominant contribution to the partition function, and we can compare this result with the semi-classical computation in \S\ref{semicl}. At large $\Omega$ the integral (\ref{eq:N2}) simplifies as the fraction in the integrand approaches a constant. We obtain
\begin{equation}
\begin{aligned}
     Z^{\text{fuzzy}}_{N=2,\Omega \to \infty} &\approx 2^5\,3^3\,5\, \Omega^6\, \pi^{31/2} \text{exp}\left(\frac{27 \Omega^4}{2^{14}}\right)\, \int_0^{\infty} du \, u^5 \left(\frac{45}{9}\right)^2 \text{exp}\left(-\frac{3 \Omega^2}{16} u^2\right)\\
     & = 2^{17} 5^3 \pi^{31/2}\text{exp}\left(\frac{27 \Omega^4}{2^{14}}\right) \,,
\end{aligned}\label{eq:N2large}
\end{equation}
which matches exactly with the perturbative calculation in \eqref{pertresult} for $N = 2$. 

The $\Omega \to 0$ limit may also be obtained from (\ref{eq:N2}) as 
\begin{equation}
     Z^{\text{fuzzy}}_{N=2,\Omega \to 0} \approx 2^{17}\, 5 \, \pi^{31/2} \,.
\end{equation}
Essentially, one sets all of the $\Omega$ terms in the integrand to zero, except the one in the exponent that controls the convergence of the integral.
It is interesting that the contribution of the fuzzy sphere saddle remains finite in this limit. In fact,
\be
Z^{\text{fuzzy}}_{N=2,\Omega \to 0} = \frac{1}{32} Z^\text{IKKT}_{N=2} \,.
\ee
Here the IKKT result, which is our model with $\Omega=0$, can be found in e.g.~\cite{Krauth:1998xh}, taking care to match conventions appropriately. Thus the fuzzy sphere saddle gives a finite part of the full $\Omega = 0$ answer. The remaining contribution is, presumably, from the collapsed saddle. The $\Omega \to 0$ limit of that saddle is subtle and will be discussed elsewhere.

\subsubsection{General $N$ at large mass}\label{sec:alln}

The large $\Omega$ matching between perturbation theory and localisation in (\ref{eq:N2large}) can be extended to all $N$. A major simplification is that in the large $\Omega$ limit, the moduli space integral is dominated by $u_i^l = v_i^l  = 0$. This is because the action (\ref{oriaction}) is proportional to $\Omega^2 [(u_i^l)^2+ (v_i^l)^2]$, strongly suppressing larger values of $u_i^l,v_i^l$. To leading order it is therefore sufficient to consider fluctuations about the saddle with $u_i^l = v_i^l  = 0$ only. This coincides with the perturbative saddle point \eqref{eom}. The one-loop determinants for this saddle point, using the localising action, are computed in appendix \ref{locdetapp}. 
The ratio \eqref{1looploc} is found to be 
\begin{align}\label{eq:largeM}
    Z^{\text{1-loop}}_{u=v=0} = 2^{-(8N+37)(N-1)/2} & (2\pi)^{(8N+5)(N-1)} \, 3^{-(N-2)(N-1)}\, \Omega^{-(N-1)(N-5)} \nonumber \\
    & \times  N^{3(N-1)} \frac{\Gamma(N+1)}{\sqrt{N} G(N+2)^2} \prod_{l=1}^{N-1} (3l+2)^3 \,.
\end{align}
This factor does not depend on the moduli (by assumption) and therefore can be taken out of the integral over the moduli. The partition function is then simply
\be
Z_{\Omega \to \infty}^{\text{fuzzy}} = \mathcal{N} I_\mathcal{M} Z^{\text{1-loop}}_{u=v=0} \,,
\ee
where $\mathcal{N}$ is given by \eqref{norm} and the integral over the moduli
\begin{equation}
\begin{aligned}
     I_{\mathcal{M}} &= \int_{-\infty}^{\infty} \left( \prod_{i,l}d u^l_i \, dv_i^l\right) e^{-S'_\Omega(u_i^l,v_i^l)} \\
    &= e^{\frac{9 \Omega^4}{2^{15}} N (N^2 - 1)} \, 2^{18(N-1)} (2\pi)^{3 (N-1)} 3^{-3 (N-1)} N^{-3 (N-1)} \Omega^{-6(N-1)} \prod_{l=1}^{N-1} \frac{1}{(3l+1)^3} \,,
\end{aligned}
\end{equation}
where $S'_\Omega(u_i^l,v_i^l)$ is given by \eqref{oriaction}. The localisation answer for the maximal fuzzy sphere saddle at large $\Omega$ is then
\begin{equation}
\begin{aligned}
     Z_{\Omega \to \infty}^{\text{fuzzy}}
     &= \, \, 2^{3(N^2-1)} (2\pi)^{(10N + 11)(N-1)/2} N^{\frac{3}{2}} \frac{ G(N)^2\, \Gamma(N)^5}{G(N+2)^3} \prod_{l=1}^{N-1} \left( \frac{3l+2}{3l+1} \right)^3 e^{\frac{9 \Omega^4}{2^{15}} N (N^2 - 1)} \\
     & = Z_{\Omega \to \infty}
\end{aligned}
\end{equation}
where $\mathcal{N}$ is given by \eqref{norm}. The final line states that, using $G(z+1) = \Gamma(z) G(z)$, the localisation result agrees precisely
with the perturbative answer \eqref{pertresult}.

\section{Discussion}
\label{sec:disc}

The polarised IKKT model is a deformation of the IKKT matrix integral that preserves all of its supersymmetries but breaks the symmetry $SO(10) \to SO(3) \times SO(7)$. We have shown that the spacetime description of this model, in the limit of large mass deformation, is 
a spherical $D1$-brane in a supersymmetric background of Euclidean type IIB string theory. We have argued that going to smaller mass deformation corresponds to backreaction of the $D1$-brane on the spacetime and we have developed a localisation method that simplifies matrix integral computations in this limit. In this discussion section we make various remarks concerning future directions.

\subsection*{Technical next steps}

Within type IIB supergravity, the backreacted Euclidean geometries appearing in Fig.~\ref{fig:phases} have the same symmetries and supersymmetries as the ellipsoidal cavity geometry of \S\ref{sec:grav}. It should be possible to obtain these geometries explicitly. Some geometries with similar ingredients have been discussed in \cite{Bobev:2018ugk}, but do not seem to be quite what is needed here. The backreacted geometries should be analogous to certain instances of the M-theory LLM geometries \cite{Lin:2004nb}, that correspond to vacua of the BMN matrix quantum mechanics.

Within the matrix integral, it remains to solve the model at all $\Omega$. We have developed a localisation procedure that has reduced this task to computing the one-loop contribution and moduli space integral of the localising saddles. These saddles are all the fuzzy spheres together with a collapsed configuration. It will surely be possible to evaluate some of these. One interesting point here is the connection of the $\Omega \to 0$ limit to the results of IKKT localisation \cite{Moore:1998et}. As we have explained in \S\ref{sec:loc}, while our localisation procedure builds in much of the structure of that work, we do not require a regularisation of the integral by shifting contour.

Both sides of the duality seem potentially very tractable in this context, suggesting that a detailed matching is within reach and possibly a corresponding insight into the emergence of spacetime in this model.

\subsection*{Finding the backreaction in the matrix integral}

\begin{figure}[h]
    \centering
    \includegraphics[width=0.8\linewidth]{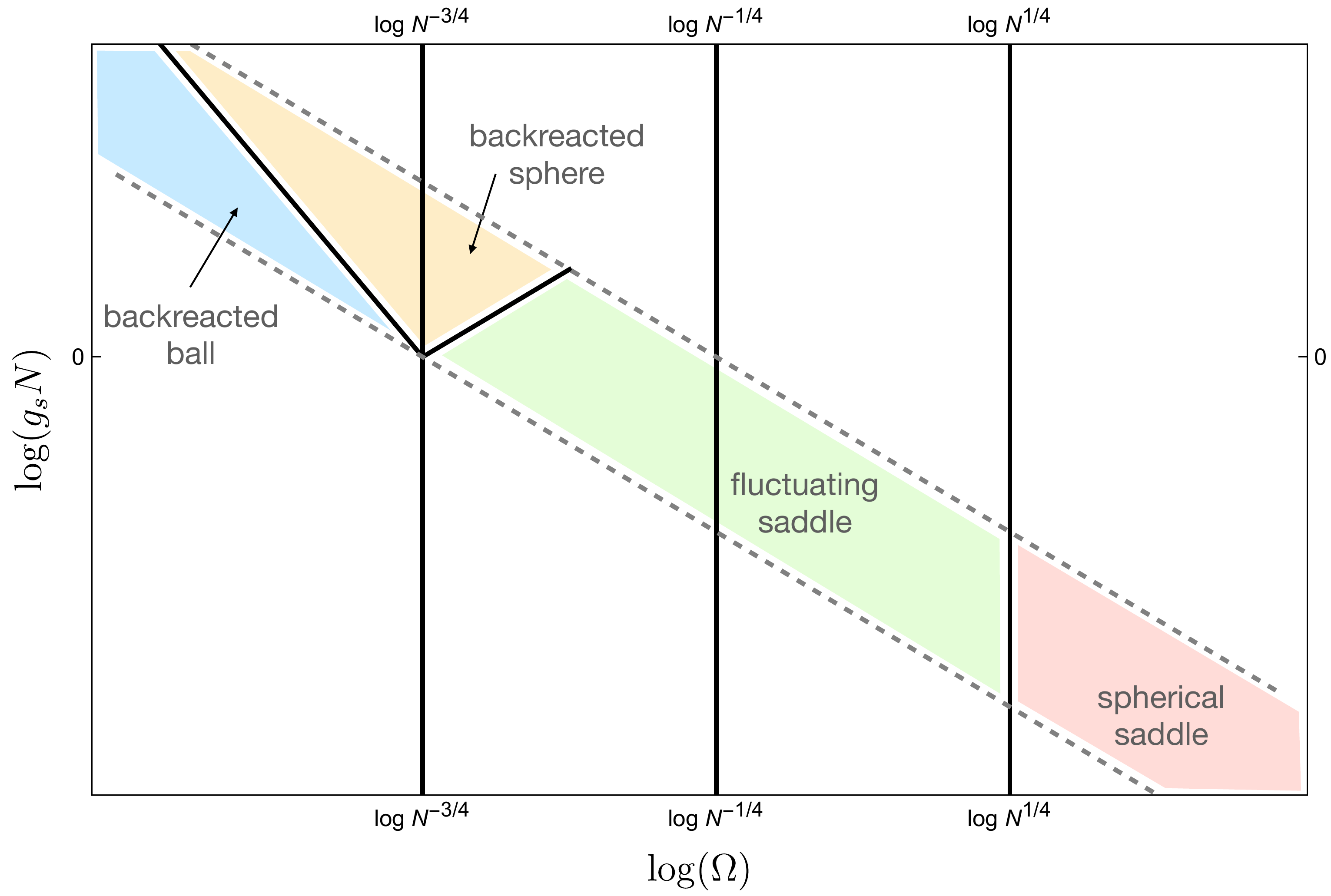}
    \caption{The phase diagram of Fig.~\ref{fig:phases} with (the logarithm of) $\Omega$ on the $x$ axis. The shaded regions, solid lines and dashed lines are in correspondence with those in Fig.~\ref{fig:phases}. The extra vertical solid line on the left shows the scale beyond which there is no suppression of subdominant saddles. The vertical line in the middle is the point at which Gaussian fluctuations compete with the action of the saddle.}
    \label{fig:phasediag}
\end{figure}

In Fig.~\ref{fig:phasediag} we have redrawn the phase diagram of Fig.~\ref{fig:phases} with the $x$ axis replaced by the matrix integral parameter $\Omega$, rather than the radius of the $D1$-brane.
In Fig.~\ref{fig:phasediag} we have added two extra solid lines, relative to Fig.~\ref{fig:phases}, at
\be\label{eq:tunnel}
\Omega^4 N^3 \sim 1 \,, \qquad \text{and} \qquad \Omega^4 N \sim 1 \,.
\ee
The first of these, the leftmost vertical line in Fig.~\ref{fig:phasediag}, sets the action of a typical fuzzy sphere saddle in (\ref{eq:act0}). Once it becomes of order one, all of the reducible fuzzy spheres contribute comparably to the integral. This is the ultimate, nonperturbative, breakdown of the saddle point approximation. It is reassuring, then, to see in Fig.~\ref{fig:phasediag} that this effect onsets in tandem with bulk backreaction. The second relation in (\ref{eq:tunnel}), the middle vertical line in Fig.~\ref{fig:phasediag}, is the scale at which Gaussian fluctuations, which are of order $e^{-N^2 \log N}$ in (\ref{pertresult}), become of comparable magnitude to the action of the saddle.

A disconcerting aspect of Fig.~\ref{fig:phasediag} is that the $y$ axis does not immediately have a corresponding quantity in the matrix integral.
The constant term (\ref{eq:DeltaS}) in the action, that does depend on $g_\text{s}$, has no influence on the dynamics. In conventional holographic dualities with time, this axis would be the 't Hooft coupling at some given energy scale \cite{Itzhaki:1998dd}. That is, this axis would correspond to the renormalisation group scale in the matrix theory and the radial location in the bulk. However, our timeless theory does not have energy scales.

The bulk, does, however have a radial direction.
In fact, there are two natural radial coordinates.
Correspondingly in the matrix theory one can consider the radial matrices
\be
R_3^2 \equiv \sum_{a=8}^{10} X_a^2 \,, \qquad R_7^2 \equiv \sum_{A=1}^7 X_A^2  \,.
\ee
We propose that $g_\text{s} N$ in Fig.~\ref{fig:phasediag} instructs us to zoom in on a certain range of these matrix radial coordinates in order to find the corresponding bulk physics. Our suggestion is that this is the analogous step to focusing on a certain window of energy scales in a QFT.

Let us start at large $\Omega$, where the bulk is described by the probe $D1$-brane embedding. In this regime the matrix $R_3$ will have an eigenvalue distribution $n(r_3)$ that is strongly peaked on the radius of the fuzzy sphere. To leading order, from the fuzzy sphere saddle (\ref{eq:background}) and the corresponding value of the $SU(2)$ Casimir,
\be\label{eq:nr3}
n(r_3) = N \delta\left(r_3 - \frac{3 \Omega N}{16} \right) \,.
\ee
This probe $D1$-brane is, from (\ref{eq:rescale}) and (\ref{eq:match}), at a corresponding bulk radial coordinate
\be\label{eq:ruler}
\frac{r_{D1}}{l_\text{s}} = (2 \pi g_\text{s})^{1/4}\, r_3 \,.
\ee
The $R_7$ matrix is instead collapsed close to the origin in this regime. From the bulk point of view, the gravitational influence of the $D1$-brane in these transverse directions is set by the backreaction scale $\ell$ introduced in \S\ref{sec:backreact}. We propose, using the same rescaling as (\ref{eq:ruler}), that the backreaction scale can be identified with a spread $\Delta r_7$ of the eigenvalue distribution of $R_7$ by
\be
\frac{\ell}{l_\text{s}} = (2 \pi g_\text{s})^{1/4}\, \Delta r_7 \,.
\ee
There is also a comparable spread $\Delta r_3 \sim \Delta r_7$.
The formula (\ref{eq:lrat}) for the backreaction scale then becomes:
\be\label{eq:delta7}
(\Delta r_7)^6  \sim \frac{1}{g_\text{s} N} \frac{1}{\Omega^2} \,.
\ee
Thus, the $y$-axis in Fig.~\ref{fig:phasediag} is instructing us to zoom in on the part of the matrix integral in which the eigenvalues of $R_7$ are within a range $\Delta r_7 \propto 1/(g_\text{s} N \Omega^2)^{1/6}$ of the origin. We should emphasise that (\ref{eq:delta7}) is not a prediction for the variance of $R_7$ in the matrix integral. It is a prediction about the part of that variance that describes the backreacted geometry in the bulk.

For the smallest values of $\Omega$ the backreaction is equally strong in all directions and swallows up the spherical brane. This is the regime that approaches the undeformed IKKT integral. From (\ref{eq:dins}), we now expect 
\be\label{eq:delta73}
(\Delta r_3)^8 \sim (\Delta r_7)^8 \sim \frac{N^2}{g_\text{s} N} \,.
\ee

It may appear surprising that $g_\text{s} N$ appears inversely in the expressions (\ref{eq:delta7}) and (\ref{eq:delta73}), because on the bulk side the backreaction length grows with $g_\text{s} N$.
This is a consequence of the factors of $g_\text{s}$ appearing in the conversion (\ref{eq:ruler}).
Thus (\ref{eq:delta7}) and (\ref{eq:delta73}) are saying that in order to find the backreaction at large $g_\text{s} N$ in the matrix integral one needs to zoom in close to the matrix origin.

\subsection*{The emergence of time}

In conventional, `timeful', versions of holography, time is given but some dimensions of space emerge.\footnote{Even within conventional models, it might be said that time emerges in black hole interiors \cite{Frenkel:2020ysx, Hartnoll:2022snh}. There is also an emergent time in dS/CFT versions of holography \cite{Strominger:2001pn, Anninos:2011ui}. The emergent time in these setups still benefits from a renormalisation group-like structure in the dual description, albeit a more subtle one.}
Because time is given, one can use the gravitational redshift in the bulk as a ruler that maps out the emergent radial direction. As we have discussed above, this ruler does not exist in timeless holography. However, conventional models of holography often also have emergent internal spaces that are not related to gravitational redshift. The dual description of
the geometry of these internal directions must be found directly within the matrix degrees of freedom. For example, the emergent $S^5$ in the holographic dual of ${\mathcal N} = 4$ SYM is closely related to the six matrix scalar fields of the QFT. Our discussion above can be paraphrased as saying that in timeless holography all emergent dimensions, including the emergent Euclidean time direction, are internal dimensions.

Euclidean geometries can define states in the gravitational theory in the spirit of Hartle and Hawking \cite{Hartle:1983ai}. The Hartle-Hawking state is just an example of a solution to the Wheeler-DeWitt equation \cite{DeWitt:1967yk}. The Wheeler-DeWitt framework involves an inherently relational notion of time that is likely well-suited to dualities in which one side has no time at all. It has proven fruitful to describe bulk evolution in the radial direction using
the Hamilton-Jacobi framework \cite{deBoer:1999tgo, Heemskerk:2010hk, Faulkner:2010jy}, which is the classical limit of the Wheeler-DeWitt equation. It should be possible to apply this structure to internal-type dimensions also. In this connection, it may be interesting to explore whether higher trace deformations of the matrix integral can be used to move in the emergent time direction, in the sense of $T^2$ deformations \cite{McGough:2016lol, Hartman:2018tkw, Araujo-Regado:2022gvw}.

Our Euclidean type IIB background does not analytically continue to a real solution of Lorentzian type IIB theory, because the axion becomes imaginary. For the purposes of generating emergent cosmological spacetimes, a Lorentzian matrix integral, in which one of the matrices is Wick rotated, may be a more promising starting point. The cosmology of the Lorentzian IKKT model has been widely studied, following \cite{Kim:2011cr}. A major theme of work on IKKT cosmology has been the possibility of dynamical spontaneous symmetry breaking. We will now make some comments on this question, from the perspective of the mass deformed model.

\subsection*{Spontaneous symmetry breaking}

The supersymmetric mass deformation provides a well-controlled framework for investigating possible spontaneous symmetry breaking in the (undeformed) Euclidean IKKT model. A large body of work has argued for spontaneous symmetry breaking $SO(10) \to SO(3)$ in the Euclidean IKKT model. Recent numerical studies include \cite{Anagnostopoulos:2015gua, Anagnostopoulos:2020xai, Kumar:2022giw, Anagnostopoulos:2022dak}, while the first hint of symmetry breaking was seen in \cite{Aoki:1998vn}. However, spontaneous symmetry breaking can only occur in a thermodynamic $N \to \infty$ limit, whereas the exact solution to the Euclidean IKKT integral \cite{Moore:1998et} does not have a well-defined large $N$ limit. It seems possible that the physics at play here may be subtle. The localisation method we have developed may allow an explicit solution of the $\Omega \to 0$ limit of the large $N$ deformed matrix integral, which may shed some light on this matter.

The previous work \cite{Kumar:2022giw} has taken a similar approach to spontaneous symmetry breaking, by studying the deformed IKKT model numerically. The numerical results presented in that paper include regimes that should be dominated by the fuzzy sphere saddle and yet show the opposite behaviour to the one expected. Namely, the 7 directions are reported to be significantly larger than the 3 directions. It seems possible to us that the numerical methods have missed the fuzzy sphere saddles.

Sponteneous symmetry breaking has also been claimed in the Lorentzian IKKT model, where it is interpreted as giving an emergent cosmology with 3+1 large spacetime dimensions \cite{Kim:2011cr, Kim:2012mw, Nishimura:2019qal, Hirasawa:2024dht, Brandenberger:2024ddi}. As we have noted above, it may be possible to study this question within a Lorentzian version of the mass-deformed model.

\subsection*{Fluctuating instanton number and the cavity}

The background cavity geometry of Fig.~\ref{fig:cavity} does not have an asymptotic region, where one could measure the $D$-instanton charge. It is possible, then, that the number $N$ of $D$-instantons inside the cavity should be allowed to fluctuate. Indeed, it is natural to promote the matrix integral (\ref{eq:Z}) to the sum over instanton number
\be\label{eq:Zsum}
Z[\Omega, g_\text{s}] = \sum_N e^{- \frac{4 \pi N}{g_\text{s}}} Z_N[\Omega] \,.
\ee
The exponent here is the instanton weighting term (\ref{eq:DeltaS}). It has been emphasised in \cite{Krauth:1998xh} that the undeformed IKKT integral is a rather non-analytic function of $N$, to the extent that a simple large $N$ limit does not exist (as we noted above), but is a nicer function of the instanton weighting parameter.\footnote{The partition function (\ref{eq:Zsum}) has a nontrivial dependence on $g_\text{s}$. One might think that this could help with the discussion above on finding the backreaction in the matrix integral, where we considered $g_\text{s} N$ as an external parameter to the integral. However, a nontrivial dependence on $g_\text{s}$ in (\ref{eq:Zsum}) has come at the cost of $N$ no longer being a free parameter. Thus the mismatch between parameters in the bulk and matrix descriptions remains.}

We are not yet in a position to evaluate (\ref{eq:Zsum}). However, it is instructive to consider, from a bulk point of view, the contribution of probe $D1$-brane configurations to the sum over $N$. This sum is necessarily truncated at a maximal worldvolume flux of $N_\text{max} \sim 1/(l_\text{s} \mu)^2 \sim 1/(\Omega^2 \sqrt{g_\text{s}})$, such that the $D1$-brane puffs up all the way to the boundary of the cavity. That is, the probe $D1$-brane contribution to (\ref{eq:Zsum}) contains a signature of the cavity:
\be\label{eq:saddlesum}
Z_{D1}[\Omega, g_\text{s}] = \sum_{N=1}^{N_\text{max}} e^{- S^\text{min}_{D1}} \,.
\ee
In appendix \ref{app:susyembed} we obtain an expression for $S^\text{min}_{D1}$ at any radius, which has the parametric form
\be\label{eq:Ng}
S^\text{min}_{D1} = \frac{N_\text{max}}{g_\text{s}} F\left(\frac{N}{N_\text{max}}\right) \,,
\ee
where $F(x)$ is a function that increases monotonically with $x$, with linear behaviour at small $x$ and a constant value at $x=1$. The behaviour of the sum in (\ref{eq:saddlesum}) is therefore seen to depend on $N_\text{max}/g_\text{s} \sim 1/\sqrt{\Omega^4 g_\text{s}^3}$. If this ratio is large, the sum is dominated by $N=1$. If the ratio is small, then all $N$ up to $N_\text{max}$ contribute equally.
This behaviour is analogous to the thermal partition function of a Debye phonon, in which the linear dispersion is cut off at a lattice wavevector.

It is interesting to note that the ratio $N_\text{max}/g_\text{s}$ can also be seen from a matrix integral description of  (\ref{eq:Zsum}), in which the cavity is not a priori manifest. Specifically, the contribution of maximal fuzzy sphere saddles to (\ref{eq:Zsum}) can be written as
\be\label{eq:saddlesum2}
Z_\text{fuz}[\Omega, g_\text{s}] \approx \sum_{N=1}^{\Omega^4} e^{- \frac{4 \pi N}{g_\text{s}}} e^{\frac{9 \Omega^4}{2^{15}} N^3} \sim \int^{\Omega^4} dN e^{-\frac{N}{g_\text{s}} + \Omega^4 N^3} \sim \int^{\frac{\Omega^4}{N_\text{max}}} dx \, e^{- \frac{N_\text{max}}{g_\text{s}} \left(x - x^3 \right)} \,.
\ee
The exponent in the integrand here has the form (\ref{eq:Ng}), although the cutoff on the integral is different to that in (\ref{eq:saddlesum}). The essential point here is that the two terms in the matrix action, $\Delta S$ and $S_\Omega$, are comparable for fuzzy spheres that correspond to $D1$-branes close to the boundary of the spherical cavity.

As discussed around (\ref{eq:conds}) above, close to the singular boundary of the cavity, the DBI worldvolume action for the $D1$-branes may not be valid and, furthermore, corrections are likely needed in the polarised IKKT model. Nonetheless, the observations just made suggest that the polarised IKKT model, summed over $N$, may capture some of the physics related to the presence of an outer boundary of the bulk spacetime.

\section*{Acknowledgements}

It is a pleasure to thank Michael Green, David Tong, Boan Zhao and Yoav Zigdon for helpful discussions. This work has been partially supported by STFC consolidated grant ST/T000694/1. SAH is partially supported by Simons Investigator award $\#$620869. JL is supported by a Harding Distinguished Postgraduate Scholarship. 

\appendix
\section{Gamma matrices and fermions}\label{gam}

A chiral basis for the $SO(10)$ gamma matrices $\Gamma^\mu$ can be built from the $SO(9)$ gamma matrices $\gamma^i$, with $i=1,\ldots, 9$:
\begin{equation}
     \Gamma^{10} = - \sigma_2 \otimes \mathbb{1}  = 
    \begin{pmatrix}
        0 & i \mathbb{1}_{16} \\
        - i \mathbb{1}_{16} & 0 
    \end{pmatrix} \,, \qquad
    \Gamma^i = \sigma_1 \otimes \gamma^i =
    \begin{pmatrix}
        0 & \gamma^i \\
        \gamma^i & 0 
    \end{pmatrix} \,,
\end{equation}
where $\gamma^i$ obeys
\begin{equation}
    \{\gamma^i, \gamma^j\} = 2 \delta^{ij} \mathbb{1}_{16}.
\end{equation}
We may therefore write the ten dimensional gamma matrices in blocks as 
\begin{equation}
    \Gamma^\mu = 
    \begin{pmatrix}
        0 & \sigma^\mu \\
        \Bar{\sigma}^\mu & 0
    \end{pmatrix} \,,
\end{equation}
where $\sigma^\mu$ and $\Bar{\sigma}^\mu$ satisfy 
\begin{equation}
    \sigma^\mu \Bar{\sigma}^\nu + \sigma^\nu \Bar{\sigma}^\mu = 2 \delta^{\mu \nu}, \hspace{15pt} \Bar{\sigma}^\mu \sigma^\nu + \Bar{\sigma}^\nu \sigma^\mu = 2 \delta^{\mu \nu}.
\end{equation}
Explicitly, from above, $\sigma^{1,\ldots,9} = \gamma^{1,\ldots,9}$ and $\sigma^{10} = -\Bar{\sigma}^{10} = i\mathbb{1}_{16}$.

An explicit construction of the $\gamma^i$ matrices is given by the octonion structure constants in the following way, with $A,B,C = 1,\ldots,7$:
\begin{equation}
        \gamma^{A} = 
        \begin{pmatrix}
        0 & t^A \\
        - t^A & 0 
        \end{pmatrix} \,, \qquad
        \gamma^8 = 
         \begin{pmatrix}
        0 & \mathbb{1}_8 \\
        \mathbb{1}_8 & 0 
        \end{pmatrix} \,, \qquad
        \gamma^9 = 
         \begin{pmatrix}
        \mathbb{1}_8 & 0 \\
        0& -\mathbb{1}_8 
        \end{pmatrix} \,,
\end{equation}
where the $t^A$ are 8 by 8 matrices given by 
\begin{equation}
    t^A_{BC} = C_{ABC} \,, 
    \qquad t^A_{B8} = - t^A_{8B} = \delta_{AB} \,, \qquad t^A_{88} = 0 \,,
\end{equation}
which are real anti-symmetric. The octonion structure constants $C_{ABC}$ are defined using the seven imaginary bases $e_{A}$ as
\begin{equation}
    e_A e_B = - \delta_{AB} + C_{ABC}e_C. 
\end{equation}
The structure constants are totally anti-symmetric among the three indices. The Euclidean gamma matrices $\Gamma^\mu$ defined in the above basis are all hermitian. 

We will be interested in real Weyl spinors. The chirality operator can be defined as
\begin{equation}
    \Gamma^{11} = -i \Gamma^{1}\ldots\Gamma^{10}, \hspace{15pt} \left(\Gamma^{11}\right)^2 = \mathbb{1}_{32} \,.
\end{equation} 
Left-handed Weyl spinors then obey
\begin{equation}
    \Gamma^{11} \Psi_{L} = \Psi_{L}.
\end{equation}
Finally, we can also define a charge conjugation matrix as
\begin{equation}
    \mathcal{C} = -i \Gamma^{10},
\end{equation}
which satisfies
\begin{equation}
    \mathcal{C} = \mathcal{C}^* = - \mathcal{C}^T, \hspace{15pt} \mathcal{C}^2 = - 1, \hspace{15pt} \Gamma^{\mu T} = - \mathcal{C}^{-1} \Gamma^\mu \mathcal{C} \,.
\end{equation}
Explicitly, in the chiral representation above, this acts on a left-handed Weyl spinor with real components $\theta$ as
\begin{equation}\label{eq:chiral}
    \Psi = 
    \begin{pmatrix}
        \theta \\
        0
    \end{pmatrix}
    \,, \qquad
    \Psi^T \mathcal{C} = 
    \begin{pmatrix}
        0 & \theta^T 
    \end{pmatrix} \,.
\end{equation}

\section{The normal modes}\label{app:modes}

This appendix describes the decomposition of $\delta X$ and $\theta$ in \S\ref{sec:quad} into normal modes, that diagonalise the kinetic operator in the various quadratic actions. We use the matrix spherical harmonics $\hat Y_{lm}$ as a convenient basis of matrices to build these normal modes. For a detailed recent discussion of matrix spherical harmonics see \cite{Han:2019wue,Frenkel:2023aft}. The following properties of these matrices will be important for us:
\begin{align}
    [J_3, \hat{Y}_{lm}] & = m \hat{Y}_{lm}, \\
    [J_{\pm}, \hat{Y}_{lm}] & = \sqrt{(l \mp m)(l \pm m +1)} \hat{Y}_{l(m\pm1)}, \\
    [J_a, [J_a, \hat{Y}_{lm}]] & = l(l+1) \hat{Y}_{lm}, \label{sphharmeigen}\\ 
    \hat{Y}_{lm} & = (-1)^m \hat{Y}^\dagger_{l(-m)}, \label{sphharmadj}\\ 
    \text{Tr} \left( \hat{Y}^\dagger_{lm} \hat{Y}_{l^\prime m^\prime} \right) & = N \delta_{l l^\prime} \delta_{m m^\prime}, \label{sphharnorm}
\end{align}
with $1 \leq l \leq N-1$ and $-l \leq m \leq l$ to be a basis for traceless $N \times N$ matrices. An explicit construction of matrix spherical harmonics is
\begin{equation}
    \hat{Y}_{l(-l)} = C J_{-}^l, \qquad \hat{Y}_{l (m+1)} = \frac{[J_+, \hat{Y}_{lm}]}{\sqrt{(l-m)(l+m+1)}} \,,
\end{equation}
where $C$ is a constant determined by the normalisation in equation \eqref{sphharnorm}.

\subsection*{Transverse bosonic modes}

Since the $X_A$ are hermitian matrices, we use equation \eqref{sphharmadj} to construct hermitian normal modes given by 
\begin{gather}\label{hermharm}
    \hat{H}^+_{lm} = \frac{1}{\sqrt{2}}\left(\hat{Y}_{lm} + (-1)^m \hat{Y}_{l(-m)} \right), \qquad 0 < m \leq l \\
    \hat{H}^-_{lm} = \frac{i}{\sqrt{2}}\left(\hat{Y}_{lm} - (-1)^m \hat{Y}_{l(-m)} \right), \qquad 0 < m \leq l  \\
    \hat{H}^0_l = \hat{Y}_{l0} \,.
\end{gather}
These are normalised so that 
\begin{equation}\label{7norm}
    \text{Tr} ( \hat H^i_{lm} \hat H^j_{l^\prime m^\prime} ) = N \delta^{ij} \delta_{l l^\prime} \delta_{m m^\prime} \,,
\end{equation}
where $i,j$ take values in $\pm$.

The result (\ref{eq:tranmodes}) for the eigenvalues then follows immediately by using the property (\ref{sphharmeigen}) of the matrix spherical harmonics in the mode equation \eqref{S7eigen}.

\subsection*{Fuzzy sphere bosonic modes}

The analysis in this case is very similar to appendix C in \cite{Han:2019wue}.
We may decompose $\hat{H}_{ia}$ into matrix spherical harmonics as
\be\label{eq:yi}
\hat H_{ia} = \sum_{lm} y^{lm}_{ia} \hat{Y}_{lm} \,.
\ee
In what follows we suppress the $i$ index, and for convenience introduce $y^{lm}_\pm = y^{lm}_1 \pm i y^{lm}_2$. Substituting \eqref{eq:yi} into the mode equation \eqref{simpeigen} gives
\begin{align}
    & \left( 3l(l+1) + 2 \right) y_3^{lm} + \sqrt{(l+m+1)(l-m)} y_+^{l(m+1)} -  \sqrt{(l-m+1)(l+m)} y_-^{l(m-1)}  = 2 \lambda \, y_3^{lm}, \nonumber \\
    & \left(3l(l+1) \mp 2 m\right) y_{\pm}^{l(m\pm1)} \pm 2 \sqrt{(l \mp m) (l \pm m +1)} y_3^{lm} = 2 \lambda \, y_\pm^{l(m\pm1)}, 
\end{align}
valid for $|m| \leq l-1$. This is an eigenvalue problem for the three variables $y_3^{lm}, y_\pm^{l(m\pm1)} $. The eigenvalues are seen to be given by the solutions to the characteristic polynomial
\begin{equation}
    \left(\lambda - \frac{3l(l+1)}{2}\right) \left(\lambda - \frac{l(3l+1)}{2}\right) \left(\lambda - \frac{(l+1)(3l+2)}{2}\right) = 0 \,.
\end{equation}
Each eigenvalue here has multiplicity $2l - 1$ (one for each possible $m$ since equations with different $m$ decouple). Correspondingly this gives $3 \sum_{l=1}^{N-1} (2l-1)  = 3 N(N-2)$ normal modes. Apart from these, there are four boundary cases with $m=l$. These are firstly,
\begin{equation}
    \begin{cases}
      \left(3l(l+1) + 2 \right) y_3^{ll} - \sqrt{2l} \, y_-^{l(l-1)} = 2\lambda \, y_3^{ll} \\
    \left(3l(l+1) + 2l\right) y_-^{l(l-1)} - 2 \sqrt{2l} \, y_3^{ll} = 2 \lambda \, y_-^{l(l-1)}
    \end{cases}, \qquad \lambda = \frac{3l(l+1)}{2}, \, \frac{(l+1)(3l+2)}{2} \,.
\end{equation}
Secondly,
\begin{equation}
    \begin{cases}
    \left(3l(l+1) + 2 \right) y_3^{ll} +  \sqrt{2l} \, y_+^{l(-l+1)} = 2 \lambda \, y_3^{ll} \\
    \left(3l(l+1) + 2l\right) y_+^{l(-l+1)} + 2 \sqrt{2l} \, y_3^{l(-l)} = 2 \lambda \, y_+^{l(-l+1)}
    \end{cases}, \qquad \lambda = \frac{3l(l+1)}{2}, \, \frac{(l+1)(3l+2)}{2}\,.
\end{equation}
Thirdly,
\begin{equation}
    (3l+2)(l+1) y_-^{ll} = 2 \lambda \, y_-^{ll}, \qquad \lambda = \frac{(l+1)(3l+2)}{2} \,.
\end{equation}
Fourthly,
\begin{equation}
    (3l+2)(l+1) y_+^{l(-l)} = 2 \lambda \, y_+^{l(-l)}, \qquad \lambda = \frac{(l+1)(3l+2)}{2} \,. 
\end{equation}
Putting the results together we obtain the following degeneracies for each eigenvalue
\begin{equation}\label{eq:lam}
\lambda =
    \begin{cases}
      \frac{3}{2}l(l+1) & \text{with degeneracy $2l+1$}\\
      \frac{1}{2}l(3l+1) & \text{with degeneracy $2l-1$}\\
      \frac{1}{2}(l+1)(3l+2) & \text{with degeneracy $2l+3$}
    \end{cases} \,.
\end{equation}
In total this gives $3(N^2-1)$ modes as expected. However, these include the zero modes which do not obey \eqref{simpeigen}. That is, the analysis we have just performed was not valid for some of the modes. We have seen in the main text that there are $N^2-1$ zero modes. From counting the degrees of freedom, we can anticipate that the modes with eigenvalue $\frac{3}{2}l(l+1)$ 
in (\ref{eq:lam}) should have been zero modes. This is indeed the case, as we now verify with an explicit construction of the modes. Thus we obtain \eqref{eq:lamresult} in the main text.

The eigenbasis can be reconstructed by inverting the eigenvalue equations above. We label the eigenbases with eigenvalues $\frac{3}{2}l(l+1), \frac{1}{2}l(3l+1), \frac{1}{2}(l+1)(3l+2)$, respectively, with subscripts $d, t, f$. We have the following general expression
\begin{equation}\label{eq:nonherm}
    \hat{H}_{a,lm}^{(d,t,f)} = 
    \begin{pmatrix}[2]
        \frac{1}{2} y^{(d,t,f);l(m+1)}_+ \hat{Y}_{l(m+1)} + \frac{1}{2}y^{(d,t,f);l(m-1)}_- \hat{Y}_{l (m-1)} \\
        \frac{1}{2 i} y^{(d,t,f);l(m+1)}_+ \hat{Y}_{l(m+1)} - \frac{1}{2 i}y^{(d,t,f);l(m-1)}_- \hat{Y}_{l (m-1)} \\
        y^{(d,t,f);lm}_3 \hat{Y}_{lm}
    \end{pmatrix}_a,
\end{equation}
where we take any $\hat{Y}_{lm}$ with $|m| > l$ to be the zero matrix. The coefficients are given by
\begin{align}
y^{d;lm}_\pm & = \sqrt{\frac{(l\mp m+1)(l \pm m)}{l(l+1)}}\,, && y^{d;lm}_3 = \frac{m}{\sqrt{l(l+1)}}\,, &&  -l \leq m \leq l \,, \\
y^{t;lm}_\pm & = \mp \sqrt{\frac{(l \pm m - 1)(l \pm m)}{l(2l+1)}}\,, && y^{t;lm}_3 = \sqrt{\frac{l^2 - m^2}{l(2l+1)}}\,, && l+1 \leq m \leq l-1 \,, \\
y^{f;lm}_\pm & = \mp \sqrt{\frac{(l \mp m + 1)(l \mp m+2)}{(l+1)(2l+1)}}\,, && y^{f;lm}_3 = -\sqrt{\frac{(l+1)^2 - m^2}{(l+1)(2l+1)}}\,, && -l-1 \leq m \leq l+1 \,.
\end{align}
With these expressions at hand one can check explicitly that the $d$-modes are zero modes of equation \eqref{S3eigen} and that they do not satisfy equation \eqref{gauge}.

We can construct a hermitian eigenbasis by taking linear combinations of modes with the same eigenvalue. For the $t$- and $f$-modes the coefficients satisfy 
\begin{equation}\label{eq:ypm}
    y^{lm}_+ = - y^{l(-m)}_-, \hspace{15pt} y^{lm}_3 = y^{l(-m)}_3.
\end{equation}
This enables one to construct the following hermitian basis:
\begin{align}
    \hat{H}^{(t,f);+}_{a,lm} &= \frac{1}{\sqrt{2}} \left(\hat{H}^{(t,f)}_{a,lm} + (-1)^m \hat{H}^{(t,f)}_{a,l(-m)} \right) \,, \label{m1} \\
    \hat{H}^{(t,f);-}_{a,lm} & = \frac{i}{\sqrt{2}} \left(\hat{H}^{(t,f)}_{a,lm} - (-1)^m \hat{H}^{(t,f)}_{a,l(-m)} \right) \,, \label{m2} \\
    \hat{H}^{(t,f);0}_{a,l0} & = \hat{H}^{(t,f)}_{a,l0} \,, 
\end{align}
with $0 < m \leq l-1$ for $t$ and $0<m \leq l+1$ for $f$. They are distinguished from the non-hermitian basis by an extra label $(\pm ,0)$ and normalised by
\begin{equation}\label{3norm}
    \sum_a \text{Tr} (\hat{H}^i_{a,lm} \hat{H}^j_{a,l^\prime m^\prime}) = N \delta^{ij} \delta_{l l^\prime} \delta_{m m^\prime},
\end{equation}
where $i,j$ indicates $(t,f);(\pm,0)$.

A similar construction produces a hermitian eigenbasis for the $d$-modes (the zero modes). For these modes, the signs must be flipped between the two expressions in (\ref{eq:ypm}), and between (\ref{m1}) and (\ref{m2}).

\subsection*{Fermionic modes}

The mode equations (\ref{fermioneigen}) are again solved by decomposing the eigenvectors into matrix spherical harmonics, $\hat{\psi}_\alpha = \sum_{lm} y^{lm}_\alpha \hat{Y}_{lm}$, where $\alpha = \{+,-\}$ labels the two components of the spinor. Then, setting $\omega = \Omega/8 \lambda $,
\begin{gather}
    \left(\frac{3}{2} m + 1 - \lambda \right) y^{lm}_- + \frac{3}{2}\sqrt{(l + m+1)(l-m)} y^{l(m+1)}_+ = 0 \,, \\
     \left(\frac{3}{2} m + \frac{1}{2} + \lambda \right) y^{l(m+1)}_+ - \frac{3}{2} \sqrt{(l + m + 1)(l-m)} y^{lm}_- = 0 \,.
\end{gather}
The problem is seen to break up into pairs of linear equations involving only $y_+^{l(m+1)}$ and $y_-^{lm}$. Nontrivial solutions only exist if the associated determinant vanishes, leading to
\begin{equation}
\lambda = 
    \begin{cases}
    -\frac{1}{2} (3l+1) & \text{with degeneracy $2l$}\\
      \frac{1}{2} (3l+2) & \text{with degeneracy $2l+2$}
    \end{cases}.
\end{equation}

The explicit eigenmodes are constructed as
\begin{equation}
    \hat{\psi}_{i,lm} = C_i (l,m) \begin{pmatrix}
        y_{i,+}^{lm} \hat{Y}_{lm} \\
        y_{i,-}^{l(m-1)} \hat{Y}_{l(m-1)}\\
    \end{pmatrix},
\end{equation}
where $i = \{1,2\}$ labels, respectively, the modes with eigenvalue $-(3l+1)/2$ or $(3l+2)/2$. For $i=1$, $-l+1 \leq m \leq l$ giving $2l$ modes; for $i = 2$, $-l \leq m \leq l+1$ giving $2l+2$ modes. The $C_i (l,m)$ are normalisation constants:
\begin{align}\label{fermnorm}
    C_1 (l,m) & = \left[(l(l+1) - m(m-1))^{1/4} \sqrt{2l+1} \right]^{-1} \,, \\
    C_2 (l, m) & = 
    \begin{cases}
        C_1(l, m) & -l < m < l+1 \\[8pt]
        \frac{1}{ \sqrt{2l+1} } & m = -l, \, l+1
    \end{cases} \,,
\end{align}
and the coefficients are given by
\begin{gather}
    y_{1,+}^{lm} = - \sqrt{l(l+1) - m(m-1)}, \hspace{20pt} y_{1,-}^{lm} = l - m , \hspace{20pt} -l+1 \leq m \leq l \,,
\end{gather}
and for the other family
\begin{equation}
    y_{2,+}^{lm} =
    \begin{cases}
       \sqrt{l(l+1) - m(m-1)}& -l < m \leq l+1\\
    1 & m = -l
    \end{cases},
    \hspace{20pt}
    y_{2,-}^{lm} = l+ m +1 \,.
\end{equation}

The relevant antisymmetrised product is
\begin{align}
    \langle \hat{\psi}^I, \hat{\psi}^J \rangle & \equiv \text{Tr} \left( \hat{\psi}^I 
    \begin{pmatrix}
        0 & 1 \\
        -1& 0
    \end{pmatrix}
    \hat{\psi}^J \right)\\
    &= \text{Tr} \left( \hat{\psi}^I_+ \hat{\psi}^J_- - \hat{\psi}^I_- \hat{\psi}^J_+ \right) \,,
\end{align}
where $I,J$ are a shorthand for $\{l,m,i\}$. The normalisation above ensures that
\begin{equation}\label{fermnormeq}
    \langle \hat{\psi}^{lm}_{i}, \hat{\psi}^{l^\prime m^\prime}_{j} \rangle = (-1)^m N \delta_{ij} \delta_{l l^\prime} \delta_{m,1-m^\prime} \,.
\end{equation}
We see that within each $l$-sector the modes are paired up such that the pair
$(m, 1-m)$ have product value $\pm N$. The sign depends on $m$ and also the ordering of the two modes. Schematically, we have established a set of modes satisfying
\begin{equation}
    \langle \hat{\psi}^I, \hat{\psi}^J \rangle = 
    \begin{pmatrix}
        0 & N & 0 & 0 & \cdots & 0 & 0 \\
        -N & 0 & 0 & 0 & \cdots & 0 & 0 \\
        0 & 0 & 0 & N & \cdots & 0 & 0\\
        0 & 0 & -N & 0 & \cdots & 0  & 0\\
        \vdots & \vdots & \vdots &  \vdots & \ddots  & \vdots  & \vdots \\
        0 & 0 & 0 & 0 & \cdots&  0 & N \\
        0 & 0 & 0 & 0 & \cdots&  -N & 0 
    \end{pmatrix}^{I J}. 
\end{equation}
The fermions are decomposed in this basis with Grassman-odd components $\alpha$ as 
\begin{equation}
    \psi = \sum_{l,m,i} \alpha^{lm}_i \hat{\psi}_{i, lm} \,.
\end{equation}
Then the fermionic action for a single copy of the two-component spinor becomes 
\begin{align}
    S^{(2)}_f &= \sum_{i=1,2} \sum_l \sum_{\text{pairs }(m,1-m)} \alpha_i^{lm} \alpha_i^{l(1-m)} 2 N \omega_i \,. \label{eq:sf2}
\end{align}
The factor of $2$ comes from each pair of modes contributing twice, once in each order. The $N$ comes from the normalisation in \eqref{fermnormeq}.
In the main text we have rescaled the Grassmann field so that the factor of $N$ in (\ref{eq:sf2}) is replaced by an additional factor of $2$. As explained in the main text, and as with the bosonic modes previously, this is to ensure that the normal modes are normalised in the same way as the generators of $SU(N)$ that define the normalisation of the original Grassmann integral.

\section{Supergravity equations of motion and Ansatz} \label{app:sugra}

The Euclidean equations of motion given by the variation of the type IIB action \eqref{eq:iibaction} are as follows. For the dilaton, axion and three-form field strengths:
\begin{align}\label{eomphi}
    \nabla^2 \phi + e^{2 \phi} \partial_\mu C_0 \partial^\mu C_0 + \frac{1}{12} e^{-\phi} H^2 + \frac{1}{12}e^{\phi} \tilde{F}^2 & = 0 \,, \\
\label{eomC}
    \nabla_{\mu}\left( e^{2 \phi} \partial^\mu C_0 \right) + \frac{1}{6} e^\phi \tilde{F}_{\mu \nu \rho} H^{\mu \nu \rho} & = 0 \,, \\\label{eomF}
    d \left( e^{\phi} \star \tilde{F} \right) & = 0 \,, \\
\label{eomH}
    d \left( e^{-\phi} \star H + e^\phi C_0 \star \tilde{F}\right) & = 0 \,,
\end{align}
while the Einstein equation is
\be
\label{eomgrav}
    R_{\mu\nu} = \frac{1}{2} \partial_\mu \phi \partial_\nu \phi - \frac{1}{2}e^{2\phi} \partial_\mu C_0 \partial_\nu C_0 + \frac{1}{4} e^{-\phi} H^2_{\mu \nu} - \frac{1}{4} e^\phi \tilde{F}^2_{\mu\nu} - \frac{1}{48} g_{\mu\nu} \left( e^{-\phi} H^2 - e^{\phi} \tilde{F}^2 \right) \,,
\ee
where we use the shorthand $A^2_{\mu \nu} = A_{\mu \rho \sigma} A_\nu^{\ \rho \sigma}$. 

We consider the Ansatz
\begin{equation}\label{ansatz}
    d \phi = e^{\phi} dC_0, \qquad \tilde{F} = e^{-\phi} H \,.
\end{equation}
With this Ansatz the axion equation becomes the same as that of the dilaton, and the RR three-form equation becomes the same as that of the NSNS three-form. The various contributions to the energy-momentum tensor in (\ref{eomgrav}) cancel precisely, so that the Einstein equation demands Ricci flatness. Thus the equations of motion simplify to
\begin{equation}\label{eq:eomsimp}
    \nabla^2 e^{\phi} = - \frac{1}{6} H^2 \,, \qquad d \star H = 0 \,, \qquad R_{\mu \nu} = 0.
\end{equation}

We may note that the first equation in (\ref{ansatz}) implies that $C_0 = - e^{-\phi} + \text{const}$. We may use an $SL(2,\R)$ transformation to set the constant to zero. The second equation in (\ref{ansatz}), recall that $\tilde{F} = F - C_0 H$, then implies that the RR three-form vanishes, $F = 0$.

The next step is to impose supersymmetry. With the Ansatz \eqref{ansatz}, the supersymmetry transformation $\delta \lambda^+$ in (\ref{eq:susy}) vanishes automatically while $\delta \lambda^- = 0$ imposes the relation
\begin{equation}\label{eq:relatepm}
    \epsilon^+ = M \epsilon^- \,, \qquad M \equiv \frac{1}{2} e^{-\phi/2} (\Gamma^\mu \partial_\mu \phi)^{-1} \Gamma^{(3)} H \,.
\end{equation}
Thus the number of independent spinor components is halved, so that sixteen supersymmetries are preserved. Substituting the relation (\ref{eq:relatepm}) into the gravitino transformation in (\ref{eq:susy}) and requiring $\delta \psi^+_\mu = 0$ we have
\begin{equation}\label{ksep}
    \nabla_\mu (M \epsilon^-) - \frac{1}{4} \partial_\mu \phi M \epsilon^- = 0 \,. 
\end{equation}
The integrability condition of (\ref{ksep}) gives
\begin{equation}
\begin{aligned}
    [\nabla_\mu - \frac{1}{4} \partial_\mu \phi, \nabla_\nu - \frac{1}{4} \partial_\nu \phi] M \epsilon^- &= \left([\nabla_\mu, \nabla_\nu] - \frac{1}{4}[\nabla_\mu, \partial_\nu \phi] - \frac{1}{4}[\partial_\mu \phi, \nabla_\nu] \right) M \epsilon^- \\
    &= \frac{1}{4} R^{ab}_{\mu \nu} \Gamma_{ab} M \epsilon^- = 0 \,.
\end{aligned}
\end{equation}
This is solved by a flat Euclidean metric
\begin{equation}\label{gsol}
    g_{\mu \nu} = \delta_{\mu \nu} \,,
\end{equation}
and thereafter \eqref{ksep} solved by
\begin{equation}\label{spinsol}
    \epsilon^- = M^{-1} e^{\phi/4} \epsilon_0,\qquad \epsilon^+ = M \epsilon^- = e^{\phi/4} \epsilon_0 \,,
\end{equation}
with $\epsilon_0$ a constant spinor.

Having obtained the flat Euclidean metric, it is now useful to make a more specific Ansatz for the NSNS three-form. Since the polarized IKKT model includes a Myers term of the form $X_8 [X_9,X_{10}]$, it is natural to align the three form only along the $\{8,9,10\}$ directions. Then \eqref{eq:eomsimp} and the Bianchi identity for $H$ imply that
\begin{equation}\label{hsol}
    H = \mu \, dx^9 \wedge dx^9 \wedge dx^{10},
\end{equation}
with $\mu$ a constant. This Ansatz breaks the $SO(10)$ symmetry of the metric down to the $SO(3)\times SO(7)$ symmetry of the mass-deformed matrix integral.

Substituting \eqref{ansatz}, \eqref{hsol}, \eqref{gsol} and \eqref{spinsol} into the final remaining supersymmetry transformation, $\delta \psi_\mu^- = 0$, we have
\begin{equation}\label{eq:kse2}
    \left(\nabla_\mu (M^{-1}) + \frac{1}{2} \partial_\mu \phi M^{-1} - \frac{1}{8} e^{-\phi/2}\left(\Gamma_\mu \Gamma^{(3)} + 2 \Gamma^{(3)} \Gamma_\mu \right) H \right) e^{\phi/4} \epsilon_0 = 0 \,.
\end{equation}
As we show explicitly below, (\ref{eq:kse2}) is solved without further constraints on $\epsilon_0$ by 
\begin{equation}\label{eq:phisol}
    e^{\phi} = 1 - \frac{\mu^2}{32}\left(\sum_{A = 1}^7 x_A^2 + 3 \sum_{a=8}^{10} x_a^2 \right). 
\end{equation}
The constant value of $e^\phi$ at the origin has been set to unity. This reference value of the string coupling can be re-instated by rescaling the fields as described in footnote \ref{foot:rescale} in the main text.

We have therefore obtained the background (\ref{eq:ds2}), (\ref{eq:dilaton}) and (\ref{eq:H}) given in the main text and shown that it preserves sixteen supersymmetries. Recall that we have vanishing RR three-form. This background solves the IIB equations of motion above.

Finally, we have promised to give some details on solving (\ref{eq:kse2}). In Cartesian coordinates \eqref{eq:kse2} requires
\begin{equation}\label{eq:kse2flat}
    \partial_\nu M^{-1} + \frac{1}{2} \partial_\nu \phi M^{-1} - \frac{1}{8} e^{-\phi/2} \left(\Gamma_\nu \Gamma^{(3)} + 2 \Gamma^{(3)} \Gamma_\nu\right) H =0 \,. 
\end{equation}
The inverse of $M$ can be found as
\begin{equation}
    M^{-1} = -12 e^{\phi/2} \frac{ \Gamma^{(3)}H}{H^2} \Gamma^\rho \partial_\rho \phi \,.
\end{equation}
Plugging in our ansatz $H = \mu dx^8 \wedge dx^9 \wedge dx^{10}$ we have
\begin{equation}
    M^{-1} =  - \frac{2}{\mu} e^{\phi/2} \Gamma^{8}\Gamma^9 \Gamma^{10} \Gamma^\rho \partial_\rho \phi \,,
\end{equation}
so that \eqref{eq:kse2flat} becomes
\begin{equation}
    \hat{N} \Gamma^\rho \partial_\nu \phi \partial_\rho \phi + \hat{N} \Gamma^\rho \partial_\nu \partial_\rho \phi + \frac{\mu^2}{16} e^{-\phi} \left(\Gamma_\nu \hat{N} + 2\hat{N} \Gamma_\nu \right) = 0 \,, \label{eq:NNN}
\end{equation}
where we used $\hat{N} = - \Gamma^8 \Gamma^9 \Gamma^{10}$. The minus sign here is to align with the definition below (\ref{eq:fullS}). One should note the similarity of the $\Gamma^\mu \hat{N} + 2 \hat{N} \Gamma^\mu$ term in (\ref{eq:NNN}) with $T^\mu$ defined below (\ref{eq:masssusy}).

Multiplying (\ref{eq:NNN}) through by $\hat{N}$ we may write the equation as
\begin{equation}\label{eq:GamN}
    \Gamma^\rho \partial_\nu \partial_\rho e^\phi = \frac{\mu^2}{16} \left( \hat{N} \Gamma_\nu \hat{N} - 2 \Gamma_\nu \right). 
\end{equation}
In this expression we may consider two cases. The first is when $\nu \neq 8,9,10$ then $\hat{N} \Gamma_\nu = -\Gamma_\nu \hat{N}$ so that the two terms on the right tend to cancel leaving only one $\Gamma_\nu$. The other case is when $\nu = 8,9,10$, so that $\hat{N} \Gamma_\nu = \Gamma_\nu \hat{N}$ and the two terms on the right add up to three times $\Gamma_\nu$:
\begin{equation}
\begin{aligned}
    \Gamma^\rho \partial_A \partial_\rho e^\phi = -\frac{\mu^2}{16} \Gamma_A, \hspace{15pt} A = 1,\ldots,7 \,, \\
    \Gamma^\rho \partial_a \partial_\rho e^\phi = -\frac{3\mu^2}{16} \Gamma_a, \hspace{15pt} a = 8,9,10 \,.
\end{aligned}
\end{equation}
Therefore (\ref{eq:GamN}) can be written
\begin{equation}
    \partial_\nu \partial_\rho e^\phi = -\frac{\mu^2}{16}
    \begin{pmatrix}
        \mathbb{1}_{7} & 0 \\
        0 & 3\mathbb{1}_3
    \end{pmatrix}_{\nu \rho} \,.
\end{equation}
The solution to this equation is the advertised (\ref{eq:phisol}). That solution is immediately recognised as the particular integral of the simplified dilaton equation of motion in (\ref{eq:eomsimp}).

We could add to (\ref{eq:phisol}) a complementary function $f$ satisfying $\partial_\nu \partial_\rho f = 0$. However, such a function is necessarily linear and can therefore be absorbed into the quadratic solution in (\ref{eq:phisol}) by a constant coordinate shift. Such a coordinate shift does not change the metric nor the NSNS three-form, since they are both homogeneous. 


\section{Supersymmetry of the brane embedding}\label{app:susyembed}

In this appendix we show that the stable brane embeddings found in \S\ref{sec:probe} preserve all of the supersymmetries of the background, discussed in appendix \ref{app:sugra}. The spherical branes with radii $r = \frac{3}{4} N \pi \alpha^\prime \mu$ and $r=0$ are supersymmetric while the brane at the maximum of the potential, $r = \frac{1}{4} N \pi \alpha^\prime \mu$, is not. These statements match the corresponding ones in the matrix model.

The spacetime supersymmetries preserved by the $D1$-brane embedding obey \cite{Marino:1999af} 
\begin{equation}\label{eq:kappa}
\begin{aligned}
    \frac{\sqrt{\text{det}\mathcal{G}}}{\sqrt{\text{det}(\mathcal{G} + \mathcal{M})}} \left(1 + \frac{1}{2}\gamma^{ij}\mathcal{M}_{ij} \right) \Gamma_{(0)} \epsilon_1 &=  \epsilon_2 \,, \\
    \frac{\sqrt{\text{det}\mathcal{G}}}{\sqrt{\text{det}(\mathcal{G} + \mathcal{M})}} \left(1 - \frac{1}{2}\gamma^{ij}\mathcal{M}_{ij} \right) \Gamma_{(0)} \epsilon_2 &= - \epsilon_1 \,.
\end{aligned}
\end{equation}
Here we have adapted the results of \cite{Marino:1999af} to Euclidean signature, by changing an $i$ on the right hand side of the equations above to $\pm$ and letting their $\eta_{1,2} \to \epsilon_{2,1}$. We will obtain a check of this procedure below when we find that solutions to (\ref{eq:kappa}) are also solutions to the full Euclidean $D1$-brane equations of motion. In (\ref{eq:kappa}) we have set $\mathcal{M} \equiv \mathcal{B} - 2 \pi \alpha^\prime \mathcal{F}$, while $\gamma^i$ are the pull-back of the spacetime $\Gamma$-matrices. We have also introduced
\begin{equation}
    \Gamma_{(0)} \equiv \frac{1}{2\sqrt{\text{det}\mathcal{G}}} \epsilon^{ij} \gamma_{ij} \,,
\end{equation}
where $\epsilon^{ij}$ is the antisymmetric matrix with $\epsilon^{12} = 1$. In (\ref{eq:kappa}) the spinors $\epsilon_{1,2}$ are related to the supersymmetries of the background by
\begin{equation}
    \epsilon_1 = \frac{1}{2}\left( \epsilon^+ + \epsilon^-\right), \qquad \epsilon_2 = \frac{1}{2}\left( \epsilon^+ - \epsilon^-\right) \,.
\end{equation}
See also the discussion below (\ref{eq:sl2r1}) in the main text.

The second equation in \eqref{eq:kappa} is equivalent to the first, so we only consider the first equation from now on. 
Then, separately from \eqref{eq:kappa}, the supersymmetry condition on the background \eqref{eq:relatepm} imposes 
\begin{equation}
    - (1 + M)^{-1} (1 - M) \epsilon_1 =   \epsilon_2 \,.
\end{equation}
Hence, for supersymmetric embeddings we must have
\begin{equation}
    \frac{1}{2\sqrt{\text{det}(\gamma + \mathcal{M})}}\left(1 - \frac{1}{2}\gamma^{ij}\mathcal{M}_{ij} \right) \epsilon^{kl} \gamma_{kl} \, \epsilon_1 = -(1 + M)^{-1} (1-M) \epsilon_1 \,.
\end{equation}
Evaluating both sides explicitly using the embedding Ansatz in \S\ref{sec:probe}, placing the brane at the origin in the $x_A$ directions but leaving $r$ unfixed, we arrive at
\be\label{eq:cond}
     \frac{-\left(\frac{\mu}{3} r^3 - N \pi \alpha^\prime \right) + e^{\phi/2}r^2 \Gamma^{\hat{\theta}} \Gamma^{\hat{\phi}} }{\sqrt{e^\phi r^4 + \left(\frac{\mu}{3} r^3 - N \pi \alpha^\prime\right)^2}}
    = \frac{m(r)^2 - 1 + 2 m(r) \Gamma^{\hat{\theta}}\Gamma^{\hat{\phi}}}{1 + m(r)^2} \,,
\ee
where we defined
\begin{equation}
    m(r) \equiv \frac{8 \sqrt{1-\frac{3 \mu ^2}{32}r^2}}{3 \mu r} \,.
\end{equation}
The hatted $\Gamma$-matrices satisfy the flat space Clifford algebra, $\{\Gamma^{\hat{\rho}}, \Gamma^{\hat{\sigma}} \} = 2 \delta^{\hat{\rho} \hat{\sigma}}$.

The supersymmetry condition (\ref{eq:cond}) is solved if the following two equations are satisfied simultaneously
\begin{gather}
    \frac{\frac{\mu}{3} r^3 - N \pi \alpha^\prime }{\sqrt{\left( 1- \frac{3 \mu^2}{32}r^2 \right) r^4 + \left(\frac{\mu}{3} r^3 - N \pi \alpha^\prime\right)^2}} = \frac{15 \mu^2 r^2 - 64}{3 \mu^2 r^2+64} \,, \label{eq:susy1}\\
    \frac{r^2}{\sqrt{\left( 1- \frac{3\mu^2}{32}r^2 \right) r^4 + \left(\frac{\mu}{3} r^3 - N \pi \alpha^\prime\right)^2}} = \frac{48 \mu r}{3\, \mu^2 r^2+64} \,. \label{eq:susy2}
\end{gather}
We can perform the same rescaling as \eqref{eq:rescale} and expand these two equations, analogously to \eqref{eq:actionexpand} in the main text. This leads to
\begin{gather}
    \frac{y^2}{2} \left(y^2 - \frac{9}{16} \right) \eta^2 + \mathcal{O}(\eta^4) = 0 \,,\\
    y \left(y - \frac{3}{4} \right) \eta + \mathcal{O} (\eta^3) = 0 \,.
\end{gather}
Thus we recover the stable solutions in \eqref{eq:ysol}:
\begin{equation}
    y = 0, \ \frac{3}{4} \,.
\end{equation}

One can verify that, without performing the small $\eta$ expansion, there are two stable solutions to the full equation of motion \eqref{eq:reom} and that these agree with the simultaneous solutions to \eqref{eq:susy1} and \eqref{eq:susy2}. Thus the agreement between the $D$1-brane probe and the matrix integral, in terms of supersymmetric minima, extends beyond the small $\eta$ limit. Explicitly, the nontrivial supersymmetric solution is at
\be
y = \frac{2}{3^{2/3} \eta} \frac{- 3^{1/3} \cdot 16 + (27 \eta + \sqrt{2^{12}3 + 3^6 \eta^2})^{2/3} }{(27 \eta + \sqrt{2^{12}3 + 3^6 \eta^2})^{1/3}} \,.
\ee
The action on this solution is given by the remarkably simple expression
\be
S^\text{min}_{D1} = \frac{4 \pi T_1}{g_\text{s}} \frac{8 \eta y}{3 \mu^2}  = \frac{4 \pi T_1}{g_\text{s}}\frac{8 r}{3 \mu} \,.
\ee

\section{Match of fluctuations about the brane and fuzzy sphere}\label{app:fieldtheory}

In this appendix we show that the quadratic action describing fluctuations about the spherical $D$1-brane is equal to the quadratic action describing fluctuations about the matrix fuzzy sphere.

\subsection*{Fluctuations about the spherical $D1$-brane}

We parametrise fluctuations of the worldvolume fields about the spherical $D1$-brane configuration in \S\ref{sec:probe} by fields $\delta r$, $X_A$ and $f$. These are all function of the angles $\theta, \phi$ on the sphere. The fluctuation $f$ of the worldvolume Maxwell field is defined by
\be
    \mathcal{F} = \left(\frac{N}{2} +  \frac{2}{3 \pi \alpha^\prime} \ f \right) \, \text{sin}\theta \, d\theta \wedge d\phi \,.
\ee
The fluctuations have the action, with $d\Omega = \sin\theta d\theta d\phi$,
\begin{align}
    S_{D1} = \frac{T_1}{g_\text{s}} & \int d\Omega \, e^{-\phi}
    \left\{ \left[\vphantom{\left(\frac{\mu}{3} (r + \d r)^3 - \pi \alpha^\prime N - \frac{4}{3 \mu} f\right)^2} e^\phi (r + \d r)^2 \left[(r + \d r)^2 + \partial_i X^A \partial^i X_A + \partial_i \d r \partial^i \d r + \left( \varepsilon^{ij}\partial_i X_A \partial_j \d r \right)^2 \right] \right. \right. \nonumber\\
    & + \left. \left. \left(\frac{\mu}{3} (r + \d r)^3 - \pi \alpha^\prime N - \frac{4}{3 \mu} f\right)^2\right]^{1/2}
     - \left(\frac{\mu}{3} (r + \d r)^3 - \pi \alpha^\prime N - \frac{4}{3 \mu} f\right) \right\} \,,
\end{align}
where 
\begin{equation}
    e^\phi = 1 - \frac{\mu^2}{32}\left(X_A X^A + 3 (r + \d r)^2\right).
\end{equation}
In these equations, $A = 1,\ldots,7$ are contracted with the flat metric $\delta_{AB}$ while $i = 1,2$ are contracted with the usual round metric on a unit sphere. 

Now we first expand to quadratic order in the fluctuations, and then take leading order in the $\eta$-expansion (as in \S\ref{sec:probe}). The quadratic action is found to be
\begin{align}
    & S_{D1}^{(2)}  = \frac{16 \mu^2 N \pi \alpha^\prime}{9} \frac{T_1}{g_\text{s}} \int d\Omega \left(\textstyle  \frac{1}{2} \partial_i X^A \partial^i X_A + \frac{1}{2} \partial_i \d r \partial^i \d r + \frac{1}{2}f^2 - \frac{4}{3} f \d r + \frac{2}{3} \d r^2 + \frac{1}{9} X_A X^A \right) \label{eq:s21} \\
    & = \frac{\Omega^2 N }{9 \pi } \frac{T_{-1}}{g_\text{s}} \int d\Omega \left(\textstyle \frac{1}{2} \partial_i X^A \partial^i X_A + \frac{1}{2} \partial_i \d r \partial^i \d r + \frac{1}{4} \mathscr{F}_{ij} \mathscr{F}^{ij} - \frac{2}{3} \d r \varepsilon^{ij} \mathscr{F}_{ij} + \frac{2}{3} \d r^2 + \frac{1}{9} X_A X^A \right) \,,\nonumber
\end{align}
where $\varepsilon = \sin\theta\, d\theta \wedge d\phi$ is the volume form on a unit sphere, and $\mathscr{F}_{ij} = f \varepsilon_{ij}$.

\subsection*{Continuum limit of matrix fluctuations}

In this section we will write the matrix theory as a noncommutative gauge theory on the fuzzy sphere saddle, following the approach of \cite{Iso:2001mg}. Taking the commutative limit (the regime of validity of which we discuss below) gives a conventional Abelian gauge theory on the sphere. This theory will match the quadratic fluctuations about the $D1$-brane, derived above.

We will focus on the bosonic part. The first step is to re-write the matrices as
\begin{equation}\label{eq:xa}
    X_\mu = X_\mu^{(0)} + L \hat{a}_\mu \,,
\end{equation}
where $L = 3 \Omega/8$, $X^{(0)}_a = L J_a$, and $X^{(0)}_A = 0$. Feeding (\ref{eq:xa}) into the bosonic part of the matrix action \eqref{eq:fullS} we obtain
\begin{multline}
    S_B = \text{Tr}\left( -\frac{1}{4} \left(L^4 \, [\hat{a}_A, \hat{a}_B]^2 + 2 L^2 \, [\hat{a}_A, L(J_a +  \hat{a}_a)]^2 + L^4 \, [J_a +  \hat{a}_a, J_b +  \hat{a}_b]^2 \right) \right.\\
    \left.+ \frac{\Omega^2}{4^3} L^2 \hat{a}_A^2 + \frac{3 \Omega^2}{4^3}L^2 (J_a +  \hat{a}_a)^2 + \frac{i \Omega}{6} L^3 \epsilon_{abc} \left(J_a +  \hat{a}_a \right) [J_b + \hat{a}_b ,J_c +  \hat{a}_c]\right) \,.
\end{multline}
We define the combination
\begin{equation}\label{eq:F}
    \hat{F}_{ab} \equiv  [J_a, \hat{a}_b] -  [J_b, \hat{a}_a] + [\hat{a}_a, \hat{a}_b] - i \epsilon_{abc} \hat{a}_c \,.
\end{equation}
The bosonic action is then
\begin{equation}\label{eq:actmat}
\begin{aligned}
    S_B = S^{(0)}_B &+ L^4 \, \text{Tr} \left( -\frac{1}{4} [\hat{a}_A, \hat{a}_B]^2 - \frac{1}{2}  \, \left([J_a, \hat{a}_A] + [\hat{a}_a, \hat{a}_A] \right)^2 + \frac{1}{9}  a_A^2 \right. \\
    &\left. - \frac{1}{4} \hat{F}_{ab} \hat{F}^{ab} - \frac{i}{6}  \epsilon_{abc} \left( J_a [\hat{a}_b, \hat{a}_c] + \frac{1}{3} \hat{a}_a \, [\hat{a}_b, \hat{a}_c] - \frac{i}{2} \epsilon_{abf} \hat{a}_f \hat{a}_c \right) \right) \,,
\end{aligned}
\end{equation}
where $S^{(0)}$ is the on-shell action of the fuzzy sphere.

As described in \cite{Iso:2001mg}, we can define a map from matrices to functions on a unit sphere using the basis of matrix spherical harmonics. This is the Moyal map. For a hermitian matrix $\hat{f}$, the corresponding function $f(\theta, \phi)$ is found by
\begin{equation}\label{eq:moyal}
    \hat{f} = \sum_{l,m} f^{lm} \hat{Y}_{lm} \;\; \to \;\; f(\theta, \phi) = \sum_{lm} f^{lm} Y_{lm}(\theta, \phi), 
\end{equation}
where $\hat{Y}_{lm}$ are the matrix spherical harmonics, see appendix \ref{app:modes}, and $Y_{lm}(\theta, \phi)$ are the usual spherical harmonics on a sphere. Recall that the matrix spherical harmonics, and hence these sums, are cut off at $l=N$. Under the map (\ref{eq:moyal}), matrix multiplication defines the non-commutative star product, so that
\begin{equation}
    \hat{a}\hat{b} \;\; \to \;\; a \star b.
\end{equation}
For more details on the star product see appendix A of \cite{Iso:2001mg}. 

Under the Moyal map we also have
\begin{equation}\label{eq:map}
\begin{aligned}
    \text{Ad}\left(J_a\right) \;\; & \to \;\; \frac{1}{i} \epsilon_{abc} x_b \partial_c \equiv \mathcal{L}_a, \\
    \frac{1}{N} \text{Tr} \;\; & \to \;\; \frac{1}{ 4\pi} \int d\Omega,
\end{aligned}
\end{equation}
where $\mathcal{L}_a$ are the usual angular momentum operators in the position basis. Applying this map to the action \eqref{eq:actmat} leads to the 
non-commutative theory
\begin{equation}
\begin{aligned}
    S_B = \frac{N L^4}{4 \pi} \int d\Omega \left(-\frac{1}{2} \left( \mathcal{L}_a a_A + [a_a, a_A]_\star \right)_\star^2 + \frac{1}{9} (a_A^2)_\star -\frac{1}{4} [a_a, a_b]^2_\star \right) \\
    - \frac{1}{4} \left(F_{ab} F^{ab} \right)_\star - \frac{i}{6} \epsilon_{abc} \left(a_a \mathcal{L}_b a_c + \frac{1}{3} a_a [a_b, a_c] - \frac{i}{2} \epsilon_{abf} a_c a_f \right)_\star. \label{eq:nc}
\end{aligned}
\end{equation}
Here we used
\begin{equation}\label{eq:Ffield}
    F_{ab} =  \mathcal{L}_a a_b - \mathcal{L}_b a_a + [a_a, a_b]_\star - i\epsilon_{abc} a_c \,,
\end{equation}
which is the non-commutative field version of \eqref{eq:F}. The noncommutative action (\ref{eq:nc}) is equal to the matrix action (\ref{eq:actmat}), with no approximations made. The Moyal map is simply a re-writing, where all the complexity of matrix multiplication is hidden in the star product.

The Moyal map is useful because in the large $N$ limit (see also comments below regarding the implicit additional necessity of the large $\Omega$ limit) the star product tends to the usual commutative product on functions. In particular, commutators vanish. In this limit we have
\be\label{eq:Scom}
    S_{B}^{N \to \infty} = \frac{N L^4}{4 \pi} \int d\Omega \left[-\frac{1}{2}\left(\mathcal{L}_a a_A\right)^2 + \frac{1}{9} a_A^2 - \frac{1}{4} F_{ab} F^{ab} - \frac{i}{12} \epsilon_{abc} a_a F_{bc} \right]. 
\ee
At this stage, the fields still carry $SO(3)$ indices $a,b,c$. Also, the differential operator $\mathcal{L}_a$, as defined in \eqref{eq:map}, acts with respect to $x_{8,9,10}$. Now we change them all to carry only worldvolume indices, by pulling them back onto the spherical worldvolume. We define $K_a^i$ in terms of the induced metric by
\begin{equation}
    \mathcal{G}^{ij} = K_a^i K_b^j \delta^{ab} \,,
\end{equation}
where $a,b \in \{8,9,10\}$ and $i,j \in \{\theta, \phi \}$.
Explicitly, $K^i_a = \epsilon_{abc} x^b \pa \theta^i/\pa x^c$. Then we also have
\begin{equation}
    \mathcal{L}_a = -i K_a^{i} \partial_{i} \,.
\end{equation}
Using this fact, the kinetic and mass terms of the transverse scalar fields $a_A$ become
\begin{equation}
    -\frac{1}{2} (\mathcal{L}_a a_A)^2+ \frac{1}{9} a_A^2 = \frac{1}{2} \partial_i \Phi_A \partial^{i} \Phi^A + \frac{1}{9} \Phi_A^2, 
\end{equation}
where we defined $\Phi_A \equiv  a_A$. We can furthermore decompose the $SO(3)$ vector $a_a$ into a radial scalar and a two-component vector on the unit sphere as
\begin{equation}\label{eq:redef}
    a_a \equiv K_a^i A_i + x_a \delta r \,.
\end{equation}
The commutative limit of \eqref{eq:Ffield} then becomes
\begin{equation}
    F_{ab} =   -i K_a^i K_b^j F_{ij} - i (K_a^i x_b - K_b^i x_a) \partial_i \delta r + i \epsilon_{abc} x_c \delta r \,,
\end{equation}
where $F_{ij} = \partial_i A_j - \partial_j A_i$. The $F_{ab}$ terms in \eqref{eq:Scom} generate kinetic terms for the vector $A_i$ and the scalar $\delta r$:
\begin{equation}
    -\frac{1}{4} F_{ab} F^{ab} = \frac{1}{4} \left(F_{ij} F^{ij} + 2 \partial_i \d r \partial^i \d r + 2 \d r^2 - 2 \varepsilon^{ij}F_{ij} \d r \right) \,,
\end{equation}
while the final term in \eqref{eq:Scom} gives us
\begin{equation}
    - \frac{i}{12} \epsilon_{abc} a_a F_{bc} = \frac{1}{6} \left( \d r^2 - \varepsilon^{ij} F_{ij} \d r \right). 
\end{equation}
Combining the above we transform \eqref{eq:Scom} into
\begin{equation}\label{eq:sbfinal}
    S_{B}^{N \to \infty} = \frac{L^4 N}{4 \pi} \int d\Omega \left[ \frac{1}{2} \partial_i \Phi^A \partial^i \Phi_A + \frac{1}{9} \Phi_A^2 + \frac{1}{4} F_{ij} F^{ij} + \frac{1}{2} \partial_i \d r \partial^i \d r + \frac{2}{3} \d r^2 - \frac{2}{3} \varepsilon^{ij} F_{ij} \d r \right].
\end{equation}
The action is seen to be in precise agreement with (\ref{eq:s21}). The overall normalisation of the action can be matched by rescaling the fields.

While the large $N$ limit is sufficient to formally obtain the continuum theory (\ref{eq:sbfinal}) from the polarised IKKT action, the large $\Omega$ limit is also needed for these manipulations to be valid inside the matrix integral. Otherwise matrices will contribute that correspond to functions with discontinuities over scales of order $1/N$, and the matrix action evaluated on such functions is not the commutative action. In fact, even in `classical' limits such as the large $\Omega$ limit  it can be necessary to further smooth the functions that appear in order to obtain a continuum theory \cite{Frenkel:2023aft}. The following subsection identifies the limit (\ref{eq:masslimit}) that controls the nonlocality induced by matrix interactions.

\subsection{Validity of perturbation theory}

The large $N$ action (\ref{eq:sbfinal}) is quadratic. Away from the strict large $N$ limit the $[a_a,a_b]_\star$ term in \eqref{eq:Ffield} is nonvanishing and introduces interactions between the fields. In this subsection we will estimate the effects of these interactions. The dominant contributions to the partition function come from the high momentum modes of the fields. In particular, consider modes with momenta that are large compared to the radius of the sphere but not so large that UV/IR mixing effects become important. For these modes, for convenience, approximate the theory by a noncommutative theory on the plane rather than the sphere. In this limit, see e.g.~\cite{Minwalla:1999px},
\be
[a_a,a_b]_\star = \frac{i}{N} \epsilon^{ij} \pa_i a_a \pa_j a_b + \cdots \,.
\ee
Where $\cdots$ refers to higher derivative terms. To move all factors of $N$ to the front of the action, it is convenient to define $a_a \equiv N \bar a_a$. The field theory partition function then takes the schematic form
\be
Z_B \sim \int {\mathcal D} \bar a \exp{-\Omega^4 N^3 \int d^2x \left[ \left(\pa \bar a \right)^2 + \left(\pa \bar a \right)^3 + \left(\pa \bar a \right)^4 + \cdots \right]} \,.
\ee
We may note that the higher derivative terms here are of the same type as those that appear in the DBI action --- for each term the number of derivatives and power of the field are the same.

The partition function can be written in terms of the Fourier transformed fields $\bar a_k$ as
\be
Z_B \sim \int {\mathcal D} \bar a_k \exp{-\Omega^4 N^3 \left[ \int d^2k \left(k \bar a_k \right)^2 + \left(\int d^2k\right)^2 \left(k \bar a_k \right)^3 + \left(\int d^2k\right)^3 \left(k \bar a_k \right)^4 + \cdots \right]} \,.
\ee
Suppose that the first, Gaussian, term dominates. Then the typical magnitude of a mode is
\be
k \bar a_k \sim \frac{1}{\Omega^2 N^{3/2}} \,.
\ee
This scaling may now be used to obtain the magnitude of the higher order terms in the partition function. Recalling that the sum over momenta is cut off at $k \sim N$ we obtain
\begin{align}
Z_B & \sim \int {\mathcal D} \bar a_k \exp{-\Omega^4 N^3 \left[ \frac{N^2}{\Omega^4 N^3} + \frac{N^4}{\Omega^6 N^{9/2}} + \frac{N^6}{\Omega^8 N^6} + \cdots \right]} \\
& \sim \int {\mathcal D} \bar a_k \exp{- N^2 \left[ 1 + \frac{N^{1/2}}{\Omega^2} + \frac{N}{\Omega^4} + \cdots \right]} \,.
\end{align}
In the final line here we see that the higher order terms can be neglected if $N/\Omega^4 \ll 1$,
as advertised in (\ref{eq:masslimit}).

The argument we have just given can also be formulated directly in terms of matrices, using matrix spherical harmonics. The steps of the argument are in precise correspondence. The field theory argument above is physically more transparent and is helpful for identifying the corresponding effect in the probe $D$-brane picture.

\section{Moduli space of localisation}\label{app:moduli}

In this appendix we will solve the conditions that $\delta_\epsilon \theta_\alpha = 0$ to obtain the moduli space (\ref{moduliformal}). Recall that the supercharge is generated by 
\begin{equation}
    \epsilon  = 
    \begin{pmatrix}
        0 & \ldots & 0 & i & 1
    \end{pmatrix}^T \,.
\end{equation}
We introduce the linear combinations of the bosonic matrices,
\begin{equation}
A_1 = X_1 + i X_6, \ \  A_2 = X_2 - i X_5, \ \ A_3 = X_3 - i X_4,\ \ A_4 = X_7 + i X_8, \ \ \phi = X_9 + i X_{10} \,,
\end{equation}
and their hermitian conjugates. We also rearrange the components of the fermion. Setting
\begin{equation}
    \theta = \begin{pmatrix}
        \psi_1 & \ldots & \psi_8 & \chi_1 & \ldots & \chi_8
    \end{pmatrix}^T,
\end{equation}
we define the linear combinations 
\begin{equation}
        \lambda_1 = \psi_1 + i \psi_6, \ \ \lambda_2 = \psi_2 - i \psi_5, \ \ \lambda_3 = \psi_3 - i \psi_4, \ \ \lambda_4 = \psi_7 + i \psi_8,
\end{equation}
\begin{equation}
        \xi_1 = \chi_1 + i \chi_6, \ \ \xi_2 = \chi_2 - i \chi_5, \ \ \xi_3 = \chi_3 - i \chi_4, \ \ \xi_4 = \chi_7 + i \chi_8. 
\end{equation}
The auxiliary fields are also re-organized as 
\begin{equation}
    h_1 = H_1 + i H_6, \ \ h_2 = H_2 - i H_5, \ \ h_3 = H_3 - i H_4. 
\end{equation}
The definition of $\phi, \phi^\dagger$ and the decomposition of the fermion components here are reminiscent of \cite{Moore:1998et}. With these variables the non-vanishing supersymmetry transformations (\ref{eq:newsusy}) of fermionic fields under our supercharge are 
\begin{equation}
    \begin{aligned}
        \delta_\epsilon \lambda_i &= 2 [\phi, A_i], \\ \delta_\epsilon \lambda_4 & = 2 [\phi,A_4] + \frac{3 i \Omega}{4} \phi, \\
        \delta_\epsilon \xi_i &= \epsilon_{ijk} [A_j, A_k]^\dagger + 4 h_i + \frac{\Omega}{4} A_i,\\
        \delta_\epsilon \xi_i^\dagger &= \epsilon_{ijk} [A_j, A_k],\\
        \delta_\epsilon \xi_4 &= - [\phi, \phi^\dagger] - 4 i H_7 - \frac{i \Omega}{4} ( 2 A_4 - A_4^\dagger), 
    \end{aligned}
\end{equation}
where $i,j,k = 1,\ldots,3$. In the above equations the $X,H,\psi,\chi$ matrices treated as hermitian when taking the hermitian conjugates that appear. Setting both the hermitian and antihermitian parts of these transformations to zero leads to the following equations for the moduli space
\begin{equation}
\begin{aligned}
    [\phi, A_i] = 0, \quad  [A_i, A_j] = 0&, \quad h_i = - \frac{\Omega}{16} A_i, \quad H_7 = - \frac{\Omega}{16} X_7 \,, \\ 
    [\phi, \phi^\dagger] = \frac{3 \Omega}{4} X_8&, \quad [\phi, A_4] = - \frac{3 i \Omega}{8} \phi \,.
\end{aligned}
\end{equation}
The last two equations are written in the original variables as 
\begin{equation}\label{fsph}
\begin{aligned}
    [X_9, X_7] - [X_{10},X_8] &= \frac{3 i \Omega}{8} X_9,  \\
    [X_{10}, X_7] + [X_9, X_8] &= \frac{3 i \Omega}{8} X_{10},  \\
    [X_9, X_{10}] &= \frac{3 i \Omega}{8} X_8. 
\end{aligned}
\end{equation}
These can be further simplified using the hermiticity of the matrices. Multiplying through the first equation with $[X_9, X_7]$ and the second by $[X_{10}, X_7]$, and taking the trace gives
\begin{equation}
    \begin{aligned}
       \text{Tr} \left( [X_9, X_7]^2 \right) = \text{Tr} \left( [ X_{9}, X_7] [X_{10}, X_8] \right) \, , \\
       \text{Tr} \left( [X_{10}, X_7]^2 \right) = -\text{Tr} \left( [ X_{10}, X_7] [X_{9}, X_8] \right) \,. \\
    \end{aligned}
\end{equation}
Summing these equations and using the Jacobi identity gives
\begin{equation}
    \text{Tr} \left( [X_9, X_7]^2 + [X_{10},X_7]^2 \right) = 0 \,.
\end{equation}
Since all $X$'s are hermitian, this requires
\begin{equation}
    [X_9, X_7] = [X_{10}, X_7] = 0 \,.
\end{equation}
With these equations and the third line of \eqref{fsph}, one can further show that $X_7$ commutes with $X_8$. Hence \eqref{fsph} becomes a fuzzy sphere equation and $X_7$ is seen to commute with the sphere:
\begin{equation}
    [X_7, X_a] = 0, \qquad [X_a, X_b] = \frac{3 i \Omega}{8} \epsilon_{abc} X_c, \qquad a,b,c = 8,\ldots,10 \,. 
\end{equation}
Thus we have the moduli space equations given in \eqref{moduli}:
\begin{equation}
\begin{aligned}
    [A_i, A_j] = 0, \ \  [\phi, A_i] = 0,& \hspace{15pt} i,j \in \{1,2,3\} \,, \\
    [X_7, X_a] = 0, \ \ [X_a, X_b] = i \frac{3 \Omega}{8} \epsilon_{abc} X_c,&   \hspace{15pt} a,b,c \in \{8,9,10\} \,, \\
    H_A = -\frac{\Omega}{16} X_A,& \hspace{15pt} A = 1,\ldots,7 \,.
\end{aligned}
\end{equation}

\section{Gauge fixing and normalisation factors for localisation}\label{app:factors}

\subsection{Gauge fixing}

Firstly, we will gauge fix the perturbations described in \S\ref{sec:locresult}.
As always, a convenient basis for the modes are the hermitian matrix spherical harmonics (\ref{hermharm}):
\begin{equation}
    \delta X_\mu = \sum_{\pm;\,l,m} \delta x_\mu^{\pm;lm} \hat{H}^{\pm}_{lm}, \qquad \delta H_A = \sum_{\pm;\,l,m} \delta h_A^{\pm;lm} \hat{H}^{\pm}_{lm}, \qquad \delta x_\mu^{\pm;lm}, \,\delta h_A^{\pm;lm} \in \mathbb{R} \,.
\end{equation}
Removing fluctuations on the moduli space corresponds to setting
\begin{equation}
    \delta x_{1,2,3}^{\pm;\,ll} = 0 \,.
\end{equation}
Gauge fixing to diagonal $X_8$ fixes
\begin{equation}\label{X8diagfluc}
    \delta X_8^{\pm;\, lm} = 0, \hspace{15pt} \text{if} \ m \neq 0 \,,
\end{equation}
since only $\hat{H}^{+}_{l0}$ are diagonal. 

To fix the residual $U(1)^{N-1}$ gauge freedom, we consider the action of this symmetry on background values of $X_{9,10}$. For $V_l = \text{exp}(i \alpha^l H^+_{l0}) \in U(1)^{N-1}$, infinitesimally we have
\begin{equation}\label{inftu1}
    \delta X_{9,10} = i \alpha^l [\hat{H}^+_{l0}, X_{9,10}] = -i \frac{3 \Omega}{8} \alpha^l [ J_{1,2}, \hat{H}^+_{l0}] \,.
\end{equation}
Using the properties
\begin{equation}
    [J_1, \hat{H}^{+}_{l0}] = -i  \sqrt{\frac{l(l+1)}{2}} \hat{H}^{-}_{l1}, \qquad  [J_2, \hat{H}^{+}_{l0}] = -i  \sqrt{\frac{l(l+1)}{2}} \hat{H}^{+}_{l1} \,,
\end{equation}\label{inftu1str}
we find that the residual gauge transformations infinitesimally act as
\begin{equation}\label{eq:str}
    \delta X_{9} = - \frac{3 \Omega}{8} \alpha^l  \sqrt{\frac{l(l+1)}{2}} \hat{H}^{-}_{l1}, \qquad \delta X_{10} = - \frac{3 \Omega}{8} \alpha^l \sqrt{\frac{l(l+1)}{2}} \hat{H}^{+}_{l1} \,.
\end{equation}
The zero-modes corresponding to these residual gauge transformations can therefore be fixed by setting
\begin{equation}\label{fixu1}
    \delta x_9^{-;\, l1} = 0 \,.
\end{equation}

\subsection{Normalisation factors}\label{app:normalization}

As for the perturbative calculation in \S\ref{semicl}, the original integration measure is normalised with respect to an orthonormal basis of hermitian traceless matrices, with $\text{Tr}(T^a T^b) = 2 \delta^{ab} $. We now list the various factors that have to appear in the localisation result.

\paragraph{Auxiliary fields} For closure of off-shell supersymmetry, we introduced auxiliary fields $D_A$ and a term $2 \text{Tr}(D_A^2)$ into the action. We have an additional integral
\begin{equation}
    \int dD_A \, e^{- 2 \text{Tr}(D_A^2)} = \left(\int_{-\infty}^{\infty} dD e^{- 4 \text{Tr}(D^2)}\right)^{7(N^2-1)} = \left( \frac{\sqrt{\pi}}{2} \right)^{7(N^2-1)},
\end{equation}
which should be normalized to unity. Thus we should divide the localisation result by this factor. 

\paragraph{Changing to $\hat{H}^{\pm}_{lm}$ basis} The Gaussian integrals are performed in 
$\hat{H}^{\pm}_{lm}$ basis. These are normalised according to (\ref{7norm}), which differs from the $T^a$ basis by a factor of $N/2$. Changing to this basis therefore results in a Jacobian factor
\begin{equation}
    \text{Jacobian} = \left(\frac{N}{2}\right)^{(N^2-1)/2} \,.
\end{equation}
The localisation result should be multiplied by this expression.

\paragraph{Diagonalisation of $X_8$} Fixing to the gauge where $X_8$ is diagonal introduces a Vandermonde determinant and a gauge volume. This gauge volume corresponds to that of the $SU(N)/U(1)^{N-1}$ quotient. This volume does not depend on the action, as long as the symmetry remains the same. A simple way to find this volume is therefore to consider a single Gaussian matrix integral and equate the gauge-fixed result with the ungauged:
\begin{equation}\label{Ndiag}
    \mathcal{N}_{\text{diag}} \int_{-\infty}^{\infty} d\lambda_1 \ldots d\lambda_{N}\, \delta(\sum_{i=1}^N \lambda_i)\, \prod_{i<j} (\lambda_i - \lambda_j)^2 \, e^{- \sum_{i=1}^N \lambda_i^2} \equiv \int dX^{\pm}_{lm} e^{- N \sum_{\pm;\, l,m} (X^{\pm}_{lm})^2}, 
\end{equation}
where $\mathcal{N}_{\text{diag}}$ is the normalisation we wish to calculate, and $X^{\pm}_{lm}$ are the components of an $N \times N$ hermitian traceless matrix in the $\hat{H}^{\pm}_{lm}$ basis. The $\delta$-function on the left imposes tracelessness. Fourier transforming the $\delta$-function, we can re-write the left hand side as a Selberg integral:
\begin{equation}
\begin{aligned}
    \mathcal{N}_{\text{diag}}\int_{-\infty}^{\infty} \frac{d\omega}{2\pi} \int_{-\infty}^{\infty} &d\lambda_1 \ldots d\lambda_{N}\, \prod_{i<j} (\lambda_i - \lambda_j)^2 \, e^{- \sum_{i=1}^N \left(\lambda_i^2 + i \omega \lambda_i\right)} \\
    &= \mathcal{N}_{\text{diag}}\int_{-\infty}^{\infty} \frac{d\omega}{2\pi} e^{- N \omega^2/4} \, \int_{-\infty}^{\infty} d\Tilde{\lambda}_1 \ldots d\Tilde{\lambda}_{N}\, \prod_{i<j} (\Tilde{\lambda}_i - \Tilde{\lambda}_j)^2 \, e^{- \sum_{i=1}^N \Tilde{\lambda}_i^2} \\
    &= \mathcal{N}_{\text{diag}} \, 2^{-N(N-1)/2} \, \pi^{(N-1)/2} \, \frac{G(N+1)}{\sqrt{N}}, 
\end{aligned}
\end{equation}
where $G$ is the Barnes $G$-function. The right hand side of \eqref{Ndiag} is just $N^2-1$ Gaussian integrals. Thus we have
\begin{equation}
    \mathcal{N}_{\text{diag}} = \frac{N^{1-N^2/2} (2 \pi)^{N(N-1)/2}}{G(N+2)} \,.
\end{equation}
The localisation result should be multiplied by this expression.

\paragraph{Saddle points of $X_8$} The Vandermonde in the previous paragraph
was used in a simple Gaussian integral in order to obtain the correct measure factor. In our actual integral, the Vandermonde determinant should be evaluated at the saddle points of localisation. We are concerned in this paper with the saddle point corresponding to $X_8 = 3\Omega/8\, J_3$. For an $N$-dimensional irreducible representation we have, up to $N!$ permutations,
\begin{equation}
    J_3 = \text{diag}\left( \frac{N-1}{2}, \frac{N-1}{2}-1, \ldots, -\frac{N-1}{2}+1, -\frac{N-1}{2} \right). 
\end{equation}
The Vandermonde determinant evaluated on this saddle then becomes
\begin{equation}
    \prod_{i<j} (\lambda_i - \lambda_j)^2 = \left(\frac{3 \Omega}{8} \right)^{N(N-1)} \prod_{l=1}^{N-1} l^{2(N-l)} \,.
\end{equation}
The localisation result should be multiplied by this expression, along with a factor of $N!$ for the permutations.

\paragraph{Diagonal fluctuations of $X_8$} The integral over the eigenvalues of $X_8$ can be conveniently done using the $\hat H^{+}_{l0}$ matrices, see  \eqref{X8diagfluc}. Integrating over the coefficients $\delta X_8^{+;\,l0}$ differs from integrating over the eigenvalues $\lambda_{1,\ldots,N-1}$ by a Jacobian
\begin{equation}
    \left| \text{det}_{il} \left(\frac{\partial \lambda_i}{\partial \delta X_8^{+;\,l0}}\right) \right| = N^{N/2-1}. 
\end{equation}
The localisation result should be multiplied by this expression.

\paragraph{Moduli space integral} The moduli space parameters $u^l$ and $v^l$, as in \eqref{moduliparm}, differ from the $H^{\pm}_{lm}$ basis coefficients by factors of $1/\sqrt{2}$. This amounts to a Jacobian of $(1/2)^{3(N-1)}$ that should be multiplied with the result.

\paragraph{Residual $U(1)^{N-1}/\mathbb{Z}^N$ freedom} The $U(1)^{N-1}$ factor is generated by the Cartan subalgebra of $SU(N)$, which is spanned by $\hat{H}^{+}_{l0}$ as described in \eqref{inftu1}. We have seen above that integrating over this volume corresponds to integrating over $\delta X_9^{-;\,l1}$. Equivalently one can integrate over $\alpha^l$ in \eqref{inftu1}, with a stretch factor of $\frac{3\Omega}{8} \, \sqrt{\frac{1}{2}l(l+1)}$ from (\ref{eq:str}). To find the range of integration, we look at how $J_1$ transforms under a finite transformation
\begin{equation}
    J_1 \to e^{-i \alpha^l \hat{H}^+_{l0}}\, J_1 \,e^{i \alpha^l \hat{H}^+_{l0}} \,.
\end{equation}
One can compute the transformed $J_1$ explicitly. It is given by the original $J_1$ with each component rotated by phases given by a sum of $\alpha^l$ with some coefficients. These phases going from $0$ to $2 \pi$ parametrise distinct transformed $J_a$. Consider a change of integration variables from $\alpha^l$ to these $N-1$ phases, this is seen (from pattern spotting) to require a Jacobian $N^{-N/2}$. The volume of $U(1)^{N-1}$ is then the product of $2\pi$ for each phase, stretch factor $\frac{3\Omega}{8} \, \sqrt{\frac{1}{2}l(l+1)}$ for each $\alpha^l$, and Jacobian $N^{-N/2}$. Computing the volume in this way automatically takes care of the $\mathbb{Z}_N$ identification. The volume is,
\begin{equation}
    \text{vol}\left(U(1)^{N-1}/\mathbb{Z}^N\right) = \left(\frac{3 \Omega}{8}\right)^{N-1} (2 \pi)^{N-1} N^{-N/2} \prod_{l = 1}^{N-1} \sqrt{\frac{l(l+1)}{2}} \,.
\end{equation}

Putting all of the above factors together, and simplifying a little, gives the total normalisation factor (\ref{norm}) in the main text.

\section{Localisation determinant at large mass}\label{locdetapp}

In this appendix we compute the localisation determinant for general $N$ in the large $\Omega$ limit, i.e.~the large mass limit, to obtain (\ref{eq:largeM}) in the main text. 

\subsection*{Bosonic part}

The leading large $\Omega$ contribution to \eqref{bosloc1} is 
\begin{equation}
    (\delta_\epsilon \theta_\alpha)^{(1)} = (\sigma^{a \nu}\epsilon)_\alpha [X^{(0)}_a, \delta X_\nu] + (v_A)_\alpha \left(\delta H_A + \frac{1}{2}[\delta X_A, X_8^{(0)}]\right) + i \frac{\Omega}{8} (T^\mu \epsilon)_\alpha \delta X_\mu \,,
\end{equation}
where as before $a = 8,\ldots,9$, $A = 1,\ldots,7$, and $\mu, \nu = 1,\ldots,10$. Recall that the background fuzzy sphere matrices scale with $\Omega$. Although there is not an explicit $\Omega$ in front of the $\delta H_A$ term, we should still include it since it is the only term containing $\delta H_A$. The localisation action is then seen to decompose as
\begin{equation}\label{eq:split}
    \text{Tr}\left( (\delta_\epsilon \Psi_\alpha)^\dagger \delta_\epsilon \Psi_\alpha \right) = S_6 (\delta X_{1,\ldots,6}, \delta H_{1,\ldots,6}) + S_4 (\delta X_7, \delta H_7, \delta X_{8,9,10}) \,,
\end{equation}
where 
\begin{multline}
    S_6 = \text{Tr} \left( -2 [X^{(0)}_a, \delta X_i] [X_a^{(0)}, \delta X_i] + 2 [X_8^{(0)}, \delta X_i][ X_8^{(0)}, \delta X_i]\right. \\
   \left. - 4 C_{ij7} [X_9^{(0)}, \delta X_i][X_{10}^{(0)}, \delta X_j]+ 8 \, \delta H_i^2 + \frac{\Omega^2}{32} \delta X_i^2 + \Omega \, \delta H_i \delta X_i \right),
\end{multline}
with $i,j = 1,\ldots,6$, and 
\begin{multline}
    S_4 =  \text{Tr} \left(-2 [X^{(0)}_a, \delta X_7] [X_a^{(0)}, \delta X_7]  + 2 [X_8^{(0)}, \delta X_7][ X_8^{(0)}, \delta X_7]\right. \\
    + 8 \delta H_7^2 + \Omega \delta H_7 \delta X_7 + \frac{\Omega^2}{32} \delta X_7^2 -2 [X^{(0)}_a, \delta X_b] [X_a^{(0)}, \delta X_b]+ 2 \sum_a \left([X_a^{(0)}, \delta X_a] \right)^2\\
    + \frac{3 i \Omega}{2} \epsilon_{abc}[X_a^{(0)}, \delta X_b] \delta X_c \left. + 4 [X_8^{(0)}, \delta X_7][\delta X_{10}, X_9^{(0)}] - 4 [X_8^{(0)}, \delta X_7][\delta X_9, X_{10}^{(0)}] \right). 
\end{multline}

The split in (\ref{eq:split}), allows us to integrate the first six matrices and the final four separately. Using the hermitian spherical harmonic basis, $\hat{H}^{\pm}_{lm}$ in (\ref{hermharm}), the action $S_6$ can be written as
\begin{align}
    S_6 = N & \sum_{\pm; l,m} \left\{ 2 \left(\frac{3 \Omega}{8}\right)^2 \left(l(l+1)-m^2 + \frac{1}{9}\right) \left(\delta X_i^{\pm;lm}\right)^2 \right. \nonumber \\
    & \qquad \qquad \left. + 8 \left(\delta H_i^{\pm;lm}\right)^2  + \Omega \left(\delta H_i^{+;lm} \delta X_i^{+;lm} + \delta H_i^{-;lm} \delta X_i^{-; lm}\right) \right\} \\
    + N & \sum_{l,m>0}\, m\, C_{ij7}  \delta X_i^{-;lm} \delta X_j^{+;lm}.  \nonumber
\end{align}
We observe that there is no mixing between different $l,m$ in the action, so we can find the determinant for each mode separately. The $m=0$ case is special because there is no $\hat{H}^{-}_{l0}$, and the $m=l$ case is special because we have held $\delta x_{1,2,3}^{\pm;\,ll} = 0$, per the discussion in appendix \ref{app:factors}. Thus we obtain:
\begin{equation}
\text{det} = 
    \begin{cases}
    \left(\frac{3}{2}\right)^24 \left[ (l^2 -m^2)((l+1)^2 - m^2) \right]^6 N^{24} \, \Omega^{24} & \quad 0 < m < l\\
    \left(\frac{3}{2}\right)^{12} \left[l(l+1)\right]^6 N^{12} \, \Omega^{12} & \quad m = 0 \\
    2^6 3^{12} l^6 N^{18}\, \Omega^{12} & \quad m = l 
    \end{cases}.
\end{equation}
Together the contribution from these six bosonic matrices is 
\begin{align}
    \text{det}[S_6] = \frac{3^{12N(N-1)}}{2^{6(N-1)(2N-3)}} & \Omega^{12 N(N-1)} N^{6(N-1)(2N+1)} \nonumber \\
& \times \prod_{l=1}^{N-1} \left(\prod_{m=1}^{l-1} \left[(l^2-m^2)((l+1)^2 - m^2))\right]^6 \right) l^{12} (l+1)^6 \,.
\end{align}

To simplify $S_4$, we integrate out $\delta X_7$ and $\delta H_7$ first. We can complete the square for $\delta H_7$ by setting $\delta \tilde{H}_7 = \delta H_7 + \frac{\Omega}{16} \delta X_7$. Then the terms in $S_4$ involving $\delta H_7$ and $\delta X_7$ are re-written as
\begin{align}
    S_4 = \text{Tr} & \left( 8 \delta \tilde{H}_7^2 - 2 [X_a^{(0)}, \delta X_7] [X_a^{(0)}, \delta X_7] + 2 [X_8^{(0)}, \delta X_7]^2 \right. \nonumber \\
    & \left.+ 4 [X_8^{(0)}, \delta X_7] \left( [\delta X_{10}, X_9^{(0)}] - [\delta X_9, X_{10}^{(0)}] \right) \right) \,.
\end{align}
Integrating out $\delta \tilde{H}_7$ and $\delta X_7$ in the $\hat{H}^\pm_{lm}$ basis gives a determinant 
\begin{equation}
    \text{det}_{\delta X_7, \delta H_7} = ( 8 N)^{N^2-1} \prod_{l=1}^{n-1}\prod_{m=-l}^l \left(2N\left(\frac{3 \Omega}{8} \right)^2 (l(l+1) - m^2) \right) \,,
\end{equation}
and also an effective action for the other matrices
\begin{equation}\label{eq:add1}
    S_{\text{eff}} = - \sum_{\pm; l,m} \frac{2}{N} \left( \frac{3\Omega}{8} \right)^2  \left(\text{Tr} \left\{\left([J_3, [J_1, \delta X_{10}]] - [J_3, [J_2, \delta X_{9}]] \right) \hat{H}^{\pm}_{lm}\right\} \right)^2.
\end{equation}

With the additional term (\ref{eq:add1}), the remaining action in the spherical harmonics $\hat{Y}_{lm}$ basis can be written as 
\begin{multline}
    \frac{S[\delta X_8, \delta X_9, \delta X_{10}]}{2 (3 \Omega /8)^2 N} = \sum_{l,m} \left[ (-1)^m \left\{ [(l(l+1)+1] y_8^{lm} y_8^{l(-m)} + [l(l+1)+1)] y_-^{lm} y_+^{lm} \right\} \right.\\
    + \frac{(-1)^m}{4}  \left\{ y_-^{l(m-1} y_-^{l(-m-1)} f^+(l, -m-1) f^+ (l,m-1) \left( 1- \frac{m^2}{l(l+1) - m^2}\right) \right\}\\
    + \frac{(-1)^m}{4} \left\{y_+^{l(m+1} y_+^{l(-m+1)} f^-(l, m+1) f^- (l,-m+1) \left( 1- \frac{m^2}{l(l+1) - m^2}\right) \right\}\\
    + \frac{(-1)^m}{4}  \left\{2 y_-^{l(m-1)} y_+^{l(-m+1)} f^+(l,m-1) f^-(l,-m+1) \left( 1+ \frac{m^2}{l(l+1) - m^2}\right) \right\} \\
    \left. (-1)^m \left\{ m y_-^{lm} y_+^{l(-m)} - f^+(l, m-1) y_8^{l(-m)}y_8^{l(-m)} - f^-(l,m+1) y_+^{l(m+1)}y_8^{l(-m)} \right\} \right],
\end{multline}
where we have defined, for convenience, $\delta X_{\pm} = \delta X_9 \pm i \delta X_{10}$, and $\delta X_a = \sum_{l,m} y_a^{lm} \hat{Y}_{lm}$. We have also defined 
\begin{equation}
    f^{\pm}(l,m) = \sqrt{(l\mp m)(l \pm m + 1)}. 
\end{equation}
With the gauge fixing \eqref{X8diagfluc}, only $y_8^{l0}$ is non-zero. Also, we have $y^{l0}$ = $\delta X^{+;l0}$ since $\hat{H}^{+}_{lm} = \hat{Y}_{lm}$. Hence we integrate out $y_8^{l0}$ first to simplify the action. This contributes a determinant
\begin{equation}
    \text{det}_{\delta X_8} = \prod_{l=1}^{N-1} \left( 2 N \left(\frac{3 \Omega}{8} \right)^2 (l(l+1)+1) \right),
\end{equation}
and a further effective action for $\delta X_9$ and $\delta X_{10}$, 
\begin{equation}
    S_{\text{eff}}[\delta X_9, \delta X_{10}] = - 2 N \left(\frac{3 \Omega}{8} \right)^2 \sum_{l=1}^{N-1} \frac{\left(f^+(l,-1) y_-^{l(-1)} - f^-(l,1) y_+^{l1}\right)^2}{4[l(l+1)+1]}. 
\end{equation}
Since the matrices $X_{9,10}$ are hermitian but the $\hat{Y}_{lm}$ are not, we have the following reality conditions on $y_\pm^{lm}$, 
\begin{equation}\label{reality}
    \Bar{(y_+^{lm})} = (-1)^m y_-^{l(-m)}, \hspace{15pt} -l \leq m \leq l.
\end{equation}
We use this relation to re-write the action as a sum over only $m \geq 0$, and such that only $y_+^{l(m+1)}$ and $y_-^{l(m-1)}$ appear, 
\begin{equation}\label{S910}
\begin{aligned}
     \frac{S[\delta X_9, \delta X_{10}]}{2 N (3\Omega/8)^2} = &\sum_{l}\left[ \frac{l(l+1)}{4} \left(y_-^{l(-1)} + y_+^{l1}\right)^2 - \frac{l(l+1)}{4 [l(l+1)+1]} \left(y_-^{l(-1)} - y_+^{l1}\right)^2 \right] \\
    + &\sum_{l;m \geq 1}^{m=l+1}\left[l(l+1)+m - \frac{1}{2}G^+(l,m) \left( 1 + \frac{m^2}{l(l+1)-m^2} \right)\right] \left|y_-^{l(m-1)}\right|^2 \\
    + &\sum_{l;m \geq 1}^{m=l-1}\left[l(l+1)-m - \frac{1}{2}G^-(l,m) \left( 1 + \frac{m^2}{l(l+1)-m^2} \right)\right] \left|y_+^{l(m+1)}\right|^2\\
    -&\frac{1}{2}\sum_{l;m \geq 1}^{m=l-1}F(l,m) \left( 1 - \frac{m^2}{l(l+1)-m^2} \right) \left(y_-^{l(m-1)} \Bar{y_+^{l(m+1)}} + c.c. \right),
\end{aligned}
\end{equation}
where we defined
\begin{equation}
    \begin{aligned}
    G^\pm(l,m) & = (l \pm m)(l \mp m +1) = f^\mp(l,m)^2 \,, \\
    F(l,m) & = \sqrt{(l^2-m^2((l+1)^2 - m^2)} = f^+(l,m)f^-(l,m) \,.
    \end{aligned}
\end{equation}

The action written as \eqref{S910} has the benefit that each sector of fixed $(l,m)$ decouples from others. It is convenient to separate $y_\pm^{lm}$ into its real and imaginary parts, 
\begin{equation}
\begin{aligned}
    y_\pm^{lm} &= a_\pm^{lm} + i b_\pm^{lm},\hspace{15pt} 0 \leq m \leq l \,. \\ 
\end{aligned}
\end{equation}
The other half, with $m<0$, are related to these by \eqref{reality} and hence they are merely different combinations of $a_\pm^{lm}$ and $b_\pm^{lm}$. The special case of $m=0$ gives $a_+^{l0}$ and $b_+^{l0}$ in terms of $a_-^{l0}$ and $b_-^{l0}$. The way we have written the action (involving only $y_-^{l(m-1)}$ and $y_+^{l(m+1)}$) ensures that $a_+^{l0}$ and $b_+^{l0}$ do not appear. Changing integration variables from $\delta X_{9,10}^{\pm;lm}$ to $a_\pm^{lm}$ and $b_\pm^{lm}$ does not introduce any Jacobian. With this redefinition, we analyze \eqref{S910} in cases of $m=0$, $1 \leq m \leq l-1$, $m=l$, and $m=l+1$. The $m=0$ part is given by
\begin{equation}
    S_{m=0} = \sum_l 2 N \left( \frac{3 \Omega}{8} \right)^2 l(l+1) \left(1 - \frac{1}{l(l+1)+1} \right) a_+^{l1},
\end{equation}
where we note that the $b_+^{l1}$ term vanishes (i.e.~it is a zero mode). This zero mode simply corresponds to the $U(1)^{N-1}$ residual gauge transformations discussed above. This can be seen as 
\begin{equation}
    \delta X_9^{-; l1} = \frac{1}{\sqrt{2}} \left( b_+^{l1} + b_-^{l1} \right), \hspace{15pt} \delta X_{10}^{+; l1} = \frac{1}{\sqrt{2}} \left( b_+^{l1} - b_-^{l1} \right),
\end{equation}
and the previous prescription \eqref{fixu1} corresponds to setting $ b_+^{l1} =- b_-^{l1}$. Thus $\delta X_{10}^{+; l1} =  -\sqrt{2} b_-^{l1}$ which introduces a Jacobian factor of $\sqrt{2}$ for each $l$, i.e. $2^{(N-1)/2}$ in total, that should be multiplied to the integration result. This is equivalent to dividing the determinant by $2^{(N-1)/2}$, since the determinant enters the denominator under a square root. Including this Jacobian factor, we write the determinant from this $m=0$ sector as
\begin{equation}
    \text{det}_{\delta X_{9,10}}^{m=0} = \prod_{l=1}^{N-1} N \left( \frac{3 \Omega}{8} \right)^2 l(l+1) \left(1 - \frac{1}{l(l+1)+1}\right).
\end{equation}
For $ 1 \leq m \leq l$, we have 
\begin{equation}
    S_{1\leq m \leq l-1} = \sum_{l; 1\leq m\leq l-1}
    \begin{pmatrix}
        a_-^{l(m-1)} & a_+^{l(m+1)}
    \end{pmatrix}
    \begin{pmatrix}
        C^+(l,m) & D(l,m)\\
        D(l,m) & C^-(l,m)
    \end{pmatrix}
    \begin{pmatrix}
        a_-^{l(m-1)} \\
        a_+^{l(m+1)}
    \end{pmatrix}
    + a \to b, 
\end{equation}
where 
\begin{equation}
    \begin{aligned}
        &C^\pm(l,m) = l(l+1) \pm m -\frac{1}{2} G^\pm(l,m) \left (1 + \frac{m^2}{l(l+1)-m^2} \right), \\
        &D(l,m) = - \frac{1}{2} F(l,m) \left(1 - \frac{m^2}{l(l+1) - m^2} \right).
    \end{aligned}
\end{equation}
Hence the contribution to the determinant from $1 \leq m \leq l-1$ is
\begin{equation}
    \text{det}^{1 \leq m \leq l-1}_{\delta X_{9,10}} = \prod_{l=1}^{N-1} \prod_{m=1}^{l-1} \left( 2 N \left(\frac{3\Omega}{8} \right)^2 \right)^4  \left[m^2 (l(l+1)-m^2)\right]^2. 
\end{equation}
The other two cases are $m = l, l+1$, and they give
\begin{equation}
    S_{m=l} = l \left(a_-^{l(l-1)}\right)^2 + l \left(b_-^{l(l-1)}\right)^2, \hspace{15pt} S_{m=l+1} = (l+1)^2 \left(a_-^{l(l-1)}\right)^2 + l \left(b_-^{l(l-1)}\right)^2.
\end{equation}
Therefore, they contribute determinants
\begin{equation}
    \text{det}^{m=l,l+1}_{\delta X_{9,10}} = \prod_{l=1}^{N-1} \left( 2 N \left(\frac{3\Omega}{8} \right)^2 \right)^4 l^2 (l+1)^4 \,.
\end{equation}
Putting together the results above, the bosonic determinant at large $\Omega$ is given by
\begin{align}
    \text{det}^\prime[(\delta_\epsilon V)^{(2)}_b] & = \text{det}[S_6] \, \text{det}_{\delta X_7, \delta H_7}\, \text{det}_{\delta X_8} \,\text{det}_{\delta X_{9,10}}^{m=0}\, \text{det}_{\delta X_{9,10}}^{1 \leq m \leq l-1} \,\text{det}_{\delta X_{9,10}}^{ml,l+1} \nonumber \\
    & =2^{-(24N -5)(N-1)}\, 3^{6(N-1)(3N+1)}\, N^{2(N-1)(8N+5)} \,\Omega^{6(N-1)(3N+1)} G(N+2)^4 \\ 
    & \times   \,\Gamma(N)^{15} \,\Gamma(N+1)^9 \,  \prod_{l=1}^{N-1} \prod_{m = 1}^{l-1} \,[l(l+1)-m^2]^4\, [(l+1)^2 - m^2]^6\, (l^2 - m^2)^6 \,. \nonumber 
\end{align}

\subsection*{Fermionic part}

In the large mass limit, \eqref{locferm} and \eqref{fermoriginal} become
\begin{equation}
S_f + t (\delta_\epsilon V)_f^{(2)} = 
    \frac{3\Omega}{8} \theta^T
    \begin{pmatrix}
        (\frac{1}{2} + 2 t)\, \text{Ad}(J_-) & A \ \text{Ad}(J_3) \\
       A^T \ \text{Ad}(J_3)  & - (\frac{1}{2} + 2 t ) \, \text{Ad}(J_+)  
    \end{pmatrix} \theta
        + 
   \theta^T \begin{pmatrix}
        0 & B \\
        -B^T & 0
    \end{pmatrix} \theta,
\end{equation}
where $A$ and $B$ are $8 \times 8$ matrices acting on spinor indices, and $Ad(J)$ the adjoint action of $J$ (i.e. the action is on gauge indices  and hence commutes with $A$ and $B$). Explicitly, we have
\begin{equation}\label{Amat}
    A = \begin{pmatrix}
        \frac{1}{2} \mathbb{1}_{6\times 6} & 0\\
        0 & \begin{pmatrix}
            1/2 + t & - i t\\
            it & 1/2 + t 
        \end{pmatrix}
    \end{pmatrix} \,,
\end{equation}
\begin{equation}\label{Bmat}
    B = \begin{pmatrix}
        (1 + 3t) \delta_{ij} - 3 i t C_{ij7} & 0 \\
        0 & (1+ 6 t) \mathbb{1}_{2 \times 2}
    \end{pmatrix} \,,
\end{equation}
with $i,j = 1,\ldots,6$ as above. 

Similarly to our perturbative analysis, we write the fermion as two $8$-component spinors:
\begin{equation}
    \theta = \begin{pmatrix}
        \psi_a \\
        \lambda_b
    \end{pmatrix},
\end{equation}
where here $a,b = 1,\ldots,8$ rather than previously $8,9,10$ for the three bosons. In contrast to pertubation theory, as in \eqref{fermioneigen}, where there is no interaction between different components of $\psi$ or $\lambda$, here different components have pair-wise interactions. This can be seen in the expressions for $A$ and $B$ in \eqref{Amat} and \eqref{Bmat}. The bottom-right block of $A$ induces interaction between $a,b = (7,8)$, and the $C_{ij7}$ in the top-right block of $B$ induces interaction between $a,b = (1,6), (2,5), (3,4)$. Hence we have four decoupled fermion systems each with four fermion matrices. For example, $\psi_1, \psi_6, \lambda_1, \lambda_6$ interact with each other, but are decoupled from $\psi_2,\psi_5,\lambda_2, \lambda_5$. 

Using the spherical harmonics basis, $ \psi_a = \sum_{lm} \psi_a^{lm} \hat{Y}_{lm}$ and similarly for $\lambda$, we can write the action as 
\begin{align}
      S_f + t (\delta_\epsilon V)_f^{(2)} = \frac{3 \Omega}{8} N \sum_{l; 1 \leq m\leq l} & \left[ (1 + 4t) \left( - f^-(l,m) \psi_a^{lm} \Bar{\psi_a^{l(m-1)}} + f^+(l,m-1) \lambda_a^{l(m-1)} \Bar{\lambda_a^{lm}} \right) \right.\nonumber \\
        &+ 2(m-1) A_{ab} \Bar{\psi_a^{l(m-1)}} \lambda_b^{l(m-1)} - 2m A_{ab} \psi_a^{lm} \Bar{\lambda_b^{lm}} \nonumber \\
        & \left. + \frac{2}{3} B_{ab} \left( \Bar{\psi_a}^{l(m-1)} \lambda_b^{l(m-1)} + \psi_a^{lm} \Bar{\lambda_b^{lm}} \right) \right]\nonumber \\
        + \frac{3 \Omega N}{8} N \sum_{l} &\left( 2l A_{ab} \Bar{\psi_a^{ll}} \lambda_b^{ll}  + \frac{2}{3} B_{ab} \Bar{\psi_a^{ll}} \lambda_b^{ll} \right). \label{fermcomp}
\end{align}
In this way, we have that for each $l,m$ only $\psi_a^{lm}, \Bar{\psi_a}^{l(m-1)}, \lambda^{l(m-1)}, \Bar{\lambda^{lm}} $ appear and are decoupled from fermions with other $l,m$. Here we are treating $\psi^{lm}$ ($\lambda^{lm}$) and $\Bar{\psi^{lm}}$ ($\Bar{\lambda^{lm}}$) with $m>0$ to be independent integration variables. Note that only $\Bar{\psi_a^{l0}}$ and $\lambda_a^{l0}$ appear in the action, which is consistent with them being real and hence not independent from their conjugates. This change of integration variables does not introduce any Jacobian factor. In the end, we only need to compute the determinant for each $l,m$ and each pair of $a,b = (1,6),(2,5),(3,4),(7,8)$. The determinant for $(1,6)$, $(2,5)$, and $(3,4)$ should be the same since they have the same form of interaction from $C_{ij7}$. At the end of the calculation we take the large $t$ limit. This limit can be taken before computing determinants for $m \leq l$, but extra care should be taken for $m=l+1$ which is the last line of \eqref{fermcomp}. Fermionic zero modes of the localising deformation are contained in this sector with $a,b = (1,6),(2,5),(3,4)$. 

For the $a,b = (7,8)$ family, we have the integration result
\begin{equation}
    \text{det}^{f}_{7,8} = t^{2 (N^2-1)} \left( \frac{3 \Omega}{8} N\right)^{2(N^2-1)} \prod_{l=1}^{N-1} 2^4 \, (l+1) \left(\prod_{m=1}^l 2^8 \, \left[l(l+1)-m^2 \right] \, \left[l(l+1) - (m-1)^2\right] \right),
\end{equation}
and for $a,b = (1,6), (2,5), (3,4)$ we have
\begin{equation}
    \text{det}^{f}_{1,6} = t^{(N-1)(2N+1)} \left( \frac{3 \Omega}{8} N\right)^{2(N^2-1)}\prod_{l=1}^{N-1} \left(\frac{8}{3} + 4l \right) \left(\prod_{m=1}^l 2^8 \, (l -m + 1)^2 \, (l+m)^2 \right).
\end{equation}
In total, we have
\begin{multline}
    2^{8(N^2-1)} \text{Pf}[S^{(2)}_f + t (\delta_\epsilon V)^{(2)}_f] = (\text{det}^f_{7,8}) (\text{det}^f_{1,6})^3 \\
    = t^{(N-1)(8N+5)} \, 3^{(N-1)(8N+5)} \, 2^{-2(N-1)(4N+7)} \, \Omega^{8(N^2-1)} \, N^{8(N^2-1)} \\
    \times \prod_{l=1}^{N-1} \left(3l+2\right)^3 (l+1) \left(\prod_{m=1}^l \left[l(l+1) - m^2 \right] \left[l(l+1) - (m-1)^2 \right] \left(l-m+1 \right)^6 \left(l+m \right)^6 \right)\\
    + \mathcal{O}\left(t^{(N-1)(8N+5)-1}\right) \,.
\end{multline}

Combining the fermionic and bosonic part we have, in the large $\Omega$ limit, 
\begin{multline}
    2^{3(N-1)} (2\pi)^{(8N + 5)(N-1)} \lim_{t\to\infty}  \frac{\text{Pf}\left[S^{(2)}_f + t (\delta_\epsilon V)^{(2)}_f\right]}{\sqrt{\text{det}'\left[t (\delta_\epsilon V)^{(2)}_b\right]}}\\
    =2^{-\frac{(N-1)(8N+37)}{2}} \, (2\pi)^{(8N + 5)(N-1)}\, 3^{-(N-2)(N-1)}\, \Omega^{-(N-1)(N-5)}  \frac{N^{3(N-1)} \Gamma(N+1)}{\sqrt{N} G(N+2)^2} \prod_{l=1}^{N-1} (3l+2)^3 .
\end{multline}
This is the result quoted in (\ref{eq:largeM}) in the main text.

\providecommand{\href}[2]{#2}\begingroup\raggedright\endgroup


\begin{thebibliography}{10}

\bibitem{tHooft:1973alw}
G.~'t~Hooft, {{A Planar Diagram Theory for Strong Interactions}},
  \href{http://dx.doi.org/10.1016/0550-3213(74)90154-0}{Nucl. Phys. B {\bf 72},
  461, 1974}.

\bibitem{deWit:1988wri}
B.~de~Wit, J.~Hoppe and H.~Nicolai, {{On the Quantum Mechanics of
  Supermembranes}}, \href{http://dx.doi.org/10.1016/0550-3213(88)90116-2}{Nucl.
  Phys. B {\bf 305}, 545, 1988}.

\bibitem{Banks:1996vh}
T.~Banks, W.~Fischler, S.~H. Shenker and L.~Susskind, {{M theory as a matrix
  model: A Conjecture}},
  \href{http://dx.doi.org/10.1103/PhysRevD.55.5112}{Phys. Rev. D {\bf 55},
  5112--5128, 1997},
  [\href{http://arxiv.org/abs/arXiv:hep-th/9610043}{{arXiv:hep-th/9610043}}].

\bibitem{Maldacena:1997re}
J.~M. Maldacena, {{The Large N limit of superconformal field theories and
  supergravity}}, \href{http://dx.doi.org/10.4310/ATMP.1998.v2.n2.a1}{Adv.
  Theor. Math. Phys. {\bf 2}, 231--252, 1998},
  [\href{http://arxiv.org/abs/arXiv:hep-th/9711200}{{arXiv:hep-th/9711200}}].

\bibitem{Witten:2022xxp}
E.~Witten, {{A note on the canonical formalism for gravity}},
  \href{http://dx.doi.org/10.4310/ATMP.2023.v27.n1.a6}{Adv. Theor. Math. Phys.
  {\bf 27}, 311--380, 2023},
  [\href{http://arxiv.org/abs/arXiv:2212.08270}{{arXiv:2212.08270 [hep-th]}}].

\bibitem{Anderson:2006lqb}
M.~T. Anderson, {{On boundary value problems for Einstein metrics}},
  \href{http://dx.doi.org/10.2140/gt.2008.12.2009}{Geom. Topol. {\bf 12},
  2009--2045, 2008},
  [\href{http://arxiv.org/abs/arXiv:math/0612647}{{arXiv:math/0612647}}].

\bibitem{Witten:2018lgb}
E.~Witten, {{A note on boundary conditions in Euclidean gravity}},
  \href{http://dx.doi.org/10.1142/S0129055X21400043}{Rev. Math. Phys. {\bf 33},
  2140004, 2021},
  [\href{http://arxiv.org/abs/arXiv:1805.11559}{{arXiv:1805.11559 [hep-th]}}].

\bibitem{Anninos:2023epi}
D.~Anninos, D.~A. Galante and C.~Maneerat, {{Gravitational observatories}},
  \href{http://dx.doi.org/10.1007/JHEP12(2023)024}{JHEP {\bf 12}, 024, 2023},
  [\href{http://arxiv.org/abs/arXiv:2310.08648}{{arXiv:2310.08648 [hep-th]}}].

\bibitem{Liu:2024ymn}
X.~Liu, J.~E. Santos and T.~Wiseman, {{New Well-Posed boundary conditions for
  semi-classical Euclidean gravity}},
  \href{http://dx.doi.org/10.1007/JHEP06(2024)044}{JHEP {\bf 06}, 044, 2024},
  [\href{http://arxiv.org/abs/arXiv:2402.04308}{{arXiv:2402.04308 [hep-th]}}].

\bibitem{Ishibashi:1996xs}
N.~Ishibashi, H.~Kawai, Y.~Kitazawa and A.~Tsuchiya, {{A Large N reduced model
  as superstring}},
  \href{http://dx.doi.org/10.1016/S0550-3213(97)00290-3}{Nucl. Phys. B {\bf
  498}, 467--491, 1997},
  [\href{http://arxiv.org/abs/arXiv:hep-th/9612115}{{arXiv:hep-th/9612115}}].

\bibitem{Kazakov:1986hu}
V.~A. Kazakov, {{Ising model on a dynamical planar random lattice: Exact
  solution}}, \href{http://dx.doi.org/10.1016/0375-9601(86)90433-0}{Phys. Lett.
  A {\bf 119}, 140--144, 1986}.

\bibitem{Anninos:2020ccj}
D.~Anninos and B.~M\"uhlmann, {{Notes on matrix models (matrix musings)}},
  \href{http://dx.doi.org/10.1088/1742-5468/aba499}{J. Stat. Mech. {\bf 2008},
  083109, 2020},
  [\href{http://arxiv.org/abs/arXiv:2004.01171}{{arXiv:2004.01171 [hep-th]}}].

\bibitem{Green:1997tn}
M.~B. Green and M.~Gutperle, {{D Particle bound states and the D instanton
  measure}}, \href{http://dx.doi.org/10.1088/1126-6708/1998/01/005}{JHEP {\bf
  01}, 005, 1998},
  [\href{http://arxiv.org/abs/arXiv:hep-th/9711107}{{arXiv:hep-th/9711107}}].

\bibitem{Moore:1998et}
G.~W. Moore, N.~Nekrasov and S.~Shatashvili, {{D particle bound states and
  generalized instantons}},
  \href{http://dx.doi.org/10.1007/s002200050016}{Commun. Math. Phys. {\bf 209},
  77--95, 2000},
  [\href{http://arxiv.org/abs/arXiv:hep-th/9803265}{{arXiv:hep-th/9803265}}].

\bibitem{Krauth:1998xh}
W.~Krauth, H.~Nicolai and M.~Staudacher, {{Monte Carlo approach to M theory}},
  \href{http://dx.doi.org/10.1016/S0370-2693(98)00557-7}{Phys. Lett. B {\bf
  431}, 31--41, 1998},
  [\href{http://arxiv.org/abs/arXiv:hep-th/9803117}{{arXiv:hep-th/9803117}}].

\bibitem{Gibbons:1995vg}
G.~W. Gibbons, M.~B. Green and M.~J. Perry, {{Instantons and seven-branes in
  type IIB superstring theory}},
  \href{http://dx.doi.org/10.1016/0370-2693(95)01565-5}{Phys. Lett. B {\bf
  370}, 37--44, 1996},
  [\href{http://arxiv.org/abs/arXiv:hep-th/9511080}{{arXiv:hep-th/9511080}}].

\bibitem{Bergshoeff:1998ry}
E.~Bergshoeff and K.~Behrndt, {{D - instantons and asymptotic geometries}},
  \href{http://dx.doi.org/10.1088/0264-9381/15/7/002}{Class. Quant. Grav. {\bf
  15}, 1801--1813, 1998},
  [\href{http://arxiv.org/abs/arXiv:hep-th/9803090}{{arXiv:hep-th/9803090}}].

\bibitem{Biggs:2023sqw}
A.~Biggs and J.~Maldacena, {{Scaling similarities and quasinormal modes of D0
  black hole solutions}}, \href{http://dx.doi.org/10.1007/JHEP11(2023)155}{JHEP
  {\bf 11}, 155, 2023},
  [\href{http://arxiv.org/abs/arXiv:2303.09974}{{arXiv:2303.09974 [hep-th]}}].

\bibitem{Bonelli:2002mb}
G.~Bonelli, {{Matrix strings in pp wave backgrounds from deformed
  superYang-Mills theory}},
  \href{http://dx.doi.org/10.1088/1126-6708/2002/08/022}{JHEP {\bf 08}, 022,
  2002},
  [\href{http://arxiv.org/abs/arXiv:hep-th/0205213}{{arXiv:hep-th/0205213}}].

\bibitem{Kumar:2022giw}
A.~Kumar, A.~Joseph and P.~Kumar, {{Complex Langevin Study of Spontaneous
  Symmetry Breaking in IKKT Matrix Model}},
  \href{http://dx.doi.org/10.22323/1.430.0213}{PoS {\bf LATTICE2022}, 213,
  2023}, [\href{http://arxiv.org/abs/arXiv:2209.10494}{{arXiv:2209.10494
  [hep-lat]}}].

\bibitem{Berenstein:2002jq}
D.~E. Berenstein, J.~M. Maldacena and H.~S. Nastase, {{Strings in flat space
  and pp waves from N=4 superYang-Mills}},
  \href{http://dx.doi.org/10.1088/1126-6708/2002/04/013}{JHEP {\bf 04}, 013,
  2002},
  [\href{http://arxiv.org/abs/arXiv:hep-th/0202021}{{arXiv:hep-th/0202021}}].

\bibitem{Vafa:1994tf}
C.~Vafa and E.~Witten, {{A Strong coupling test of S duality}},
  \href{http://dx.doi.org/10.1016/0550-3213(94)90097-3}{Nucl. Phys. B {\bf
  431}, 3--77, 1994},
  [\href{http://arxiv.org/abs/arXiv:hep-th/9408074}{{arXiv:hep-th/9408074}}].

\bibitem{Polchinski:2000uf}
J.~Polchinski and M.~J. Strassler, {{The String dual of a confining
  four-dimensional gauge theory}},  2000,
  [\href{http://arxiv.org/abs/arXiv:hep-th/0003136}{{arXiv:hep-th/0003136}}].

\bibitem{Myers:1999ps}
R.~C. Myers, {{Dielectric branes}},
  \href{http://dx.doi.org/10.1088/1126-6708/1999/12/022}{JHEP {\bf 12}, 022,
  1999},
  [\href{http://arxiv.org/abs/arXiv:hep-th/9910053}{{arXiv:hep-th/9910053}}].

\bibitem{Lin:2004kw}
H.~Lin, {{The Supergravity dual of the BMN matrix model}},
  \href{http://dx.doi.org/10.1088/1126-6708/2004/12/001}{JHEP {\bf 12}, 001,
  2004},
  [\href{http://arxiv.org/abs/arXiv:hep-th/0407250}{{arXiv:hep-th/0407250}}].

\bibitem{Itzhaki:1998dd}
N.~Itzhaki, J.~M. Maldacena, J.~Sonnenschein and S.~Yankielowicz,
  {{Supergravity and the large N limit of theories with sixteen supercharges}},
  \href{http://dx.doi.org/10.1103/PhysRevD.58.046004}{Phys. Rev. D {\bf 58},
  046004, 1998},
  [\href{http://arxiv.org/abs/arXiv:hep-th/9802042}{{arXiv:hep-th/9802042}}].

\bibitem{Pestun:2007rz}
V.~Pestun, {{Localization of gauge theory on a four-sphere and supersymmetric
  Wilson loops}}, \href{http://dx.doi.org/10.1007/s00220-012-1485-0}{Commun.
  Math. Phys. {\bf 313}, 71--129, 2012},
  [\href{http://arxiv.org/abs/arXiv:0712.2824}{{arXiv:0712.2824 [hep-th]}}].

\bibitem{Dasgupta:2002hx}
K.~Dasgupta, M.~M. Sheikh-Jabbari and M.~Van~Raamsdonk, {{Matrix perturbation
  theory for M theory on a PP wave}},
  \href{http://dx.doi.org/10.1088/1126-6708/2002/05/056}{JHEP {\bf 05}, 056,
  2002},
  [\href{http://arxiv.org/abs/arXiv:hep-th/0205185}{{arXiv:hep-th/0205185}}].

\bibitem{Iso:2001mg}
S.~Iso, Y.~Kimura, K.~Tanaka and K.~Wakatsuki, {{Noncommutative gauge theory on
  fuzzy sphere from matrix model}},
  \href{http://dx.doi.org/10.1016/S0550-3213(01)00173-0}{Nucl. Phys. B {\bf
  604}, 121--147, 2001},
  [\href{http://arxiv.org/abs/arXiv:hep-th/0101102}{{arXiv:hep-th/0101102}}].

\bibitem{Azuma:2004zq}
T.~Azuma, S.~Bal, K.~Nagao and J.~Nishimura, {{Nonperturbative studies of fuzzy
  spheres in a matrix model with the Chern-Simons term}},
  \href{http://dx.doi.org/10.1088/1126-6708/2004/05/005}{JHEP {\bf 05}, 005,
  2004},
  [\href{http://arxiv.org/abs/arXiv:hep-th/0401038}{{arXiv:hep-th/0401038}}].

\bibitem{Anagnostopoulos:2005cy}
K.~N. Anagnostopoulos, T.~Azuma, K.~Nagao and J.~Nishimura, {{Impact of
  supersymmetry on the nonperturbative dynamics of fuzzy spheres}},
  \href{http://dx.doi.org/10.1088/1126-6708/2005/09/046}{JHEP {\bf 09}, 046,
  2005},
  [\href{http://arxiv.org/abs/arXiv:hep-th/0506062}{{arXiv:hep-th/0506062}}].

\bibitem{Boya:2003km}
L.~J. Boya, G.~Sudarshan and T.~E. Tilma, {{Volumes of compact manifolds}},
  \href{http://dx.doi.org/10.1016/S0034-4877(03)80038-1}{Rept. Math. Phys. {\bf
  52}, 401--422, 2003},
  [\href{http://arxiv.org/abs/arXiv:math-ph/0210033}{{arXiv:math-ph/0210033}}].

\bibitem{Emparan:1997rt}
R.~Emparan, {{Born-Infeld strings tunneling to D-branes}},
  \href{http://dx.doi.org/10.1016/S0370-2693(98)00107-5}{Phys. Lett. B {\bf
  423}, 71--78, 1998},
  [\href{http://arxiv.org/abs/arXiv:hep-th/9711106}{{arXiv:hep-th/9711106}}].

\bibitem{Bergshoeff:1996ui}
E.~Bergshoeff, M.~de~Roo, M.~B. Green, G.~Papadopoulos and P.~K. Townsend,
  {{Duality of type II 7 branes and 8 branes}},
  \href{http://dx.doi.org/10.1016/0550-3213(96)00171-X}{Nucl. Phys. B {\bf
  470}, 113--135, 1996},
  [\href{http://arxiv.org/abs/arXiv:hep-th/9601150}{{arXiv:hep-th/9601150}}].

\bibitem{Polchinski:1996na}
J.~Polchinski, {{Tasi lectures on D-branes}},  in \emph{{Theoretical Advanced
  Study Institute in Elementary Particle Physics (TASI 96): Fields, Strings,
  and Duality}}, pp.~293--356, 1996.
\newblock
  [\href{http://arxiv.org/abs/arXiv:hep-th/9611050}{{arXiv:hep-th/9611050}}].

\bibitem{Green:1997tv}
M.~B. Green and M.~Gutperle, {{Effects of D instantons}},
  \href{http://dx.doi.org/10.1016/S0550-3213(97)00269-1}{Nucl. Phys. B {\bf
  498}, 195--227, 1997},
  [\href{http://arxiv.org/abs/arXiv:hep-th/9701093}{{arXiv:hep-th/9701093}}].

\bibitem{Berkovits_1993}
N.~Berkovits, {A ten-dimensional super-yang-mills action with off-shell
  supersymmetry}, \href{http://dx.doi.org/10.1016/0370-2693(93)91791-k}{Physics
  Letters B {\bf 318}, 104–106, 1993}.

\bibitem{Pestun:2009nn}
V.~Pestun, {{Localization of the four-dimensional N=4 SYM to a two-sphere and
  1/8 BPS Wilson loops}}, \href{http://dx.doi.org/10.1007/JHEP12(2012)067}{JHEP
  {\bf 12}, 067, 2012},
  [\href{http://arxiv.org/abs/arXiv:0906.0638}{{arXiv:0906.0638 [hep-th]}}].

\bibitem{Asano:2012zt}
Y.~Asano, G.~Ishiki, T.~Okada and S.~Shimasaki, {{Exact results for
  perturbative partition functions of theories with SU(2|4) symmetry}},
  \href{http://dx.doi.org/10.1007/JHEP02(2013)148}{JHEP {\bf 02}, 148, 2013},
  [\href{http://arxiv.org/abs/arXiv:1211.0364}{{arXiv:1211.0364 [hep-th]}}].

\bibitem{Austing:2000rm}
P.~Austing, {{The Cohomological supercharge}},
  \href{http://dx.doi.org/10.1088/1126-6708/2001/01/009}{JHEP {\bf 01}, 009,
  2001},
  [\href{http://arxiv.org/abs/arXiv:hep-th/0011211}{{arXiv:hep-th/0011211}}].

\bibitem{Bobev:2018ugk}
N.~Bobev, P.~Bomans and F.~F. Gautason, {{Spherical Branes}},
  \href{http://dx.doi.org/10.1007/JHEP08(2018)029}{JHEP {\bf 08}, 029, 2018},
  [\href{http://arxiv.org/abs/arXiv:1805.05338}{{arXiv:1805.05338 [hep-th]}}].

\bibitem{Lin:2004nb}
H.~Lin, O.~Lunin and J.~M. Maldacena, {{Bubbling AdS space and 1/2 BPS
  geometries}}, \href{http://dx.doi.org/10.1088/1126-6708/2004/10/025}{JHEP
  {\bf 10}, 025, 2004},
  [\href{http://arxiv.org/abs/arXiv:hep-th/0409174}{{arXiv:hep-th/0409174}}].

\bibitem{Frenkel:2020ysx}
A.~Frenkel, S.~A. Hartnoll, J.~Kruthoff and Z.~D. Shi, {{Holographic flows from
  CFT to the Kasner universe}},
  \href{http://dx.doi.org/10.1007/JHEP08(2020)003}{JHEP {\bf 08}, 003, 2020},
  [\href{http://arxiv.org/abs/arXiv:2004.01192}{{arXiv:2004.01192 [hep-th]}}].

\bibitem{Hartnoll:2022snh}
S.~A. Hartnoll, {{Wheeler-DeWitt states of the AdS-Schwarzschild interior}},
  \href{http://dx.doi.org/10.1007/JHEP01(2023)066}{JHEP {\bf 01}, 066, 2023},
  [\href{http://arxiv.org/abs/arXiv:2208.04348}{{arXiv:2208.04348 [hep-th]}}].

\bibitem{Strominger:2001pn}
A.~Strominger, {{The dS/CFT correspondence}},
  \href{http://dx.doi.org/10.1088/1126-6708/2001/10/034}{JHEP {\bf 10}, 034,
  2001},
  [\href{http://arxiv.org/abs/arXiv:hep-th/0106113}{{arXiv:hep-th/0106113}}].

\bibitem{Anninos:2011ui}
D.~Anninos, T.~Hartman and A.~Strominger, {{Higher Spin Realization of the
  dS/CFT Correspondence}},
  \href{http://dx.doi.org/10.1088/1361-6382/34/1/015009}{Class. Quant. Grav.
  {\bf 34}, 015009, 2017},
  [\href{http://arxiv.org/abs/arXiv:1108.5735}{{arXiv:1108.5735 [hep-th]}}].

\bibitem{Hartle:1983ai}
J.~B. Hartle and S.~W. Hawking, {{Wave Function of the Universe}},
  \href{http://dx.doi.org/10.1103/PhysRevD.28.2960}{Phys. Rev. D {\bf 28},
  2960--2975, 1983}.

\bibitem{DeWitt:1967yk}
B.~S. DeWitt, {{Quantum Theory of Gravity. 1. The Canonical Theory}},
  \href{http://dx.doi.org/10.1103/PhysRev.160.1113}{Phys. Rev. {\bf 160},
  1113--1148, 1967}.

\bibitem{deBoer:1999tgo}
J.~de~Boer, E.~P. Verlinde and H.~L. Verlinde, {{On the holographic
  renormalization group}},
  \href{http://dx.doi.org/10.1088/1126-6708/2000/08/003}{JHEP {\bf 08}, 003,
  2000},
  [\href{http://arxiv.org/abs/arXiv:hep-th/9912012}{{arXiv:hep-th/9912012}}].

\bibitem{Heemskerk:2010hk}
I.~Heemskerk and J.~Polchinski, {{Holographic and Wilsonian Renormalization
  Groups}}, \href{http://dx.doi.org/10.1007/JHEP06(2011)031}{JHEP {\bf 06},
  031, 2011}, [\href{http://arxiv.org/abs/arXiv:1010.1264}{{arXiv:1010.1264
  [hep-th]}}].

\bibitem{Faulkner:2010jy}
T.~Faulkner, H.~Liu and M.~Rangamani, {{Integrating out geometry: Holographic
  Wilsonian RG and the membrane paradigm}},
  \href{http://dx.doi.org/10.1007/JHEP08(2011)051}{JHEP {\bf 08}, 051, 2011},
  [\href{http://arxiv.org/abs/arXiv:1010.4036}{{arXiv:1010.4036 [hep-th]}}].

\bibitem{McGough:2016lol}
L.~McGough, M.~Mezei and H.~Verlinde, {{Moving the CFT into the bulk with $
  T\overline{T} $}}, \href{http://dx.doi.org/10.1007/JHEP04(2018)010}{JHEP {\bf
  04}, 010, 2018},
  [\href{http://arxiv.org/abs/arXiv:1611.03470}{{arXiv:1611.03470 [hep-th]}}].

\bibitem{Hartman:2018tkw}
T.~Hartman, J.~Kruthoff, E.~Shaghoulian and A.~Tajdini, {{Holography at finite
  cutoff with a $T^2$ deformation}},
  \href{http://dx.doi.org/10.1007/JHEP03(2019)004}{JHEP {\bf 03}, 004, 2019},
  [\href{http://arxiv.org/abs/arXiv:1807.11401}{{arXiv:1807.11401 [hep-th]}}].

\bibitem{Araujo-Regado:2022gvw}
G.~Araujo-Regado, R.~Khan and A.~C. Wall, {{Cauchy slice holography: a new
  AdS/CFT dictionary}}, \href{http://dx.doi.org/10.1007/JHEP03(2023)026}{JHEP
  {\bf 03}, 026, 2023},
  [\href{http://arxiv.org/abs/arXiv:2204.00591}{{arXiv:2204.00591 [hep-th]}}].

\bibitem{Kim:2011cr}
S.-W. Kim, J.~Nishimura and A.~Tsuchiya, {{Expanding (3+1)-dimensional universe
  from a Lorentzian matrix model for superstring theory in (9+1)-dimensions}},
  \href{http://dx.doi.org/10.1103/PhysRevLett.108.011601}{Phys. Rev. Lett. {\bf
  108}, 011601, 2012},
  [\href{http://arxiv.org/abs/arXiv:1108.1540}{{arXiv:1108.1540 [hep-th]}}].

\bibitem{Anagnostopoulos:2015gua}
K.~N. Anagnostopoulos, T.~Azuma and J.~Nishimura, {{Monte Carlo studies of
  dynamical compactification of extra dimensions in a model of nonperturbative
  string theory}}, \href{http://dx.doi.org/10.22323/1.251.0307}{PoS {\bf
  LATTICE2015}, 307, 2016},
  [\href{http://arxiv.org/abs/arXiv:1509.05079}{{arXiv:1509.05079 [hep-lat]}}].

\bibitem{Anagnostopoulos:2020xai}
K.~N. Anagnostopoulos, T.~Azuma, Y.~Ito, J.~Nishimura, T.~Okubo and
  S.~Kovalkov~Papadoudis, {{Complex Langevin analysis of the spontaneous
  breaking of 10D rotational symmetry in the Euclidean IKKT matrix model}},
  \href{http://dx.doi.org/10.1007/JHEP06(2020)069}{JHEP {\bf 06}, 069, 2020},
  [\href{http://arxiv.org/abs/arXiv:2002.07410}{{arXiv:2002.07410 [hep-th]}}].

\bibitem{Anagnostopoulos:2022dak}
K.~N. Anagnostopoulos, T.~Azuma, K.~Hatakeyama, M.~Hirasawa, Y.~Ito,
  J.~Nishimura, S.~K. Papadoudis and A.~Tsuchiya, {{Progress in the numerical
  studies of the type IIB matrix model}},
  \href{http://dx.doi.org/10.1140/epjs/s11734-023-00849-x}{Eur. Phys. J. ST
  {\bf 232}, 3681--3695, 2023},
  [\href{http://arxiv.org/abs/arXiv:2210.17537}{{arXiv:2210.17537 [hep-th]}}].

\bibitem{Aoki:1998vn}
H.~Aoki, S.~Iso, H.~Kawai, Y.~Kitazawa and T.~Tada, {{Space-time structures
  from IIB matrix model}}, \href{http://dx.doi.org/10.1143/PTP.99.713}{Prog.
  Theor. Phys. {\bf 99}, 713--746, 1998},
  [\href{http://arxiv.org/abs/arXiv:hep-th/9802085}{{arXiv:hep-th/9802085}}].

\bibitem{Kim:2012mw}
S.-W. Kim, J.~Nishimura and A.~Tsuchiya, {{Late time behaviors of the expanding
  universe in the IIB matrix model}},
  \href{http://dx.doi.org/10.1007/JHEP10(2012)147}{JHEP {\bf 10}, 147, 2012},
  [\href{http://arxiv.org/abs/arXiv:1208.0711}{{arXiv:1208.0711 [hep-th]}}].

\bibitem{Nishimura:2019qal}
J.~Nishimura and A.~Tsuchiya, {{Complex Langevin analysis of the space-time
  structure in the Lorentzian type IIB matrix model}},
  \href{http://dx.doi.org/10.1007/JHEP06(2019)077}{JHEP {\bf 06}, 077, 2019},
  [\href{http://arxiv.org/abs/arXiv:1904.05919}{{arXiv:1904.05919 [hep-th]}}].

\bibitem{Hirasawa:2024dht}
M.~Hirasawa, K.~N. Anagnostopoulos, T.~Azuma, K.~Hatakeyama, J.~Nishimura,
  S.~Papadoudis and A.~Tsuchiya, {{The effects of SUSY on the emergent
  spacetime in the Lorentzian type IIB matrix model}},
  \href{http://dx.doi.org/10.22323/1.463.0257}{PoS {\bf CORFU2023}, 257, 2024},
  [\href{http://arxiv.org/abs/arXiv:2407.03491}{{arXiv:2407.03491 [hep-th]}}].

\bibitem{Brandenberger:2024ddi}
R.~Brandenberger and J.~Pasiecznik, {{On the Origin of the $SO(9) \rightarrow
  SO(3) \times SO(6)$ Symmetry Breaking in the IKKT Matrix Model}},  2024,
  [\href{http://arxiv.org/abs/arXiv:2409.00254}{{arXiv:2409.00254 [hep-th]}}].

\bibitem{Han:2019wue}
X.~Han and S.~A. Hartnoll, {{Deep Quantum Geometry of Matrices}},
  \href{http://dx.doi.org/10.1103/PhysRevX.10.011069}{Phys. Rev. X {\bf 10},
  011069, 2020},
  [\href{http://arxiv.org/abs/arXiv:1906.08781}{{arXiv:1906.08781 [hep-th]}}].

\bibitem{Frenkel:2023aft}
A.~Frenkel and S.~A. Hartnoll, {{Emergent area laws from entangled matrices}},
  \href{http://dx.doi.org/10.1007/JHEP05(2023)084}{JHEP {\bf 05}, 084, 2023},
  [\href{http://arxiv.org/abs/arXiv:2301.01325}{{arXiv:2301.01325 [hep-th]}}].

\bibitem{Marino:1999af}
M.~Marino, R.~Minasian, G.~W. Moore and A.~Strominger, {{Nonlinear instantons
  from supersymmetric p-branes}},
  \href{http://dx.doi.org/10.1088/1126-6708/2000/01/005}{JHEP {\bf 01}, 005,
  2000},
  [\href{http://arxiv.org/abs/arXiv:hep-th/9911206}{{arXiv:hep-th/9911206}}].

\bibitem{Minwalla:1999px}
S.~Minwalla, M.~Van~Raamsdonk and N.~Seiberg, {{Noncommutative perturbative
  dynamics}}, \href{http://dx.doi.org/10.1088/1126-6708/2000/02/020}{JHEP {\bf
  02}, 020, 2000},
  [\href{http://arxiv.org/abs/arXiv:hep-th/9912072}{{arXiv:hep-th/9912072}}].

\end{thebibliography}
\end{document}